\documentclass[twocolumn,trackchanges]{aastex631}

\usepackage{amsmath}

\usepackage{comment}
\usepackage{hyperref}
\usepackage{placeins}
\shorttitle{Nuclear SEDs of (U)LIRGs}
\shortauthors{Gao et al.}
\graphicspath{{./}{figures/}}
\usepackage{subfigure}
\turnoffeditone
\turnoffedittwo

\begin{document}

\title{Nuclear Spectral Energy Distributions of Luminous Infrared Galaxies}

\correspondingauthor{Tianmu Gao}
\email{tianmu.gao@anu.edu.au}

\author[0000-0002-1158-6372]{Tianmu Gao}
\affiliation{Research School of Astronomy and Astrophysics, Australian National University, Weston Creek, ACT 2611, Australia}
\affiliation{ARC Centre of Excellence for All Sky Astrophysics in 3 Dimensions (ASTRO 3D); Australia}

\author[0000-0002-1912-0024]{Vivian U}
\affiliation{IPAC, Caltech, 1200 E. California Blvd., Pasadena, CA 91125, USA}
\affiliation{Department of Physics and Astronomy, 4129 Frederick Reines Hall, University of California, Irvine, CA 92697, USA}

\author[0000-0002-5504-8752]{Connor W. Auge}
\affiliation{Institute for Astronomy, University of Hawaii, 2680 Woodlawn Drive, Honolulu, HI 96822, USA}

\author[0000-0002-3139-3041]{Yiqing Song}
\affiliation{European Southern Observatory, Alonso de Córdova, 3107, Vitacura, Santiago, 763-0355, Chile}
\affiliation{Joint ALMA Observatory, Alonso de Córdova, 3107, Vitacura, Santiago, 763-0355, Chile}

\author[0000-0002-1000-6081]{Sean T. Linden}
\affiliation{Steward Observatory, University of Arizona, 933 N Cherry Avenue, Tucson, AZ 85721, USA}

\author[0000-0002-4923-3281]{Kazushi Iwasawa}
\affiliation{Institut de Ci\`encies del Cosmos (ICCUB), Universitat de Barcelona (IEEC-UB), Mart\'i i Franqu\`es, 1, 08028 Barcelona, Spain}
\affiliation{ICREA, Pg. Llu\'is Companys 23, 08010 Barcelona, Spain}

\author[0000-0003-2196-3298]{Alessandro Peca}
\affiliation{Department of Physics, University of Miami, Coral Gables, FL 33124, USA}

\author[0000-0003-3474-1125]{George C. Privon}
\affiliation{National Radio Astronomy Observatory, 520 Edgemont Rd, Charlottesville, VA, 22903, USA}
\affiliation{Department of Astronomy, University of Virginia, 530 McCormick Road, Charlottesville, VA 22903, USA}
\affiliation{Department of Astronomy, University of Florida, P.O. Box 112055, Gainesville, FL 32611, USA}

\author[0000-0002-1233-9998]{David B. Sanders}
\affiliation{Institute for Astronomy, University of Hawaii, 2680 Woodlawn Drive, Honolulu, HI 96822, USA}

\author[0000-0003-3638-8943]{N\' uria Torres-Alb\`a}
\affiliation{Department of Physics and Astronomy, Clemson University, Kinard Lab of Physics, 140 Delta Epsilon Ct, Clemson, SC 29634, USA}

\author[0000-0003-0057-8892]{Loreto Barcos-Mu\~noz}
\affiliation{National Radio Astronomy Observatory, 520 Edgemont Rd, Charlottesville, VA, 22903, USA}
\affiliation{Department of Astronomy, University of Virginia, 530 McCormick Road, Charlottesville, VA 22903, USA}

\author[0000-0002-8122-3032]{James Agostino}
\affiliation{Department of Physics \& Astronomy and Ritter Astrophysical Research Center, University of Toledo, Toledo, OH 43606, USA}

\author[0000-0001-7421-2944]{Anne M. Medling}
\affiliation{Department of Physics \& Astronomy and Ritter Astrophysical Research Center, University of Toledo, Toledo, OH 43606, USA}
\affiliation{ARC Centre of Excellence for All Sky Astrophysics in 3 Dimensions (ASTRO 3D); Australia}



\begin{abstract}

 We present nuclear (100-150\,pc) spectral energy distributions (SEDs) for a sample of 23 nearby luminous infrared galaxies hosting a total of 28 nuclei. We gather aperture photometry from high-resolution X-ray to submillimeter data for each nuclear region localized by ALMA observations of the dust continuum. We model the  broadband SEDs using X-CIGALE. Binning the merging systems by interaction class, we find that the AGN fraction (fraction of AGN infrared luminosity to total infrared luminosity) appears enhanced in the late- and post-merger stages compared to early-stage mergers. Examining the relationship between X-ray emission and infrared emission of the nuclear regions, we find that the infrared emission in the nucleus is dominated by dust and AGN, with minimal contribution from stars. We also find that nuclear regions have higher X-ray hardness ratios than the host galaxies globally among both the AGN and non-AGN population. We highlight the similarities and differences in the SEDs of dual nuclei in five closely separated late-stage merging systems: Arp 220 ($d_\mathrm{nuc} \sim$\,0.5\,kpc), NGC 6240 ($d_\mathrm{nuc} \sim$\,1\,kpc), IRAS 07251$-$0248 ($d_\mathrm{nuc} \sim$\,2\,kpc), IRAS F12112+0305 ($d_\mathrm{nuc} \sim$\,4\,kpc), and IRAS F14348+1447 ($d_\mathrm{nuc} \sim$\,6\,kpc). The SEDs for these resolved pairs are distinct, suggesting that the AGN state is much more susceptible to the stellar and dust content within the immediate circumnuclear ($<$150 pc) environment than to the host's global infrared luminosity or merger stage. 
\end{abstract}

\keywords{galaxies: active – galaxies: interactions – galaxies: nuclei – galaxies: photometry – infrared: galaxies}


\section{Introduction} \label{sec:intro}

Much of a supermassive black hole's (SMBH) growth takes place in an obscured state, such as that inside of the dusty nuclei of merging gas-rich galaxies \citep[e.g.][]{1988Sanders,1991Barnes,2005DiMatteo,2006Hopkins}. During the collision event between the progenitor galaxies, a fraction of the interstellar gas is funneled toward the center of the system. The infall of a massive gas reservoir onto the growing SMBH can trigger the active galactic nucleus (AGN) phase \citep{1988Sanders, 2005Springel, 2006Hopkins, 2007Sijacki}. However, the presence of dust and nearby star clusters confounded with the AGN duty cycle could partially account for the difficulty of AGN detection in single-epoch observations or in wavelengths susceptible to extinction \citep[e.g.][]{1989Armus, 2012Mazzarella, 2017RamosAlmeida, 2017Ricci, 2022Ub}. Many AGN may either be heavily obscured, or their emission may be contaminated with surrounding star formation activity. To document the growth pathway of SMBHs in merging galaxies would therefore prompt a resolved study of their nuclear regions using multiwavelength sub-arcsecond data sets.

Thankfully, adaptive optics (AO) on large ground-based optical and near-infrared telescopes as well as submillimeter interferometry arrays, have pushed the angular resolution limit ($\lesssim$\,0\farcs03) to match or improve the resolving capability of \emph{Hubble Space Telescope (HST)}, enabling such resolved studies at $\sim$100\,pc or better for nearby galaxies \citep[e.g.][]{2014Medling, 2020Treister}. The power of AO-assisted observations in resolving remnant nuclei in galaxy mergers down to sub-kiloparsec scales has been demonstrated in a number of cases \citep{2007Max, 2011Muller, 2011Medling, 2013U,2015Mcgurk,2018Koss, 2021Crespo}. The high-resolution capability, particularly in the near-infrared and submillimeter wavelength regimes where photons can bypass heavy dust obscuration, is critical for locating the true nucleus and isolating its luminosity contribution from that of surrounding star formation~\citep{2017BarcosMunoz,2021Song,2022Song}. 

Hard X-ray observations present another important piece of information that constrains the radiation from the putative AGN. Despite the relative limitation in angular resolution afforded by state-of-the-art X-ray facilities, high-energy photons sample light from the corona around the accretion disk and constrain the line-of-sight column density. The combination of X-ray observations and high-resolution optical to submillimeter data sets is particularly useful for generating resolved spectral energy distributions (SEDs) and complementing studies of AGN bolometric luminosity, radiation mechanisms, and intrinsic structure \citep{1986Edelson,1987Ward,1994Elvis,2006Buchanan,2014Harrison}. 

A prime example of how small-scale (FWHM $\sim$ 0\farcs2 in $1-5$\,$\mu$m and $<0\farcs5$ in $11-20$\,$\mu$m) SEDs can provide vital insights about AGN is the \cite{2010Prieto} study of 10 local Seyferts, which computed AGN bolometric luminosities with two orders of magnitude more precision than those obtained from lower-resolution work. The resulting SED templates enabled a no-contamination distinction of the two Seyfert types within the nuclear regions: Type 2 AGN exhibit a sharp drop shortward of 2\,$\rm\mu m$, with the optical emission from the nuclei being fully obscured; Type 1s instead show a gentle 2\,$\rm\mu m$ drop along with a partially absorbed optical to UV bump at about 1\,$\rm\mu m$. Furthermore, the study shows that the shape of the high-resolution nuclear AGN SED is largely distinct from those derived from large-aperture data that include emission from the circumnuclear region. Extending this multiwavelength study to better data (factor of 4$-$5 enhancement in resolution at $\lambda > 1\,\mu$m) and a larger, carefully chosen sample of interacting galaxies harboring growing SMBHs will enlighten the detailed perceived growth of SMBH pairs and the temporal evolution of the nuclear environments in mergers.

To address the questions of when and how AGN get triggered via the merger mechanism, we target local luminous and ultraluminous infrared galaxies ((U)LIRGs: $L_{\rm IR}\ge 10^{11}L_{\odot}$), most of which appear to be consequences of mergers between gas-rich spirals and display high levels of star formation and AGN activities \citep[e.g.][]{1996Sanders,2009Armus}. The re-emission from the dust heated by starbursts and/or AGN is a key process that is responsible for the high infrared-luminosity of (U)LIRGs.

X-ray studies of (U)LIRGs indicate that both the AGN fraction and the amount of materials around SMBHs increase as the merger sequence advances. Analyzing \emph{Chandra} observations, \cite{2011Iwasawa} found AGN signatures in 16 of 44 ($38\pm 7\%$) local (U)LIRGs ($L_{\rm IR} \ge 10^{11.7-12.5}L_{\odot}$) using X-ray color and the detection of the 6.4 keV Fe K$\alpha$ line. \edit1{A fluorescent Fe K$\alpha$ line, accompanied by a Compton reflection component peaking at 20-30 keV, is produced when X-ray continuum emission is reflected by relatively cold circumnuclear material, such as dust torus or the outer regions of an accretion disk. Therefore, the detection of a strong Fe K$\alpha$ line at 6.4 keV is often used as a diagnostic of heavily obscured AGN.} Subsequently, \cite{2018Torres} showed that less-luminous (U)LIRGs with $L_{\rm IR} \ge 10^{11.0-11.7}L_{\odot}$ host fewer AGN ($31\pm 5\%$), and quantified the rarity ($29\pm 14\%$) of double AGN in systems that host at least one AGN. More recently, \cite{2021Ricci} combined data from the \emph{NuSTAR}, \emph{Chandra}, and \emph{XMM-Newton} X-ray telescopes and found that the fraction of Compton-thick (CT, $N_{\rm H}\ge 10^{24}\rm cm^{-2}$) AGN in late-stage mergers ($69^{+8}_{-9}\%$) is higher than that found among early mergers ($33 \pm 12\%$) or local hard X-ray-selected sources ($27 \pm 4\%$). These studies confirm that AGN are more often triggered in the final merger stages, when much of the gas has been transported inward to fuel the central kiloparsec region. Thus, (U)LIRGs are the optimal objects to study the most rapidly growing and obscured phase of SMBHs.

The Great Observatories All-sky LIRGs Survey \citep[GOALS;][]{2009Armus} comprises of 21 ULIRGs and 180 LIRGs, forming a large, statistically complete sample of infrared-luminous galaxies with $L_{\rm IR}\ge 10^{11-12.5}L_{\odot}$ in the local universe. It features an extensive X-ray to radio data repository from both space- and ground-based telescopes --- ideal for conducting detailed, multiwavelength studies. With the primary objective of charting the maturation of SMBHs through quiescent and obscured states in galaxy mergers, here we select 23 objects from the GOALS sample and generate highly-resolved ($r$ = 100$-$150 pc), broadband SEDs of the precisely located galactic nuclei, using sub-arcsecond resolution archival data from ALMA, Keck, \emph{HST}, and \emph{Chandra}.

This paper is organized as follows. The sample and multiwavelength data sets are introduced and described in \S \ref{sec:sample} and \S\ref{sec:data}, respectively. In \S \ref{sec:analysis}, we illustrate the aperture photometry extraction and SED fitting processes. Features of our nuclear SEDs and the relationship between several derived properties are assessed in \S \ref{sec:discussion}. Finally, our conclusions are summarized in \S \ref{sec:conclusion}. Throughout this paper, we adopt \edit1{$H_{0}=70.4\,\rm km\,s^{-1}\,Mpc^{-1}$, $\Omega_{\rm m}=0.272$ and $\Omega_{\Lambda}=0.728$.}

\section{Sample} \label{sec:sample}

In order to generate resolved, nuclear SEDs of galaxy mergers with sufficient multiwavelength data points, we employed a selection criterion based on the availability of high-resolution archival data ($<$\,$0\farcs5$/pixel) across the electromagnetic spectrum within the GOALS sample. Specifically, we focused on the subset of sources  with, at minimum, high quality data from \emph{Chandra}, \emph{HST}/ACS, Keck/NIRC2, and ALMA for X-ray to submillimeter wavelength coverage. 

\begin{figure}[htbp]
	\centering
  	\includegraphics[width=0.45\textwidth]{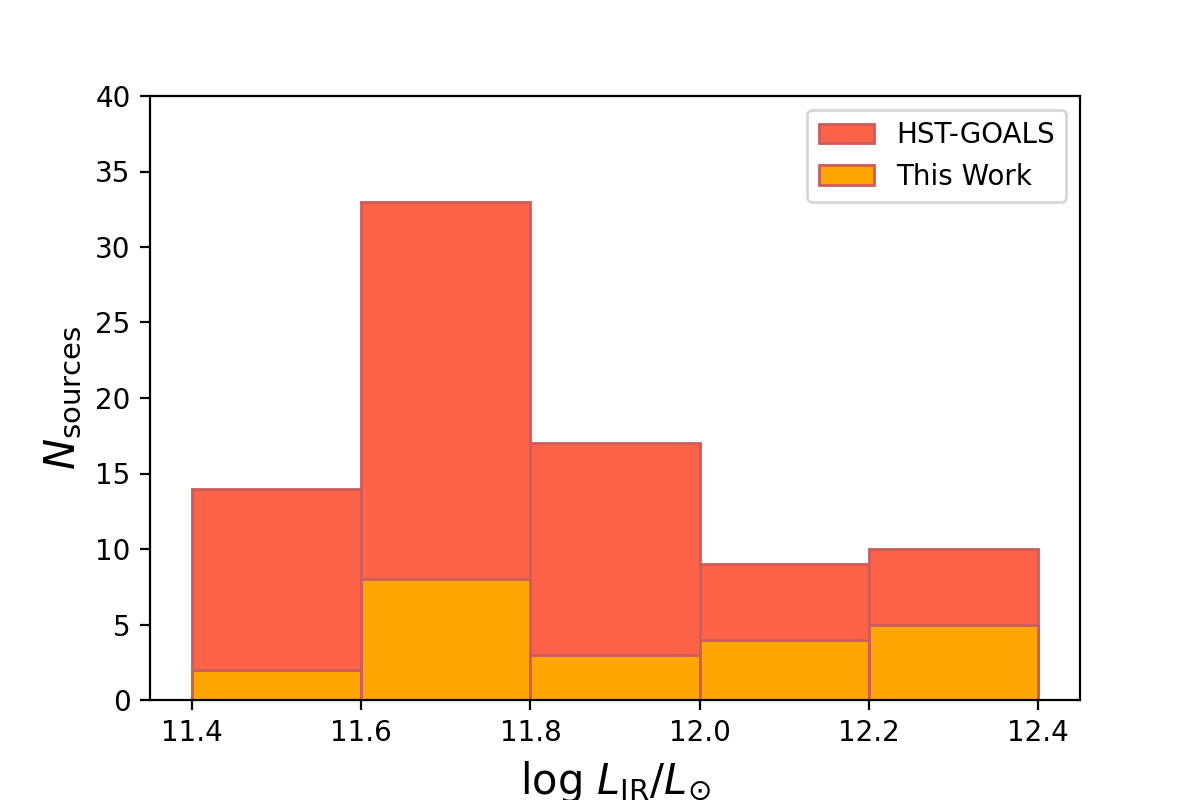}
    \includegraphics[width=0.45\textwidth]{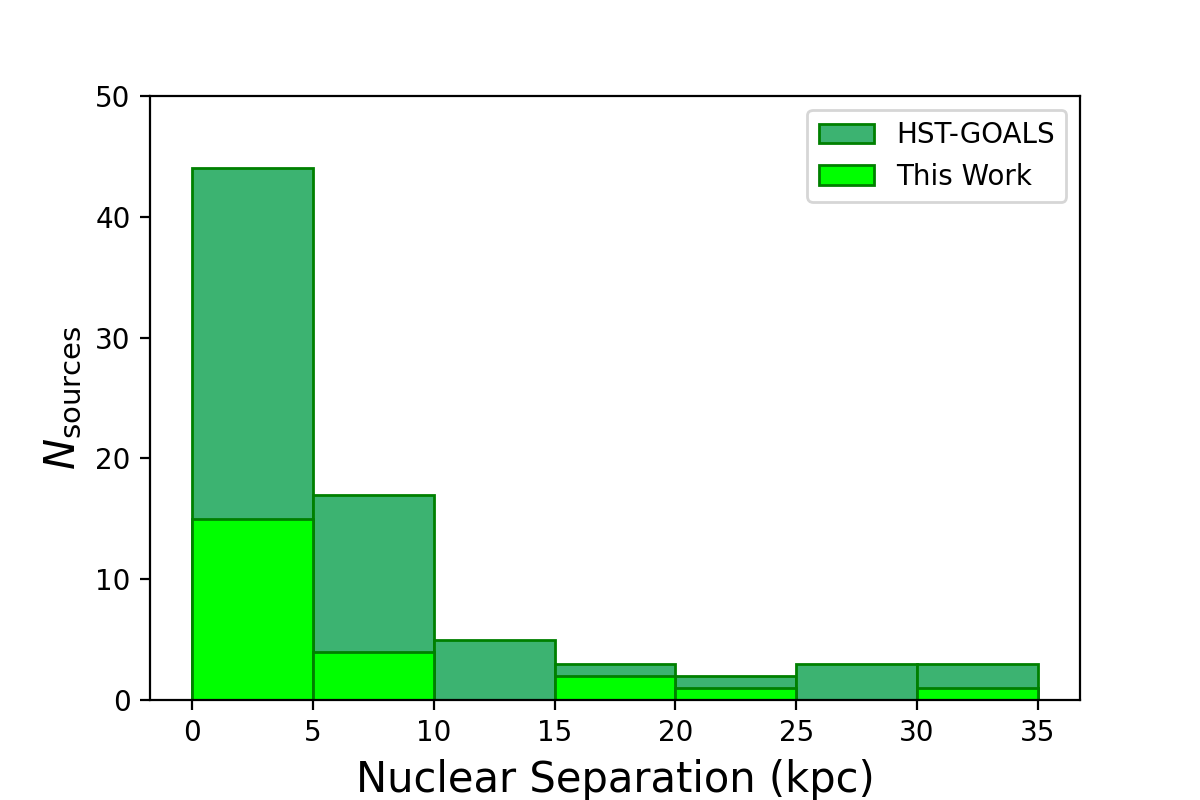}
    \includegraphics[width=0.45\textwidth]{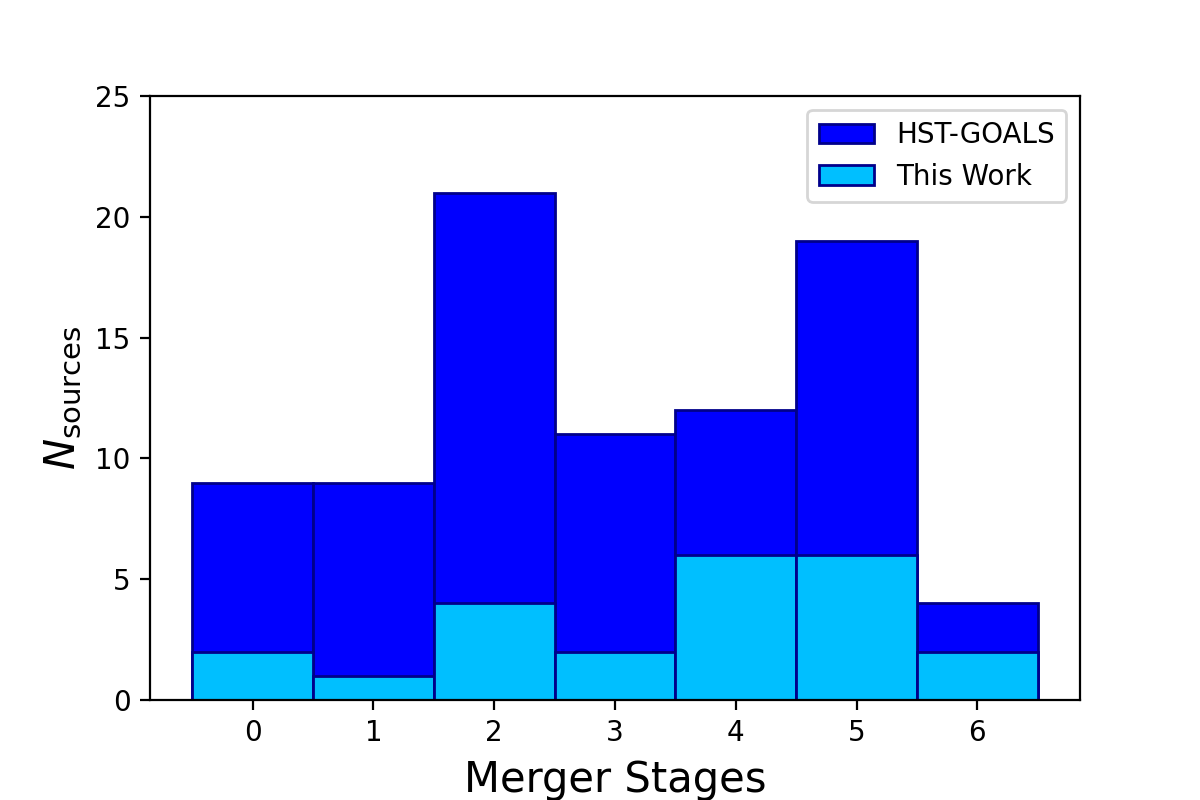}
	\caption{Histograms of the host properties ($L_{\rm IR}$, nuclear separations, merger stages) for our sample shown against those for the \emph{HST}-GOALS parent sample~\citep{2013Kim}. See text for merger stage classifications.} 
	\label{fig:sample}
\end{figure}

Our final sample of 23 sources (harboring 28 resolved nuclei with \edit1{more than 9 data points covering the full wavelength range from X-ray to submillimeter}) spans the full range of properties within the \edit1{\emph{HST-}GOALS (U)LIRGs \citep[e.g.][]{2013Kim}} with $L_{\rm IR}=10^{11.4-12.5}L_{\odot}$ (Figure \ref{fig:sample}), while favoring the high-luminosity, late-stage mergers that best map the phase when SMBHs are expected to grow most rapidly \citep[e.g.][]{2010Yuan, 2011Petric}. Basic properties of our sample are listed in Table \ref{tab:sample}. The merger stage classification is adopted from \cite{2011Haan} and \cite{2013Kim}, where stage 0 refers to a single undisturbed galaxy with no signs of interaction; stage 1 refers to separate galaxies with symmetric disks and no tails; stage 2 refers to systems whose progenitor galaxies are distinguishable with asymmetric disks and/or tidal tails; stage 3 refers to galaxies with two nuclei in a common envelope; stage 4 refers to galaxies with dual nuclei and tidal tails; stage 5 refers to galaxies with single or obscured nucleus and long prominent tails, and stage 6 refers to galaxies with single or obscured nucleus plus disturbed central morphology and short faint tails.  The nuclear separation $d_\mathrm{nuc}$ is the distance between the two nuclei if both are detected and resolved in \emph{HST}/ACS images as adopted from \cite{2013Kim}. A separation distance of 0 for a merger stage 5 galaxy indicates a coalesced nucleus, while for a merger stage 0 galaxy, it points to a single SMBH pre-merging. \edit1{For dual nuclei systems, we only include sources with ALMA detection; otherwise, we cannot determine the precise location of the obscured nuclei and accurately measure their emission in the submillimeter wavelengths.}

\begin{deluxetable*}{lcccccc}
\tablenum{1}
\tablecaption{Sample Properties\label{tab:sample}}
\tablewidth{0pt}
\tablehead{
\colhead{Galaxy} & \colhead{RA} & \colhead{Dec} &
\colhead{$D_{L}$ (Mpc)} & \colhead{Merger Stage} & \colhead{$d_\mathrm{nuc}$ (kpc)} & \colhead{${\rm log}(L_{\rm IR}/L_{\odot})$}
}
\startdata
CGCG 436$-$030 & 01h20m02.7s & +14d21m43s & 138.3 & 2 & 33.0 & 11.69 \\
IRAS F01364$-$1042 & 01h38m52.9s & $-$10d27m11s & 214.1 & 5 & 0.0 & 11.85 \\
III Zw 035 & 01h44m30.5s & +17d06m05s & 120.0 & 3 & 4.7 & 11.64 \\
NGC 0695 & 01h51m14.2s & +22d34m57s & 142.5 & 0 & 0.0 & 11.68 \\
NGC 1614 & 04h33m59.8s & $-$08d34m44s & 69.1 & 5 & 0.0 & 11.65 \\
IRAS F05189$-$2524 & 05h21m01.4s & $-$25d21m45s & 188.2 & 6 & 0.0 & 12.16 \\
IRAS F06076$-$2139 & 06h09m45.8s & $-$21d40m24s & 165.0 & 3 & 6.0 & 11.65 \\
IRAS 07251$-$0248 & 07h27m37.5s & $-$02d54m55s & 399.7 & 4 & \edit1{2.0} & 12.39 \\
NGC 2623 & 08h38m24.1s & +25d45m17s & 80.4 & 5 & 0.0 & 11.60 \\
IRAS F08572+3915 & 09h00m25.4s  & +39d03m54s & 260.3 & 4 & 6.0 & 12.16 \\
IRAS F10565+2448 & 10h59m18.1s & +24d32m34s & 190.7 & 2 & 22.9 & 12.08 \\
IRAS F12112+0305 & 12h13m46.0s & +02d48m38s & 331.4 & 4 & 4.4 & 12.36 \\
UGC 08387 & 13h20m35.3s & +34d08m22s & 100.5 & 5 & 0.0 &  11.73 \\
NGC 5331 & 13h52m16.3s & +02d06m11s & 145.1 & 2 & 17.4 & 11.66 \\
IRAS F14348$-$1447 & 14h37m38.4s & $-$15d00m20s & 374.3 & 4 & 6.2 & 12.39 \\
Arp  220 & 15h34m57.2s & +23d30m11s & 79.9 & 4 & \edit1{0.5} & 12.28 \\
NGC 6240 & 16h52m58.9s & +02d24m03s & 106.1 & 4 & 0.8 & 11.93 \\
IRAS F17138$-$1017 & 17h16m35.8s & $-$10d20m39s & 75.2 & 6 & 0.0 & 11.49 \\
IRAS F17207$-$0014 & 17h23m21.9s & $-$00d17m01s & 189.3 & 5 & 0.0 & 12.46 \\
IRAS 19542+1110 & 19h56m35.4s & +11d19m03s & 291.8 & 0 & 0.0 & 12.12 \\
II Zw 096 & 20h57m23.9s & +17d07m39s & 158.9 & 2 & 7.9 & 11.94 \\
IRAS F22491$-$1808 & 22h51m49.2s & $-$17d52m23s & 352.5 & 5 & 0.0 & 12.20 \\
NGC 7674 & 23h27m56.7s & +08d46m45s & 127.1 & 1 & 18.9 & 11.56 \\
\enddata
\tablecomments{Basic properties of our sample. Column 1: Galaxy name; Column 2: Right ascension (J2000); Column 3: Declination (J2000); Column 4: Luminosity distance in Mpc ($H_{0}=70\,\rm km\,s^{-1}\,Mpc^{-1}$); Column 5: Merger stage \citep{2011Haan,2013Kim}; Column 6: Nuclear separation $d_\mathrm{nuc}$ in kpc \citep{2013Kim}; Column 7: Infrared luminosity \citep{2009Armus}}
\end{deluxetable*}

\section{Archival Data} \label{sec:data}

The multiwavelength data have been \edit1{collected} from the \emph{Chandra} Data Archive (CDA), the Barbara A. Mikulski Archive for Space Telescopes (MAST), the Keck Observatory Archive (KOA) and the ALMA Science Archive. Here we describe any additional processing steps taken to enhance the data products delivered by these archives.

\subsection{Chandra} \label{sec:chandra}

\emph{Chandra} provides 0\farcs5-resolution X-ray data at the $\sim$0.5$-$10\,keV energy bands. The majority of the sources in our sample were observed in Cycles 7, 8, and 14 with a uniform exposure time of 15 ks. The data reduction was performed using the \emph{Chandra} data analysis package CIAO version 4.11 \citep{2006CIAO}, and HEASOFT version 6.25 \edit1{, including XSPEC version 12.10.1} \citep{1996Xspec}. Due to the resolution limitation, we choose to extract nuclear photometry from a $1^{\prime\prime}$-radius aperture rather than using pre-determined apertures (see \S \ref{sec:phot} for details). We performed spectral fitting on the data as follows: the soft X-ray spectra were fitted using single temperature plasma model APEC \citep{2001Smith}, while the hard X-ray spectra were fitted with a power-law model. \edit1{For 14 out of the 28 sources in our sample,} where the data quality was subpar \edit1{(with only $\sim10-20$ photons in the spectrum)}, we fixed the temperature of the plasma model to a typical value $kT \sim 0.7$ keV. 
\edit1{In the cases of NGC 6240N and NGC 6240S}, we added a Gaussian line profile to fit the Fe K\edit1{$\alpha$} line feature at around 6.4 keV. \edit1{The line-of-sight absorption caused by Galactic neutral hydrogen is modeled using a \textit{phabs} model, while the column density was provided by the \emph{Swift} Science Data Center \citep{2013Swift}\footnote{\edit1{The models used to fit the soft and hard X-ray bands can be expressed as \textit{phabs$\times$apec} and \textit{phabs$\times$(powerlaw+zgauss)}, respectively. }}}. We derived the Galactic absorption-corrected 0.5$-$2 keV and 2$-$10 keV fluxes for each of our nuclei in the sample, allowing for a comparison with the global values reported by \cite{2011Iwasawa} and \cite{2018Torres}. Furthermore, in order to meet the requirements of the SED fitting code, we obtained the intrinsic 2$-$10 keV X-ray luminosity for some of our galaxies from \cite{2021Ricci}, using data from the \emph{NuSTAR}. For the rest of our galaxies that were not included in \cite{2021Ricci}, we added a neutral hydrogen absorption component (\textit{zphabs}) with the same redshift as the galaxy during the fitting and measured the 2$-$10 keV X-ray intrinsic luminosity without the effect of this absorption component. For high-quality spectra, the neutral hydrogen column density of this redshifted absorption component was allowed to vary freely between $10^{22.5}<N_{\rm H}<10^{25}\,\rm cm^{-2}$, while for spectra with $\sim$10 counts, the column density was fixed at $10^{23}\,\rm cm^{-2}$. \edit1{All 0.5–10 keV X-ray spectra extracted from the nuclear regions of galaxies in our sample are presented in Figure \ref{fig:xray-spectra}, while the X-ray spectral properties are listed in Table \ref{tab:x-ray_derived-properties}. All the {\it Chandra} datasets employed in this paper are contained in~\dataset[10.25574/cdc.333]{https://doi.org/10.25574/cdc.333}}.

\subsection{HST} \label{sec:hst}

The mostly science-ready \emph{HST} data we gathered from MAST comprise of $F435W$ and $F814W$ images taken with ACS as part of the GOALS-\emph{HST} program (Program ID: 10592, PI: A. Evans) and of F110W and F160W images taken with NICMOS and WFC3 (Program ID: 11235, PI: J. Surace; Program ID: 13690, PI: T. D\'iaz-Santos). All of the images from ACS and WFC3 were re-drizzled using DrizzlePac version 3.1.8~\citep{2012DrizzlePac} and aligned to the \emph{GAIA} catalog \citep{2018GAIADR2} to achieve 0\farcs05-precision astrometry. This is particularly crucial for the short-wavelength images where the central SMBH may be obscured by dust and difficult to locate. We also included additional \emph{HST} data in other UV and optical filters where the data are available to substantiate the wavelength coverage. All the {\it HST} data used in this paper can be found in MAST: \dataset[10.17909/4x0z-gt45]{http://dx.doi.org/10.17909/4x0z-gt45}.

\subsection{Keck} \label{sec:keck}

From KOA, we downloaded publicly available, high-resolution (0\farcs01$-$0\farcs04/px), near-infrared \emph{(J, H, K$_s$, {\rm{and}} K$_p$)}, raw NIRC2 science and calibration frames for our galaxies. The observations were procured between 2009 and 2019. We reduced the raw data partially using the Keck AO Imaging (KAI) data reduction pipeline \citep{2022lu}  as follows. We first masked out the bad pixels and subtracted the dark frames, and then we applied flat-fielding to each science frame using an image processing code developed specifically for NIRC2 data\footnote{\url{https://github.com/logan-pearce/NIRC2-Image-Processing}}. Images obtained during the same night were co-added to increase the signal-to-noise ratio. We used the 2MASS Extended Source Catalog~\citep{20002MASS} to flux calibrate the NIRC2 images. Finally, We aligned the images to the WCS-calibrated \emph{HST} images to correct the astrometry.

\subsection{ALMA} \label{sec:alma}

All of the ALMA data used in this project were taken between 2016 and 2019. The angular resolution of an ALMA data set varies depending on its observed frequency band and array baseline. In order to generate internally-consistent SEDs using data sets of comparable resolutions, we selected those observations with angular scales of 0\farcs5 (matching that of \emph{Chandra}) or better for our study. Our archival data set mainly consisted of ALMA cubes taken in Band 7 (275-373 GHz), Band 6 (211-275 GHz), Band 5 (163-211 GHz) and Band 3 (84-116 GHz), with a few galaxies also observed in Band 10 (787-950 GHz), Band 9 (602-720 GHz), Band 8 (385-500 GHz), and Band 4 (125-163 GHz). Specifically, we used the beam-corrected 880- and 1300-$\mu$m~dust continuum images to locate the SMBHs in the galaxies. Given the precision of the calibrated submillimeter data sets and with visual inspection of all the multiwavelengh images for all the individual systems, the ALMA locations were taken as the absolute coordinates for the nuclei to be measured across all multiwavelengths.

\subsection{Miscellaneous Mid-infrared Data} \label{sec:MIR}

Eight galaxies in our sample (NGC 1614, NGC 2623, UGC 08387, NGC 6240, III Zw 035, NGC 7674, IRAS F05189-2524 and IRAS F08572+3915) have archival mid-infrared N band ($\sim$11\,$\rm \mu$m) and/or Q band ($\sim$18\,$\rm \mu$m) imaging data from \cite{2014Asmus}. These observations were conducted with mid-infrared instruments on the Very Large Telescope (VLT), Gemini, and Subaru Telescopes. In \cite{2014Asmus}, nuclear fluxes were measured using 2D Gaussian fitting and are treated as upper limits relative to the aperture size adopted in this work. For the other sources, mid-infrared fluxes from \emph{IRAS} and \emph{Spitzer}/MIPS at 12, 24 and 25 $\rm\mu m$ \citep{2003Sanders,2012U} have been incorporated as upper limits in our analysis. For II Zw 096, we have incorporated the flux measurements from \emph{JWST} MIRI F560W, F770W, and F1500W provided by \cite{2022Inami}.

\section{Analysis}
\label{sec:analysis}

\subsection{Aperture Photometry} 
\label{sec:phot}

\begin{figure*}[!htb]

	\centering
	\includegraphics[width=0.9\textwidth]{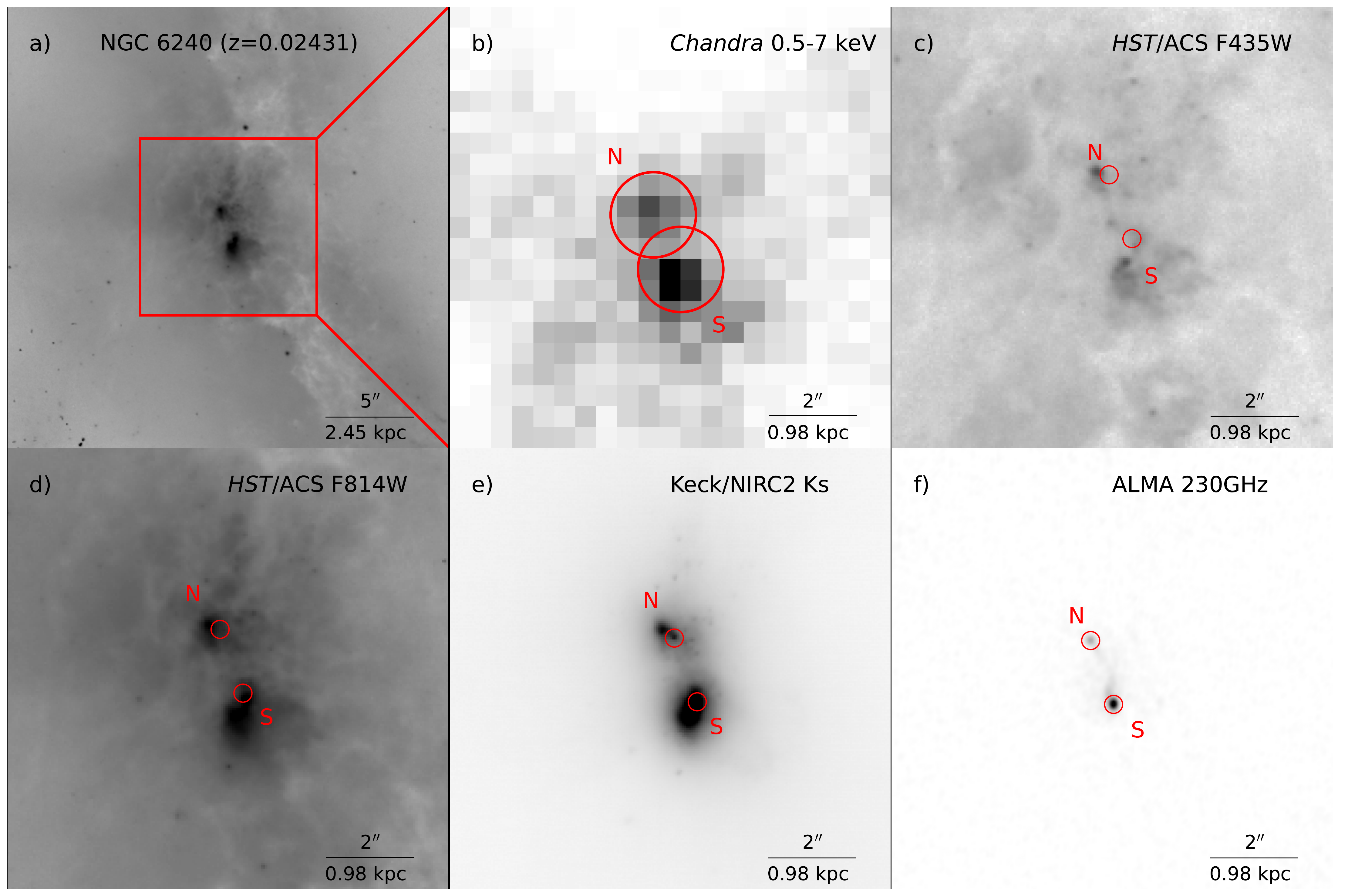}
	\caption{Example apertures used for extracting nuclear photometry from our multiwavelength data set for NGC 6240: a) \emph{HST}/ACS $F814W$ zoomed-out image of the entire galaxy; north is up and east is to the left. The \edit1{red} box highlights the inset for the remaining panels focusing on the nuclear region. b) \emph{Chandra} 0.5$-$7 keV image overlaid with two apertures for the N and S nuclei (carried over to the later panels as well); c) \emph{HST}/ACS $F435W$ image; d) \emph{HST}/ACS $F814W$ image; e) Keck/NIRC2 $K_s$ image; and f) ALMA 230\,GHz dust continuum image. These images are representative of how our aperture photometry was extracted across the electromagnetic spectrum but not all the available data are shown. See Figure \ref{fig:append_photometry} for the rest of the sample.}
	\label{fig:photometry}

\end{figure*}

We extracted photometry from the flux-calibrated, multiwavelength images using the Python \verb|photutils| package~\citep{2020photutils}. The aperture size used for measurements in the \emph{HST}, Keck/NIRC2, and ALMA data of each nucleus varied depending on the redshift of the galaxy: for sources at $z<0.05$, we adopted an aperture size of $r =0\farcs106-0\farcs284$, equivalent to a physical scale of 100\,pc; for galaxies at $z>0.05$, we adopted $r =0\farcs102-0\farcs133$ instead, which is equivalent to a physical size of 150\,pc. These apertures were selected to be larger than the full-width-half-maximum (FWHM) of both the \emph{HST} and Keck images, but the ALMA data span a range of primary beam sizes up to 0\farcs5, depending on the baseline and frequency of the observations. If the chosen aperture size was smaller than the beam size of ALMA, then we instead used its FWHM for the size of the aperture and therefore reported the resulting flux as an upper limit of the nuclear emission. 
The representative apertures used to extract photometry from the multiwavelength images are shown for one system NGC 6240 in Figure \ref{fig:photometry} and for the whole sample in Figure \edit1{\ref{fig:append_photometry}}. The measured nuclear fluxes are reported in mJy in Table \ref{tab:photometry1}.

For our most distant object IRAS 07251$-$0248 at $z=0.088$, since the angular size corresponding to 150 pc was smaller than the point spread function (PSF) of the \emph{HST/WFC3} F160W image, we elected to use an aperture size of $r = 0\farcs1$, equivalent to 164 pc for this galaxy. For point-like sources unresolved in \emph{HST} images, we determined the appropriate aperture corrections using TinyTim-simulated PSFs \citep{1993Krist}. As for NIRC2 AO images, the shape of the PSF varies across the the field of view (FOV) depending on the position of the target relative to the tip-tilt star. It is difficult to perform accurate aperture corrections and thus, we used instead the bright stars in NIRC2 images as our PSF references to calculate approximate encircled energy fractions.

\begin{longrotatetable}
\begin{deluxetable*}{lcccccccccc}
\tablecaption{Nuclear Photometry (mJy, $\lambda < 0.8\,\rm\mu m$)\label{tab:photometry1}}
\tablewidth{1000pt}
\tabletypesize{\scriptsize}
\tablenum{2}
\tablehead{
\colhead{Galaxy} & \colhead{HX intrinsic} & 
\colhead{HX observed} & \colhead{SX} & 
\colhead{\emph{HST} F140LP} & \colhead{\emph{HST} F225W} & 
\colhead{\emph{HST} F275W} & \colhead{\emph{HST} F330/336W} & 
\colhead{\emph{HST} F435W} & \colhead{\emph{HST} F555W} & \colhead{\emph{HST} F621M}\\ 
\colhead{} & \colhead{2$-$10 keV} & \colhead{2$-$10 keV} & \colhead{0.5$-$2 keV} & 
\colhead{1518\AA} & \colhead{2372\AA} & \colhead{2710\AA} &
\colhead{3350\AA} & \colhead{4329\AA} & \colhead{5361\AA} & \colhead{6219\AA}\\
\colhead{} & \colhead{(28)} & \colhead{(25)} & \colhead{(27)} & 
\colhead{(7)} & \colhead{(6)} & \colhead{(3)} &
\colhead{(14)} & \colhead{(28)} & \colhead{(4)} & \colhead{(4)}
} 
\startdata
CGCG 436-030 & 8.14E-06 & 3.85E-06 & 3.81E-06 & $\dots$ & $\dots$ & $\dots$ & $\dots$ & 1.08E-03 & $\dots$ & $\dots$ \\
& (2.24E-06) & (1.22E-06) & (1.17E-06) &  &  &  &  & (1.80E-05) &  &  \\
IRAS F01364-1042 & 4.19E-06 & 3.38E-06 & 2.32E-06 & $\dots$ & $\dots$ & $\dots$ & 8.75E-05 & 3.30E-04 & $\dots$ & $\dots$ \\
& (1.10E-06) & (1.10E-06) & (7.15E-07) &  &  &  & (4.14E-06) & (8.38E-06) &  &  \\
III Zw 035 & 2.04E-06 & 1.18E-06 & 2.37E-06 & $\dots$ & $\dots$ & $\dots$ & 1.29E-03 & 9.00E-03 & $\dots$ & $\dots$ \\
& (6.80E-07) & (4.49E-07) & (6.13E-07) &  &  &  & (1.65E-05) & (3.82E-05) &  & \\
NGC 0695 & 1.90E-05 & 1.31E-05 & 5.01E-06 & 5.32E-05 & $\dots$ & $\dots$ & $\dots$ & 1.67E-02 & $\dots$ & $\dots$ \\
& (3.90E-06) & (2.93E-06) & (1.12E-06) & (5.77E-06) &  &  &  & (5.15E-05) &  & \\
NGC 1614 & 1.24E-05 & 8.98E-06 & 2.90E-05 & 6.69E-04 & 3.18E-03 & 1.15E-02 & 7.17E-02 & 0.38 & 0.98 & $\dots$ \\
& (2.70E-06) & (1.30E-06) & (1.93E-06) & (2.37E-05) & (4.48E-05) & (6.36E-05) & (7.32E-05) & (2.45E-04) & (3.43E-04) & \\
IRAS F05189-2524 & 4.53E-04 & 2.82E-04 & 3.32E-05 & $\dots$ & 3.26E-02 & $\dots$ & $\dots$ & 6.02E-02 & $\dots$ & $\dots$ \\
& (1.34E-05) & (8.33E-06) & (1.49E-06) &  & (1.41E-04) &  &  & (9.73E-05) &  & \\
IRAS F06076-2139N & 7.44E-06 & 1.38E-06 & 2.25E-06 & 2.11E-05 & $\dots$ & $\dots$ & $\dots$ & 5.07E-03 & $\dots$ & $\dots$ \\
& (4.41E-06) & (5.06E-07) & (6.57E-07) & (3.63E-06) &  &  &  & (2.84E-05) &  & \\
IRAS 07251-0248E & $<$2.90E-06 & $\dots$ & 3.31E-06 & $\dots$ & $\dots$ & $\dots$ & $\dots$ & 9.66E-04 & $\dots$ & $\dots$ \\
& (2.27E-06) &  & (9.13E-07) &  &  &  &  & (1.29E-05) &  & \\
IRAS 07251-0248W & $<$2.90E-06 & $\dots$ & 3.31E-06 & $\dots$ & $\dots$ & $\dots$ & $\dots$ & 4.64E-03 & $\dots$ & $\dots$ \\
& (2.27E-06) &  & (9.13E-07) &  &  &  &  & (2.71E-05) &  & \\
NGC 2623 & 7.33E-06 & 1.23E-05 & 2.70E-06 & 3.61E-04 & $\dots$ & 8.66E-04 & 2.28E-03 & 9.90E-03 & 2.07E-02 & $\dots$ \\
& (1.51E-06) & (2.54E-06) & (5.55E-07) & (1.74E-05) &  & (1.71E-05) & (1.87E-05) & (4.08E-05) & (5.66E-05) & \\
IRAS F08572+3915NW & $<$1.24E-06 & 9.52E-07 & $\dots$ & $\dots$ & $\dots$ & $\dots$ & 1.23E-04 & 4.15E-04 & $\dots$ & $\dots$ \\
& (6.45E-07) & (4.94E-07) &  &  &  &  & (5.53E-06) & (9.58E-06) &  & \\
IRAS F10565+2448 & $<$2.11E-06 & 2.77E-06 & 5.49E-06 & $\dots$ & $\dots$ & $\dots$ & 3.46E-03 & 4.85E-03 & $\dots$ & $\dots$ \\
& (4.19E-07) & (5.49E-07) & (5.97E-07) &  &  &  & (2.70E-05) & (2.81E-05) &  &  \\
IRAS F12112+0305N & $<$3.08E-06 & 9.13E-07 & 7.15E-07 & $\dots$ & 1.48E-05 & $\dots$ & 2.22E-04 & 7.87E-04 & $\dots$ & $\dots$ \\
& (1.40E-06) & (4.16E-07) & (2.74E-07) &  & (4.45E-06) &  & (7.83E-06) & (1.21E-05) &  & \\
IRAS F12112+0305S & $<$3.08E-06 & 2.10E-06 & 5.63E-07 & $\dots$ & 2.12E-05 & $\dots$ & 1.44E-04 & 5.50E-04 & $\dots$ & $\dots$ \\
& (1.40E-06) & (1.18E-06) & (2.24E-07) &  & (3.83E-06) &  & (6.89E-06) & (1.04E-05) &  & \\
UGC 08387 & 7.26E-06 & 3.54E-06 & 3.92E-06 & $\dots$ & $\dots$ & $\dots$ & $\dots$ & 4.22E-03 & $\dots$ & $\dots$ \\
& (1.75E-06) & (8.72E-07) & (8.05E-07) &  &  &  &  & (2.65E-05) &  & \\
NGC 5331S & 1.73E-06 & 6.71E-07 & 1.33E-06 & 1.09E-05 & $\dots$ & $\dots$ & $\dots$ & 5.94E-04 & $\dots$ & $\dots$ \\
& (7.59E-07) & (3.23E-07) & (4.51E-07) & (3.03E-06) &  &  &  & (1.34E-05) &  & \\
IRAS F14348-1447N & $<$3.16E-06 & 1.40E-06 & 1.01E-06 & $\dots$ & 2.99E-05 & $\dots$ & $\dots$ & 2.48E-04 & $\dots$ & $\dots$ \\
& (1.50E-06) & (6.66E-07) & (3.59E-07) &  & (4.60E-06) &  &  & (1.01E-05) &  & \\
IRAS F14348-1447S & $<$3.16E-06 & 1.92E-06 & 1.81E-06 & $\dots$ & 1.52E-04 & $\dots$ & $\dots$ & 9.97E-04 & $\dots$ & $\dots$ \\
& (1.50E-06) & (6.16E-07) & (4.38E-07) &  & (9.53E-06) &  &  & (1.49E-05) &  &  \\
Arp 220E & 2.83E-04 & 5.70E-06 & 7.67E-07 & $\dots$ & $\dots$ & $\dots$ & 2.10E-04 & 1.45E-03 & 3.38E-03 & 5.01E-03 \\
& (5.19E-05) & (5.25E-07) & (1.14E-07) &  &  &  & (7.89E-06) & (1.92E-05) & (1.87E-05) & (1.16E-04) \\
Arp 220W & 2.83E-04 & 5.70E-06 & 7.67E-07 & $\dots$ & $\dots$ & $\dots$ & 2.92E-04 & 1.74E-03 & 4.29E-03 & 7.78E-03 \\
& (5.19E-05) & (5.25E-07) & (1.14E-07) &  &  &  & (8.85E-06) & (2.04E-05) & (2.04E-05) & (1.34E-04) \\
NGC 6240N & 7.66E-04 & 4.00E-05 & 8.98E-06 & $\dots$ & $\dots$ & $\dots$ & 5.13E-04 & 5.01E-03 & $\dots$ & 2.88E-02\\
& (2.43E-05) & (1.27E-06) & (2.04E-07) &  &  &  & (1.09E-05) & (2.97E-05) &  & (2.25E-04) \\
NGC 6240S & 2.01E-03 & 7.60E-05 & 1.30E-05 & $\dots$ & $\dots$ & $\dots$ & 6.83E-04 & 4.90E-03 & $\dots$ & 5.57E-03\\
& (5.94E-05) & (2.24E-06) & (2.95E-07) &  &  &  & (1.23E-05) & (2.94E-05) &  & (3.10E-04) \\
IRAS F17138-1017 & $<$1.75E-05 & 1.45E-06 & 3.37E-06 & $\dots$ & $\dots$ & $\dots$ & $\dots$ & 3.63E-04 & $\dots$ & $\dots$ \\
& (6.96E-06) & (5.75E-07) & (1.97E-06) &  &  &  &  & (1.97E-05) &  & \\
IRAS F17207-0014 & $<$4.09E-06 & 3.45E-06 & 1.25E-06 & $\dots$ & $\dots$ & $\dots$ & 2.23E-05 & 1.77E-04 & $\dots$ & $\dots$ \\
& (7.88E-07) & (6.65E-07) & (2.79E-07) &  &  &  & (3.50E-06) & (7.50E-06) &  &  \\
IRAS 19542+1110 & 7.21E-05 & 5.31E-05 & 8.62E-06 & $\dots$ & $\dots$ & $\dots$ & $\dots$ & 2.08E-02 & $\dots$ & $\dots$ \\
& (7.24E-06) & (5.22E-06) & (1.61E-06) &  &  &  &  & (5.70E-05) &  & \\
II Zw 096E & $<$4.40E-06 & 1.63E-06 & 3.32E-06 & 2.47E-05 & $\dots$ & $\dots$ & $\dots$ & 2.15E-04 & $\dots$ & $\dots$ \\
& (1.69E-06) & (6.24E-07) & (1.13E-06) & (4.54E-06) &  &  &  & (8.72E-06) &  & \\
IRAS F22491-1808 & 7.80E-06$^{(a)}$ & $\dots$ & 1.43E-06 & $\dots$ & $\dots$ & $\dots$ & $\dots$ & 7.49E-04 & $\dots$ & $\dots$ \\
& (2.81E-06) &  & (3.69E-07) &  &  &  &  & (1.25E-05) &  &  \\
NGC 7674 & 1.07E-03 & 4.53E-05 & 4.43E-05 & 6.23E-02 & $\dots$ & 0.14 & 0.21 & 0.34 & $\dots$ & $\dots$ \\
& (2.92E-04) & (1.24E-05) & (1.07E-05) & (1.98E-04) &  & (2.62E-04) & (3.96E-04) & (2.54E-04) &  & \\
\enddata
\tablecomments{Flux uncertainties, in units of mJy, are given in parantheses. The numbers in parentheses under each wavelength indicate the number of galaxies observed using the corresponding filter. $^{(a)}$ intrinsic soft X-ray flux.}
\end{deluxetable*}
\end{longrotatetable}

\begin{longrotatetable}
\begin{deluxetable*}{lcccccccccc}
\tablecaption{Nuclear Photometry (mJy, $0.8\,\rm\mu m < \lambda < 6.0\,\rm\mu m$)\label{tab:photometry2}}
\tablewidth{1000pt}
\tabletypesize{\scriptsize}
\tablenum{2}
\tablehead{
\colhead{Galaxy} & \colhead{\emph{HST} F814W} & 
\colhead{\emph{HST} F110W} & \colhead{Keck J} & \colhead{\emph{HST} F160W} & 
\colhead{Keck H} & \colhead{Keck $\rm K_{s}/K_{p}$}  & \colhead{\emph{HST} F222M} & \colhead{\emph{HST} F237M} & 
\colhead{L} & \colhead{M} \\ 
\colhead{} & \colhead{8045\AA} & \colhead{$1.1\,\mu{\rm m}$} & \colhead{$1.2\,\mu{\rm m}$} & 
\colhead{$1.5\,\mu{\rm m}$} & \colhead{$1.6\,\mu{\rm m}$} & \colhead{$2.1\,\mu{\rm m}$} & \colhead{$2.2\,\mu{\rm m}$} &
\colhead{$2.4\,\mu{\rm m}$} & \colhead{$3.0\,\mu{\rm m}$} & \colhead{$5.0\,\mu{\rm m}$}\\
\colhead{} & \colhead{(28)} & \colhead{(22)} & \colhead{(1)} & 
\colhead{(27)} & \colhead{(1)} & \colhead{(19)} &
\colhead{(9)} & \colhead{(1)} & \colhead{(1)} & \colhead{(3)}
}
\startdata
CGCG 436-030 & 1.67E-02 & 0.11 & $\dots$ & 0.36 & $\dots$ & 0.71 & $\dots$ & $\dots$ & $\dots$ & $\dots$ \\
& (6.06E-05) & (1.57E-04) &  & (1.99E-04) &  & (5.20E-04) &  &  &  & \\
IRAS F01364-1042 & 4.41E-03 & $\dots$ & 3.66E-02 & 6.68E-02 & $\dots$ & 0.31 & $\dots$ & $\dots$ & $\dots$ & $\dots$ \\
& (3.16E-05) &  & (1.86E-04) & (8.60E-05) &  & (3.89E-04) &  &  &  & \\
III Zw 035 & 8.38E-02 & 0.27 & $\dots$ & 0.54 & $\dots$  & 1.30 & $\dots$ & $\dots$ & $\dots$ & $\dots$ \\
& (1.34E-04) & (3.28E-04) &  & (6.94E-04) &  & (2.43E-03) &  &  &  & \\
NGC 0695 & 0.13 & $\dots$ & $\dots$ & $<$1.79 & $\dots$ & 1.17 & $\dots$ & $\dots$ & $\dots$ & $\dots$ \\
& (1.68E-04) &  &  & (9.40E-04) &  & (6.94E-04) &  &  &  & \\
NGC 1614 & 3.29 & 9.17 & $\dots$ & 13.68 & $\dots$ & $\dots$ & 19.72 & 15.30 & $\dots$ & 7.13 \\
& (8.39E-04) & (1.45E-03) &  & (9.31E-03) &  &  & (6.54E-03) & (4.43E-03) &  & (0.64) \\
IRAS F05189-2524 & 0.27 & $<$3.60 & $\dots$ & $<$13.02 & $\dots$ & $\dots$ & $<$36.26 & $\dots$ & $\dots$ & $\dots$ \\
& (2.40E-04) & (6.40E-03) &  & (1.39E-02) &  &  & (3.59E-02) &  &  &  \\
IRAS F06076-2139N & 4.90E-02 & $\dots$ & $\dots$ & 0.26 & $\dots$ & 0.56 & $\dots$ & $\dots$ & $\dots$ & $\dots$ \\
& (1.02E-04) &  &  & (1.29E-04) &  & (5.20E-04) &  &  &  & \\
IRAS 07251-0248E & 4.72E-03 & $\dots$ & $\dots$ & 2.14E-02 & $\dots$ & 0.17 & $\dots$ & $\dots$ & $\dots$ & $\dots$ \\
& (3.25E-05) &  &  & (3.72E-05) &  & (4.63E-04) &  &  &  & \\
IRAS 07251-0248W & 1.34E-02 & $\dots$ & $\dots$ & 4.07E-02 & $\dots$ & 0.29 & $\dots$ & $\dots$ & $\dots$ & $\dots$ \\
& (5.36E-05) &  &  & (5.12E-05) &  & (4.92E-04) &  &  &  & \\
NGC 2623 & 0.12 & 0.72 & $\dots$ & 2.19 & $\dots$ & 13.23 & $\dots$ & $\dots$ & $\dots$ & $\dots$ \\
& (1.58E-04) & (4.07E-04) &  & (3.74E-04) &  & (4.92E-03) &  &  &  & \\
IRAS F08572+3915NW & 7.14E-03 & 2.71E-02 & $\dots$ & 0.41 & $\dots$ & 2.93 & $\dots$ & $\dots$ & $\dots$ & $\dots$ \\
& (3.95E-05) & (2.03E-04) &  & (8.28E-04) &  & (2.49E-03) &  &  &  & \\
IRAS F10565+2448 & 0.10 & 0.30 & $\dots$ & 1.65 & $\dots$ & $\dots$ & 2.54 & $\dots$ & $\dots$ & $\dots$ \\
& (1.46E-04) & (4.63E-04) &  & (1.75E-03) &  &  & (3.29E-03) &  &  & \\
IRAS F12112+0305N & 4.71E-03 & 9.89E-03 & $\dots$ & 2.63E-02 & $\dots$ & 8.45E-02 & $\dots$ & $\dots$ & $\dots$ & $\dots$ \\
& (3.28E-05) & (1.33E-04) &  & (2.27E-04) &  & (2.05E-04) &  &  &  & \\
IRAS F12112+0305S & 1.44E-02 & 0.18 & $\dots$ & 0.48 & $\dots$ & 0.95 & $\dots$ & $\dots$ & $\dots$ & $\dots$ \\
& (5.59E-05) & (9.44E-04) &  & (1.58E-03) &  & (5.28E-04) &  &  &  & \\
UGC 08387 & 3.80E-02 & 0.16 & $\dots$ & 0.46 & $\dots$ & 1.37 & $\dots$ & $\dots$ & $\dots$ & $\dots$ \\
& (8.96E-05) & (2.71E-04) &  & (7.03E-04) &  & (7.88E-04) &  &  &  & \\
NGC 5331S & 1.43E-02 & 4.99E-02 & $\dots$ & $\dots$ & 0.22 & 0.82 & $\dots$ & $\dots$ & $\dots$ & $\dots$ \\
& (5.79E-05) & (1.74E-04) &  &  & (4.16E-04) & (1.67E-03) &  &  &  & \\
IRAS F14348-1447N & 3.84E-03 & 1.33E-02 & $\dots$ & 4.27E-02 & $\dots$ & $\dots$ & $<$0.30 & $\dots$ & $\dots$ & $\dots$ \\
& (3.24E-05) & (9.71E-05) &  & (1.83E-04) &  &  & (1.28E-03) &  &  & \\
IRAS F14348-1447S & 1.39E-02 & 3.70E-02 & $\dots$ & 9.72E-02 & $\dots$ & $\dots$ & $<$0.57 & $\dots$ & $\dots$ & $\dots$ \\
& (5.66E-05) & (1.62E-04) &  & (2.76E-04) &  &  & (1.76E-03) &  &  & \\
Arp 220E & 1.69E-02 & 0.12 & $\dots$ & 0.37 & $\dots$ & $\dots$ & 1.03 & $\dots$ & $\dots$ & $\dots$ \\
& (6.43E-05) & (2.17E-04) &  & (1.88E-04) &  &  & (9.84E-04) &  &  & \\
Arp 220W & 4.64E-02 & 0.35 & $\dots$ & 0.96 & $\dots$ & $\dots$ & 2.07 & $\dots$ & $\dots$ & $\dots$ \\
& (1.02E-04) & (3.72E-04) &  & (3.04E-04) &  &  & (1.39E-03) &  &  & \\
NGC 6240N & 0.14 & 0.44 & $\dots$ & 0.97 & $\dots$ & 6.81 & $\dots$ & $\dots$ & $<$16.65 & $<$18.47 \\
& (1.71E-04) & (5.11E-04) &  & (1.38E-03) &  & (3.09E-03) &  &  & (0.56) & (0.68) \\
NGC 6240S & 0.23 & 1.36 & $\dots$ & 3.46 & $\dots$ & 28.57 & $\dots$ & $\dots$ & $\dots$ & $\dots$ \\
& (2.19E-04) & (8.98E-04) &  & (2.60E-03) &  & (5.33E-03) &  &  &  & \\
IRAS F17138-1017 & 2.99E-02 & 0.15 & $\dots$ & 0.54 & $\dots$ & 1.12 & $\dots$ & $\dots$ & $\dots$ & $\dots$ \\
& (8.44E-05) & (3.77E-04) &  & (7.91E-04) &  & (8.36E-04) &  &  &  & \\
IRAS F17207-0014 & 6.95E-03 & 4.01E-02 & $\dots$ & 0.15 & $\dots$ & $\dots$ & 0.31 & $\dots$ & $\dots$ & $\dots$ \\
& (3.96E-05) & (2.47E-04) &  & (5.07E-04) &  &  & (1.01E-03) &  &  & \\
IRAS 19542+1110 & 0.13 & $\dots$ & $\dots$ & $<$4.70 & $\dots$ & 3.58 & $\dots$ & $\dots$ & $\dots$ & $\dots$ \\
& (1.66E-04) &  &  & (1.82E-03) &  & (3.32E-03) &  &  &  & \\
II Zw 096E & 1.38E-02 & 4.85E-02 & $\dots$ & 6.79E-02 & $\dots$ & 0.26 & $\dots$ & $\dots$  & $\dots$ & $<$6.17$^{(b)}$\\
& (6.51E-05) & (1.32E-04) &  & (2.48E-04) &  & (4.01E-04) &  &  &  & (0.04)\\
IRAS F22491-1808 & 5.93E-03 & 1.20E-02 & $\dots$  & 2.65E-02 & $\dots$ & $\dots$ & 3.68E-02 & $\dots$ & $\dots$ & $\dots$ \\
& (3.74E-05) & (9.24E-05) &  & (1.44E-04) &  &  & (2.43E-04) &  &  & \\
NGC 7674 & 0.54 & $<$2.69 & $\dots$ & $<$10.96 & $\dots$ & $<$14.15 & $\dots$ & $\dots$ & $\dots$ & $\dots$ \\
& (4.25E-04) & (1.66E-03) &  & (2.28E-03) &  & (1.89E-03) &  &  &  & \\
\enddata
\tablecomments{$^{(b)}$ value from \emph{JWST}/MIRI F560W.}
\end{deluxetable*}
\end{longrotatetable}

\begin{longrotatetable}
\begin{deluxetable*}{lcccccccccc}
\tablecaption{Nuclear Photometry (mJy, $6.0\,\rm\mu m < \lambda < 20.0\,\rm\mu m$)\label{tab:photometry3}}
\tablewidth{1000pt}
\tabletypesize{\scriptsize}
\tablenum{2}
\tablehead{
\colhead{Galaxy} & \colhead{\emph{JWST} F770W} & 
\colhead{VLT PAH1} & \colhead{Gemini Si2} & 
\colhead{VLT ArIII} & \colhead{N} & 
\colhead{VLT PAH2} & \colhead{\emph{IRAS} IRAS1} & 
\colhead{VLT NeII} & \colhead{\emph{JWST} F1500W} & \colhead{Q}\\ 
\colhead{} & \colhead{$7.6\,\mu{\rm m}$} & \colhead{$8.6\,\mu{\rm m}$} & \colhead{$8.7\,\mu{\rm m}$} & 
\colhead{$9.0\,\mu{\rm m}$} & \colhead{$10.0\,\mu{\rm m}$} & \colhead{$11.5\,\mu{\rm m}$} & \colhead{$12.0\,\mu{\rm m}$} &
\colhead{$12.8\,\mu{\rm m}$} & \colhead{$15.0\,\mu{\rm m}$} & \colhead{$18.0\,\mu{\rm m}$}\\
\colhead{} & \colhead{(1)} & \colhead{(1)} & \colhead{(1)} & 
\colhead{(1)} & \colhead{(1)} & \colhead{(4)} &
\colhead{(23)} & \colhead{(1)} & \colhead{(1)} & \colhead{(7)}
}
\startdata
CGCG 436-030 & $\dots$ & $\dots$ & $\dots$ & $\dots$ & $\dots$ & $\dots$ & $<$210.0 & $\dots$  & $\dots$ & $\dots$ \\
&  &  &  &  &  &  & (43.0) &  &  & \\
IRAS F01364-1042 & $\dots$ & $\dots$ & $\dots$ & $\dots$ & $\dots$ & $\dots$ & $<$160.0 & $\dots$ & $\dots$ & $\dots$ \\
&  &  &  &  &  &  &  &  &  & \\
III Zw 035 & $\dots$ & 32.3 & $\dots$ & $\dots$ & $\dots$ & $\dots$ & $<$60.0 & $\dots$ & $\dots$ & $\dots$ \\
&  & (0.9) &  &  &  &  &  &  &  & \\
NGC 0695 & $\dots$ & $\dots$ & $\dots$ & $\dots$ & $\dots$ & $\dots$ & $<$500.0 & $\dots$  & $\dots$ & $\dots$ \\
&  &  &  &  &  &  & (23.0) &  &  & \\
NGC 1614 & $\dots$ & $\dots$ & $<$89.9 & $\dots$ & $\dots$ & $<$200.2 & $\dots$ & $\dots$ & $\dots$ & $<$654.2\\
&  &  & (9.0) &  &  & (20.02) &  &  &  & (65.5)\\
IRAS F05189-2524 & $\dots$ & $\dots$ & $\dots$ & $\dots$ & $<$426.8 & $\dots$ & $\dots$ & $\dots$ & $\dots$ & $\dots$ \\
&  &  &  &  & (6.8) &  &  &  &  & \\
IRAS F06076-2139N & $\dots$ & $\dots$ & $\dots$ & $\dots$ & $\dots$ & $\dots$ & $<$60.0 & $\dots$ & $\dots$ & $\dots$ \\
&  &  &  &  &  &  & (23.0) &  &  & \\
IRAS 07251-0248E & $\dots$ & $\dots$ & $\dots$ & $\dots$ & $\dots$ & $\dots$ & $<$70.0 & $\dots$ & $\dots$ & $\dots$ \\
&  &  &  &  &  &  &  &  &  & \\
IRAS 07251-0248W & $\dots$ & $\dots$ & $\dots$ & $\dots$ & $\dots$ & $\dots$ & $<$70.0 & $\dots$ & $\dots$ & $\dots$ \\
&  &  &  &  &  &  &  &  &  & \\
NGC 2623 & $\dots$ & $\dots$ & $\dots$ & $\dots$ & $\dots$ & $\dots$ & $<$210.0 & $\dots$ & $\dots$ & $<$183.4\\
&  &  &  &  &  &  & (23.0) &  &  & (12.9) \\
IRAS F08572+3915NW & $\dots$ & $\dots$ & $\dots$ & $\dots$ & $\dots$ & $\dots$ & $<$330.0 & $\dots$ & $\dots$ & $<$491.8 \\
&  &  &  &  &  &  & (31.0) &  &  & (49.7) \\
IRAS F10565+2448 & $\dots$ & $\dots$ & $\dots$ & $\dots$ & $\dots$ & $\dots$ & $<$200.0 & $\dots$ & $\dots$ & $\dots$ \\
&  &  &  &  &  &  & (30.0) &  &  & \\
IRAS F12112+0305N & $\dots$ & $\dots$ & $\dots$ & $\dots$ & $\dots$ & $\dots$ & $<$110.0 & $\dots$ & $\dots$ & $\dots$ \\
&  &  &  &  &  &  &  &  &  & \\
IRAS F12112+0305S & $\dots$ & $\dots$ & $\dots$ & $\dots$ & $\dots$ & $\dots$ & $<$110.0 & $\dots$ & $\dots$ & $\dots$ \\
&  &  &  &  &  &  &  &  &  & \\
UGC 08387 & $\dots$ & $\dots$ & $\dots$ & $\dots$ & $\dots$ & $\dots$ & $<$250.0 & $\dots$ & $\dots$ & $<$302.6\\
&  &  &  &  &  &  & (29.0) &  &  & (30.26) \\
NGC 5331S & $\dots$ & $\dots$ & $\dots$ & $\dots$ & $\dots$ & $\dots$ & $<$290.0 & $\dots$ & $\dots$ & $\dots$ \\
&  &  &  &  &  &  & (27.0) &  &  &  \\
IRAS F14348-1447N & $\dots$ &$\dots$ & $\dots$ & $\dots$ & $\dots$ & $\dots$ & $<$100.0 & $\dots$ & $\dots$ & $\dots$ \\
&  &  &  &  &  &  &  &  &  & \\
IRAS F14348-1447S & $\dots$ &$\dots$ & $\dots$ & $\dots$ & $\dots$ & $\dots$ & $<$100.0 & $\dots$ & $\dots$ & $\dots$ \\
&  &  &  &  &  &  &  &  &  & \\
Arp 220E & $\dots$ & $\dots$ & $\dots$ & $\dots$ & $\dots$ & $\dots$ & $<$610.0 & $\dots$ & $\dots$ & $\dots$ \\
&  &  &  &  &  &  & (21.0) &  &  & \\
Arp 220W & $\dots$ & $\dots$ & $\dots$ & $\dots$ & $\dots$ & $\dots$ & $<$610.0 & $\dots$ & $\dots$ & $\dots$ \\
&  &  &  &  &  &  & (21.0) &  &  & \\
NGC 6240N & $\dots$ & $\dots$ & $\dots$ & $\dots$ & $\dots$ & $<$27.6 & $\dots$ & $\dots$ & $\dots$ & $<$75.0 \\
&  &  &  &  &  & (6.9) &  &  &  & (7.5) \\
NGC 6240S & $\dots$ & $\dots$ & $\dots$ & $\dots$ & $\dots$ & $<$99.5 & $\dots$ & $\dots$ & $\dots$ & $<$370.0 \\
&  &  &  &  &  & (8.4) &  &  &  & (72.6) \\
IRAS F17138-1017 & $\dots$ & $\dots$ & $\dots$ & $\dots$ & $\dots$ & $\dots$ & $<$630.0 & $\dots$ & $\dots$ & $\dots$ \\
&  &  &  &  &  &  & (28.0) &  &  & \\
IRAS F17207-0014 & $\dots$ & $\dots$ & $\dots$ & $\dots$ & $\dots$ & $\dots$ & $<$200.0 & $\dots$ & $\dots$ & $\dots$ \\
&  &  &  &  &  &  & (25.0) &  &  & \\
IRAS 19542+1110 & $\dots$ & $\dots$ & $\dots$ & $\dots$ & $\dots$ & $\dots$ & $<$80.0 & $\dots$ & $\dots$ & $\dots$ \\
&  &  &  &  &  &  &  &  &  & \\
II Zw 096E & $<25.4$ & $\dots$ & $\dots$ & $\dots$ & $\dots$ & $\dots$ & $<$260.0 & $\dots$ & $<$155.39 & $\dots$ \\
& (0.21) &  &  &  &  &  & (35.0) &  & (1.62) & \\
IRAS F22491-1808 & $\dots$ & $\dots$ & $\dots$ & $\dots$ & $\dots$ & $\dots$ & $<$90.0 & $\dots$ & $\dots$ & $\dots$\\
&  &  &  &  &  &  &  &  &  & \\
NGC 7674 & $\dots$ & $\dots$ & $\dots$ & $<153.6$ & $\dots$ & $<426.2^{(c)}$ & $\dots$ & $<374.3$ & $\dots$ & $<840.8$\\
&  &  &  & (9.8) &  & (35.9) &  &  (15.2) &  & (17.1)\\
\enddata
\tablecomments{$^{(c)}$ value from VLT/VISIR $\mathrm{PAH2\_2}$.}
\end{deluxetable*}
\end{longrotatetable}

\begin{longrotatetable}
\begin{deluxetable*}{lcccccccccc}
\tablecaption{Nuclear Photometry (mJy, $\lambda > 20.0\,\rm\mu m$)\label{tab:photometry4}}
\tablewidth{1000pt}
\tabletypesize{\scriptsize}
\tablenum{2}
\tablehead{
\colhead{Galaxy} & \colhead{\emph{Spitzer} MIPS1} & 
\colhead{\emph{IRAS} IRAS2} & \colhead{ALMA Band 10} & 
\colhead{Band 9} & \colhead{Band 8} & 
\colhead{Band 7} & \colhead{Band 6} & 
\colhead{Band 5} & \colhead{Band 4} & \colhead{Band 3}\\ 
\colhead{} & \colhead{$24.0\,\mu{\rm m}$} & \colhead{$25.0\,\mu{\rm m}$} & \colhead{0.3 mm} & 
\colhead{0.4 mm} & \colhead{0.7 mm} & \colhead{1.0 mm} &
\colhead{1.2 mm} & \colhead{1.6 mm} & \colhead{2.1 mm} & \colhead{3.1 mm}\\
\colhead{} & \colhead{(18)} & \colhead{(28)} & \colhead{(2)} & 
\colhead{(4)} & \colhead{(3)} & \colhead{(20)} &
\colhead{(23)} & \colhead{(10)} & \colhead{(5)} & \colhead{(14)}
}
\startdata
CGCG 436-030 & $<$1230.0 & $<$1540.0 & $\dots$ & $\dots$ & $\dots$ & 3.54 & $\dots$ & $\dots$ & $\dots$ & 2.47 \\
& (62.0) & (48.0) &  &  &  & (0.36) &  &  &  & (0.27) \\
IRAS F01364-1042 & $<$255.0 & $<$440.0 & $\dots$ & $\dots$ & $\dots$ & 6.46 & 1.08 & $\dots$ & $\dots$ & 2.09 \\
& (13.0) & (36.0) &  &  &  & (0.66) & (0.11) &  &  & (0.23) \\
III Zw 035 & $<$761.0 & $<$1030.0 & $\dots$ & $\dots$ & $\dots$& 26.46 & 7.43 & $\dots$ & $\dots$ & 4.92 \\
& (38.0) & (59.0) &  &  &  &  (2.66) & (0.75) &  &  & (0.51) \\
NGC 0695 & $<$722.0 & $<$830.0 & $\dots$ & $\dots$ & $\dots$ & 0.20 & $\dots$ & $\dots$ & $\dots$ & $\dots$ \\
& (36.0) & (41.0) &  &  &  & (6.11E-02) &  &  &  &  \\
NGC 1614 & $<$6552.0 & $<$7500.0 & $\dots$ & $\dots$ & $\dots$ & 0.66 & 0.51 & 0.56 & $\dots$ & $\dots$ \\
& (328.0) & (25.0) &  &  &  & (0.10) & (6.83E-02) & (6.62E-02) &  & \\
IRAS F05189-2524 & $<$2546.0 & $<$3470.0 & $\dots$ & 78.88 & $\dots$ & $<$6.17 & 4.45 & $<$1.20 & $\dots$ & $\dots$\\
& (127.0) & (18.0) &  & (19.59) &  & (0.62) & (0.65) & (0.12) &  & \\
IRAS F06076-2139N & $\dots$ & $<$630.0 & $\dots$ & $\dots$ & $\dots$ & 4.86 & $\dots$ & $\dots$ & $\dots$ & $\dots$ \\
&  & (16.0) &  &  &  & (0.50) &  &  &  &  \\
IRAS 07251-0248E & $\dots$ & $<$660.0 & $\dots$ & $\dots$ & $\dots$ & 16.47 & $<$2.85 & $<$2.69 & $\dots$ & $\dots$ \\
&  & (33.0) &  &  &  & (1.67) & (0.29) & (0.27) &  & \\
IRAS 07251-0248W & $\dots$ & $<$660.0 & $\dots$ & $\dots$ & $\dots$ & 2.42 & $<$0.28 & $<$0.30 & $\dots$ & $\dots$ \\
&  & (33.0) &  &  &  & (0.35) & (3.08E-02) & (3.25E-02) &  & \\
NGC 2623 & $<$1399.0 & $<$1810.0 & $\dots$ & $\dots$ & $\dots$ & $\dots$ & 6.63 & $\dots$ & $\dots$ & $\dots$ \\
& (70.0) & (41.0) &  &  &  &  & (0.69) &  &  & \\
IRAS F08572+3915NW & $<$1444.0 & $<$1760.0 & $\dots$ & $\dots$ & $\dots$ & 2.87 & 0.89 & $<$0.67 & $\dots$ & $\dots$ \\
& (72.0) & (33.0) &  &  &  & (0.29) & (8.97E-02) & (6.84E-02) &  & \\
IRAS F10565+2448 & $<$976.0 & $<$1270.0 & $\dots$ & $\dots$ & $\dots$ & $\dots$ & 0.67 & $\dots$ & $\dots$ & $<$0.21\\
& (49.0) & (31.0) &  &  &  &  & (6.82E-02) &  &  & (2.18E-02)\\
IRAS F12112+0305N & $<$364.0 & $<$660.0 & $\dots$ & $\dots$ & $\dots$ & 7.46 & 3.14 & $\dots$ & $\dots$ & $<$1.28\\
& (18.0) & (54.0) &  &  &  & (0.75) & (0.31) &  &  & (0.14)\\
IRAS F12112+0305S & $<$364.0 & $<$660.0 & $\dots$ & $\dots$ & $\dots$ & 0.85 & 0.31 & $\dots$ & $\dots$ & $<$0.30\\
& (18.0) & (54.0) &  &  &  & (0.11) & (3.37E-02) &  &  & (7.00E-02)\\
UGC 08387 & $<$1070.0 & $<$1420.0 & $\dots$ & $\dots$ & $\dots$ & 6.61 & $\dots$ & $\dots$ & $\dots$ & $\dots$ \\
& (54.0) & (40.0) &  &  &  & (0.67) &  &  &  & \\
NGC 5331S & $\dots$ & $<$590.0 & $\dots$ & $\dots$ & $\dots$ & $\dots$ & 0.33 & $\dots$ & $\dots$ & $<$8.10E-02\\
&  & (51.0) &  &  &  &  & (4.91E-02) &  &  & (1.21E-02) \\
IRAS F14348-1447N & $<$393.0 & $<$550.0 & $\dots$ & $\dots$ & $\dots$ & $\dots$ & 1.36 & $<$0.63 & $\dots$ & $\dots$ \\
& (20.0) & (62.0) &  &  &  &  & (0.24) & (6.45E-02) &  & \\
IRAS F14348-1447S & $<$393.0 & $<$550.0 & $\dots$ & $\dots$ & $\dots$ & $\dots$ & 1.83 & $<$0.90 & $\dots$ & $\dots$ \\
& (20.0) & (62.0) &  &  &  &  & (0.27) & (9.09E-02) &  & \\
Arp 220E & $<$4010.0 & $<$8000.0 & 2293.03 & 916.37 & 300.56 & 139.76 & 37.10 & $\dots$ & 7.46 & 12.01\\
& (201.0) & (34.0) & (250.97) & (143.51) & (31.74) & (15.89) & (9.78) &  & (0.75) & (1.33)\\
Arp 220W & $<$4010.0 & $<$8000.0 & 2922.62 & 1497.40 & 607.58 & 316.70 & 120.24 & $\dots$ & 19.88 & 26.33\\
& (201.0) & (34.0) & (309.55) & (186.06) & (61.61) & (32.46) & (15.05) &  & (1.99) & (2.69)\\
NGC 6240N & $\dots$ & $<$3550.0 & $\dots$ & $\dots$ & $\dots$ & 2.12 & 1.19 & $\dots$ & $<$0.54 & 0.75\\
&  & (20.0) &  &  &  & (0.22) & (0.12) &  & (6.05E-02) & (7.67E-02)\\
NGC 6240S & $\dots$ & $<$3550.0 & $\dots$ & $\dots$ & $\dots$ & 4.99 & 3.72 & $\dots$ & $<$1.45 & 2.48\\
&  & (20.0) &  &  &  & (0.50) & (0.37) &  & (0.15) & (0.25)\\
IRAS F17138-1017 & $\dots$ & $<$2120.0 & $\dots$ & $\dots$ & $\dots$ & $\dots$ & 0.82 & $\dots$ & $\dots$ & $\dots$ \\
&  & (38.0) &  &  &  &  & (0.12) &  &  & \\
IRAS F17207-0014 & $\dots$ & $<$1610.0 & $\dots$ & 358.33 & 57.11 & 51.17 & 27.17 & 6.91 & $<$2.97 & $<$2.57\\
&  & (29.0) &  & (37.69) & (5.76) & (5.16) & (3.07) & (0.69) & (0.30) & (0.26)\\
IRAS 19542+1110 & $\dots$ & $<$770.0 & $\dots$ & $\dots$ & $\dots$ & $\dots$ & 1.35 & $<$0.96 & $\dots$ & 0.77\\
&  & (37.0) &  &  &  &  & (0.14) & (9.70E-02) &  & (8.34E-02)\\
II Zw 096E & $\dots$ & $<$2300.0 & $\dots$ & $\dots$ & $\dots$ & 6.64 & $\dots$ & $\dots$ & $\dots$ & 0.48\\
&  & (26.0) &  &  &  & (0.67) &  &  &  & (6.57E-02)\\
IRAS F22491-1808 & $<$433.0 & $<$540.0 & $\dots$ & $\dots$ & $\dots$ & 5.38 & 3.58 & 1.44 & $\dots$ & $\dots$ \\
& (22.0) & (67.0) &  &  &  & (0.55) & (0.36) & (0.15) &  & \\
NGC 7674 & $<$1600.0 & $<$1920.0 & $\dots$ & $\dots$ & $\dots$ & $\dots$ & 0.68 & $\dots$ & $\dots$ & $\dots$ \\
& (80.0) & (34.0) &  &  &  &  & (1.00E-02) &  &  &  \\
\enddata
\end{deluxetable*}
\end{longrotatetable}

\subsection{SED Modeling} \label{sec:modeling}

In order to quantify the nuclear properties of our sample, we generated SEDs with the nuclear photometry and modeled them using X-CIGALE~\citep{2022Yang} \edit1{, the newly developed branch of the Code Investigating GALaxy Emission ~\citep[CIGALE;][]{2019Boquien} that incorporates a new X-ray module. CIGALE is one of the SED fitting algorithms that include an AGN component in the fitting process. CIGALE allows flexible combinations of modules responsible for different physical components (e.g., stars, dust, AGN). A basic principle of the code is the conservation of energy between the amount of energy absorbed by the dust and that re-emitted by the dust at infrared and submillimeter wavelengths~\citep{2014Roehlly}. Users can input a set of model parameters, and the code generates the model SED and model fluxes for each filter based on each possible combination of these parameters. To incorporate all the filters involved into X-CIGALE, we obtain the transmission curve for each filter from the SVO Filter Profile Service \citep{2012Rodrigo,2020Rodrigo,2024Rodrigo}.}

\edit1{The calculation of $\chi^{2}$ for each model follows Equation (14) from \cite{2019Boquien}. The code then computes the likelihood for each model as $L=\exp(-\chi^{2}/2)$ by comparing the model fluxes with the observed fluxes. CIGALE can perform two types of analyses, i.e. maximum likelihood and Bayesian-like. In the maximum-likelihood analysis, CIGALE selects the model with the largest $L$ value and derives physical properties, such as stellar mass ($M_{*}$) and star formation rate (SFR), from this model. In the Bayesian-like analysis, CIGALE calculates the marginalized probability distribution function (PDF) for each physical property using the $L$ values of all models, and derives the probability-weighted mean and standard deviation from this PDF, which are output as the estimated value and uncertainty for each physical property.} The additional X-ray module added to CIGALE~\citep{2019Boquien} allows us to better account for the X-ray emission while determining SED properties and constraining the AGN nature of the SMBH.

While we restrict the physical size of our nuclear region of interest to be 100$-$150\,pc around the central point source to minimize light contamination from the host galaxy, we still have to account for emission from star clusters and dust in the circumnuclear region or along the line of sight.
The stellar component is built based on the \cite{2003Bruzual} population synthesis model with the \cite{1955Salpeter} initial mass function (IMF). In addition, we use a delayed star formation history (SFH) model, allowing for an optional exponential burst \citep{2018Malek}. The overall star formation rate (SFR) can be described as follows:

\begin{equation}
    \resizebox{.45\textwidth}{!}{${\rm SFR}\,(t)\propto
    \begin{cases}
    t\times {\rm exp}(-t/\tau_{0})/\tau_{0}^{2}\,, & 0\leq t\leq t_{1}\\
    t\times {\rm exp}(-t/\tau_{0})/\tau_{0}^{2}+k\times{\rm exp}[-(t-t_{1})/\tau_{1}]\,, & t\geq t_{1}
    \end{cases}$}
\end{equation}

\noindent where $t_{1}$ represents the time of the onset of the starburst, while $\tau_{0}$ and $\tau_{1}$ are the e-folding times of the main population and the starburst population, respectively. The $k$ denotes the relative amplitude of the starburst component, which is calculated based on the fraction of stars formed in the starburst relative to the total mass of stars. We set the possible values for the age of the burst to 1.0 and 5.0 Myrs, as mentioned in \cite{2013Inami}.

As for dust attenuation, a modified \cite{2000Calzetti} starburst attenuation law is applied \citep{2009Noll}. The normalization of the curve affecting the continuum is determined by the color excess of the emission lines and by the reduction factor between the color excess of the stellar continuum and that of the emission lines. For dust emission properties we adopt the \cite{2014Dale} model. In this set of templates, the emission from the dust heated by star-forming activity is parameterized by only one parameter $\alpha$ defined in the following \edit1{equation}:

\begin{equation}
 {\rm d}M_{\rm dust}(U)\propto U^{-\alpha}\,{\rm d}U   
\end{equation}

\noindent where $M_{\rm dust}$ is the total mass of dust, and $U$ is the radiation field intensity. The parameter $\alpha$ is linked to the 60 to 100 $\mu$m color. The main advantage of this model is its simplicity, featuring only \edit1{two parameters, $\alpha$ and AGN fraction($f_{\mathrm{AGN}}$). Additionally, since we have also incorporated an AGN emission model, we set the $f_{\mathrm{AGN}}$ to 0, as recommended.} Therefore, we can minimize the numbers of free parameters while modeling the SEDs.

To accurately account for AGN emission, we incorporate the SKIRTOR template \citep{2012Stalevski,2016Stalevski} into the modeling procedure, \edit1{as it is the recommended UV-to-IR SED model for AGN when using X-CIGALE \citep{2020Yang}}. SKIRTOR assumes a two-phase clumpy torus that consists of high-density dust clumps \edit1{(mass fraction=97\%)} and a small amount of smoothly distributed dust, as suggested by recent theoretical and observational works \citep{2009Nikutta,2012Ichikawa,2012Stalevski,2019Tanimoto}. \edit1{The torus density is proportional to $r^{-p}e^{-q|\cos(\theta)|}$, where $r$ is the radius, $\theta$ is the viewing angle, and both $p$ and $q$ are fixed to 1 in X-CIGALE.}
In addition, X-CIGALE includes extinction by dust along the polar directions of the AGN structure to account for a non-negligible amount of obscuration in some Type 1 AGN \citep{2012Bongiorno,2012Elvis,2017Stalevski,2019Stalevski,2018Lyu}. \edit1{The SMC extinction curve \citep{1984Prevot}, which is favored by AGN observations \citep[e.g.,][]{2004Hopkins}, is used as the extinction curve for the polar dust.} However, it has been found that X-CIGALE cannot accurately constrain the extinction along the polar direction $E(B-V)_{\rm polar}$ \citep{2021Mountrichas}. Therefore, we followed the treatment presented in \cite{2021Paspaliaris}, which modeled a similar sample of local (U)LIRGs and adopted 0 and 0.8 as the values for this parameter. \edit1{The re-emission from the polar dust is modeled as a "gray body with a temperature of 100 K and an emissivity index of 1.6. This emission is assumed to be isotropic, contributing to the total IR luminosity of both type 1 and type 2 AGN.}

\edit1{The emission of the accretion disk is updated with the SED template of \cite{2012Feltre}, while the angular dependence of the UV/optical radiation can be approximated as $L(\theta)\propto\cos\theta(1+2\cos\theta)$, where $\theta$ is the viewing angle. Here, we adopt the same approach as \cite{2020Yang}, considering two viewing angles: $30^{\circ}$ for type 1 AGN and $70^{\circ}$ for type 2 AGN. The contribution of AGN is quantified by $f_{\mathrm{AGN}}$, defined as the fraction of AGN IR luminosity to total IR luminosity. The X-ray emission originating from re-processed UV/optical photons via inverse Compton scattering is connected to AGN emission at other wavelengths via the $\alpha_{\mathrm{OX}}-L_{2500\mathrm{\AA}}$ relation: 
\begin{equation}
    \alpha_{\mathrm{OX}}=-0.3838\log(L_{2500\mathrm{\AA}}/L_{2\,\mathrm{keV}})
\end{equation}
where $L_{2500\mathrm{\AA}}$ is the intrinsic UV luminosity per frequency at $2500\mathrm{\AA}$, and $\alpha_{\mathrm{OX}}$ is the spectral slope between UV ($2500\mathrm{\AA}$) and X-ray  (2 keV).} The improvements in CIGALE v2022.1 are mainly related to the anisotropy of AGN X-ray emission and the shape of the SED of AGN accretion disk \citep{2022Yang}. \edit1{The disk continuum from the SKIRTOR model can now be expressed as:}
\begin{equation}
\resizebox{.45\textwidth}{!}{$\lambda L_{\lambda}\propto
    \begin{cases}
    \lambda^{2}, & 0.008\leq\lambda\leq0.05\,[\mu m]\\
    \lambda^{0.8}\,, & 0.05<\lambda\leq0.125\,[\mu m]\\
    \lambda^{-0.5+\delta_{\mathrm{AGN}}}\,, & 0.125<\lambda\leq10\,[\mu m]\\
    \lambda^{-3}\,, & 10<\lambda\leq1000\,[\mu m]\\
    \end{cases}$}
\end{equation}
\edit1{where $\delta_{\mathrm{AGN}}$ is a free parameter introduced to allow for deviations from the default optical spectral slope.} In addition to the default value ($-0.36$), we include an option for a slightly steeper optical spectral slope ($-0.5$) for the disk, to better accommodate some luminous AGN. The final input parameter grid is presented in Table \ref{tab:parameters}. 

\begin{deluxetable*}{lccc}
\tablenum{3}
\tablecaption{Summary of the Models and Input Parameters in the SED Modeling\label{tab:parameters}}
\tablewidth{0pt}
\tablehead{
\colhead{Model} & \colhead{Module} & \colhead{Parameters} & \colhead{Value}
}
\startdata
Star formation history & sfhdelayed & $e$-folding time of the main stellar population (Myr) & 100, 500, 1000, 5000\\
& & Age of the main stellar population (Myr)& 500, 1000, 3000, 5000\\
& & & 7000, 9000, 12000\\
& & $e$-folding time of the late starburst population (Myr) & 50.0 \\
& & Age of the late burst (Myr)& 1.0, 5.0\\
& & Mass fraction of the late burst population & 0.0, 0.1, 0.3, 0.5, 0.7, 0.9\\
& & Normalise the SFH to produce 1 $M_{\odot}$ & True \\
Single stellar population & bc03 & Initial mass function & 0 (Salpeter) \\
& & Metalicity ($Z_{\odot}$) & 0.02 \\
& & Age of the separation between the young and the old star populations (Myr) & 10.0 \\
Nebular emission & nebular & Ionization parameter & -2.0 \\
& & Gas metallicity ($Z_{\odot}$) & 0.02 \\
& & Electron density ($\mathrm{cm}^{-3}$) & 100 \\
& & Fraction of Lyman continuum photons escaping the galaxy/absorbed by dust & 0.0/0.0 \\
& & Lines width ($\mathrm{km\,s^{-1}}$) & 300.0 \\
& & Include nebular emission & True \\
Dust attenuation & dustatt\_modified\_starburst & Colour excess of the nebular lines light (mag) & 0.5, 1.0, 1.5, 2.0\\
& & & 2.5, 3.0, 3.5\\
& & Reduction factor to compute the stellar continuum attenuation & 0.44, 0.66\\
& & UV bump wavelength (nm) & 217.5 \\
& & UV bump width (nm) & 35.0 \\
& & UV bump amplitude & 3.0 \\
& & Slope delta of the power law modifying the attenuation curve & 0.0 \\
& & Extinction law for emission lines & 1 (MW) \\
& & Ratio of total to selective extinction for emission lines & 3.1 \\
& & Filters for which the attenuation will be computed & B\_B90, V\_B90, FUV \\
Dust emission & dale2014 & AGN fraction & 0.0 \\
& & $\alpha$ slope & 0.5, 1.0, 1.5, 2.0\\
& & & 2.5, 3.0, 3.5, 4.0\\
AGN & skirtor2016 & Average edge-on optical depth at 9.7 $\mathrm{\mu m}$ & 7 \\
& & Power-law exponent that sets radial gradient of dust density & 1.0 \\
& & Index that sets dust density gradient with polar angle & 1.0 \\
& & Angle measured between the equatorial plane and edge of the torus & $40^{\circ}$ \\
& & Ratio of outer to inner radius of the torus & 20 \\
& & Fraction of total dust mass inside clumps & 0.97 \\
& & AGN viewing angle & $30^{\circ}$(for type I Seyfert),\\
& & & $70^{\circ}$(for type II Seyfert)\\
& & Disk type & 1 (Schartmann spectrum)\\
& & Index that modifys the optical slope of the disk & -0.36, -0.5\\
& & AGN fraction & 0, 0.01, 0.025, 0.05, 0.075\\
& &  & 0.1, 0.25, 0.5, 0.75, 0.99\\
& & Wavelength range where to compute the AGN fraction & 0/0 \\
& & Extinction law of the polar dust & 0 (SMC) \\
& & Colour excess for the extinction in the polar direction (mag) & 0, 0.8\\
& & Temperature of the polar dust (K) & 100.0 \\
& & Emissivity index of the polar dust & 1.6 \\
X-ray & xray & Photon index of the AGN intrinsic X-ray spectrum & 1.8 \\
& & Exponential cutoff energy of the AGN spectrum (keV) & 300 \\
& & Power-law slope $\alpha_{\rm OX}$ connecting $L_{\nu}$ at rest-frame 2500$\mathrm{\AA}$ and 2 keV & -1.9, -1.8, -1.7, -1.6 \\
& & &-1.5, -1.4, -1.3, -1.2, -1.1 \\
& & Maximum allowed deviation of $|\Delta\,\alpha_{\rm OX}|_{\rm max}$ & 0.5\\
& & Angle coefficients of the AGN accretion disk X-ray emission & 0.5, 0 \\
& & Deviation from the expected low-/high-mass X-ray binary & 0, 0\\
\enddata
\end{deluxetable*}

Figure \ref{fig:allseds} shows the results of the SED fits to the 28 individual nuclei in all 23 galaxies. 
All of our SEDs feature 9 to 21 photometry points, constraining the multi-component models particularly in the optical, near-infrared, and submilliter wavelengths. 
For 20 out of the 23 galaxies in our sample, we derive both the soft and hard X-ray photometry data points that enable the computation of the X-ray colors (see discussion in \S \ref{sec:xray vs ir}) and leverage the X-ray module of X-CIGALE, anchoring the high-energy portion of the SED. It's noteworthy that both Arp 220 and IRAS 07251-0248 are observed to host dual nuclei in submillimeter observations. Despite this, \emph{Chandra} does not resolve the X-ray emission from their dual nuclei. Therefore, we use the integrated X-ray flux of these two galaxies in both SED modeling and X-ray analysis. 

The submillimeter photometry points from ALMA are crucial in constraining the dust components for these sources, but excess emission is detected in several cases. \edit1{As the data points that deviate the most from the models are those at the longest wavelengths, this deviation could be attributed to a non-negligible contribution of synchrotron emission, which may originate from supernovae, AGN, or a strong magnetic field. However, this component cannot be robustly constrained at ALMA wavelengths.}


\edit1{In many cases, the mid-infrared data points are largely missed by the fits.} This issue stems from that the spatial resolution provided by \emph{IRAS} and \emph{Spitzer} falls significantly short of our ability to resolve the inner $\sim200$ pc region. Consequently, the mid-infrared data points from these two instruments encompass the total fluxes of the galaxies. Thus, they represent upper limits to the mid-infrared fluxes of the resolved nuclear regions, making it impossible to precisely determine how much of the light comes from our regions of interest. This issue is exacerbated in the far-infrared (40$-$400\,$\mu$m) range, given the lack of a large, far-infrared observatory in the foreseeable future. The emission in that range is likely dominated by the diffuse, cold dust, dwarfing the intrinsic contribution from the AGN. Higher-resolution mid- and far-infrared imaging data would enable us to more accurately determine the locations and sizes of the infrared emitting regions, thereby better estimating the energy output of the obscured nuclei and constraining the shapes of our nuclear SEDs. 

Additionally, the near-infrared morphologies of IRAS F05189-2524 and IRAS F08572+3915 are dominated by PSFs that are larger than the apertures used in our analysis. This implies that the $\sim100$ pc nuclear regions cannot be resolved, which also poses challenges for accurately measuring the nuclear fluxes. 
\edit1{Although the factors mentioned above lead to significant discrepancies between some model fluxes and the corresponding observed fluxes, we notice that for some fits, the reduced $\chi^{2}$ values are relatively small. This is because when flux upper limits are present, X-CIGALE adjusts the way $\chi^{2}$ is calculated \citep[see Equations (15) in][]{2019Boquien}. As a result, the reduced $\chi^{2}$ may not accurately reflect the quality of the fit in these cases. However, considering that X-CIGALE provides acceptable fitting results for the high-resolution data points, and given our relatively small sample size, we have retained the SEDs from all sources for subsequent analysis.} Derived properties for each galaxy in our sample are presented in Appendix \ref{sec:derived_properties}.

\edit1{To verify the reliability of the physical properties obtained through SED fitting, we also conduct a so-called mock analysis, the results of which are presented in Appendix \ref{sec:appendix_mock_analysis}. In brief, the results of the mock analysis tell us that the physical properties derived from SED fitting based on the existing data sets have significant uncertainties. This suggests that in the absence of mid- to far-infrared data, we need to be cautious when interpreting the results from the SED fitting.}

\begin{figure*}[htbp]
	\centering
	\includegraphics[width=0.8\textwidth]{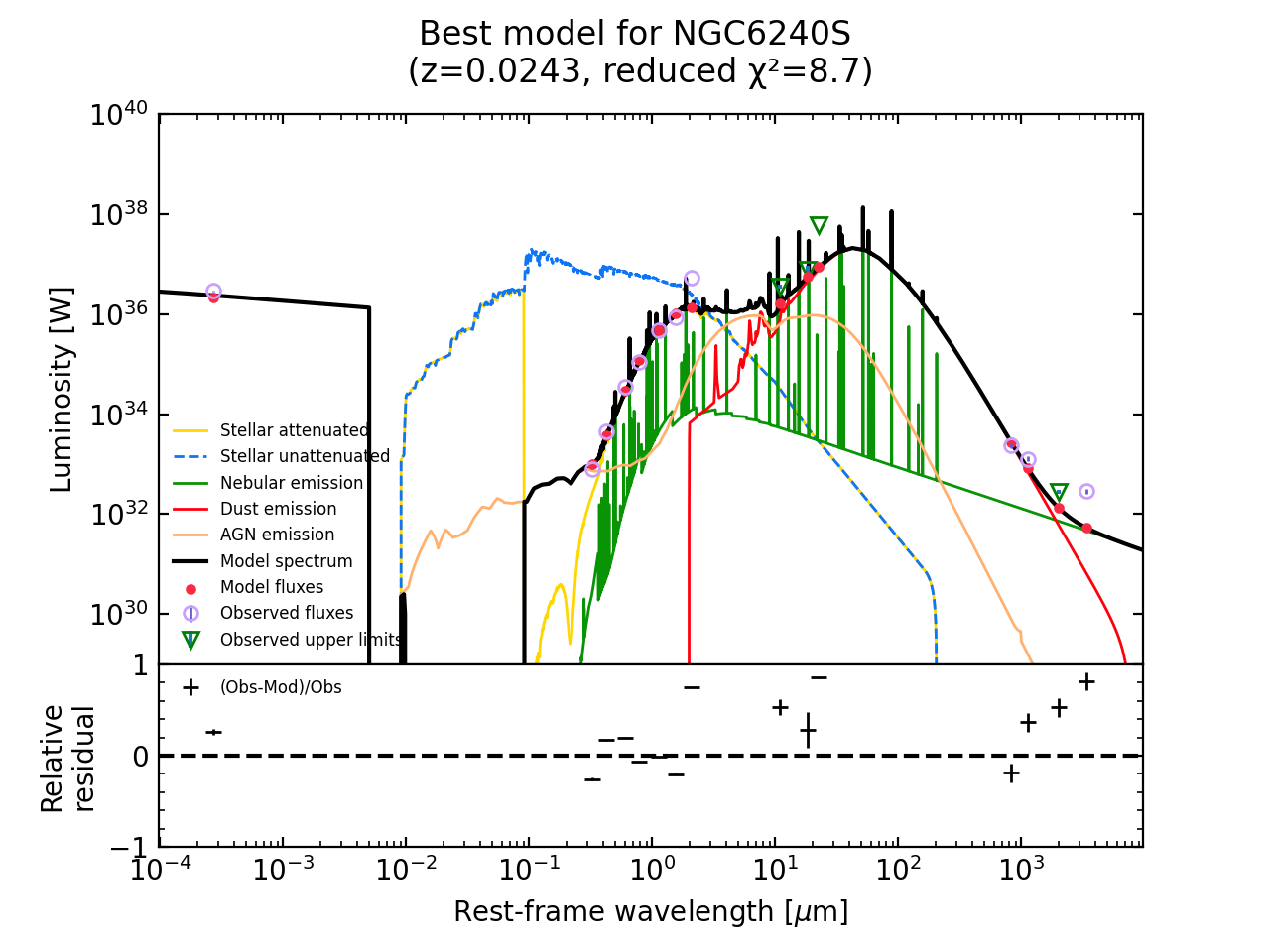}
        \includegraphics[width=0.45\textwidth]{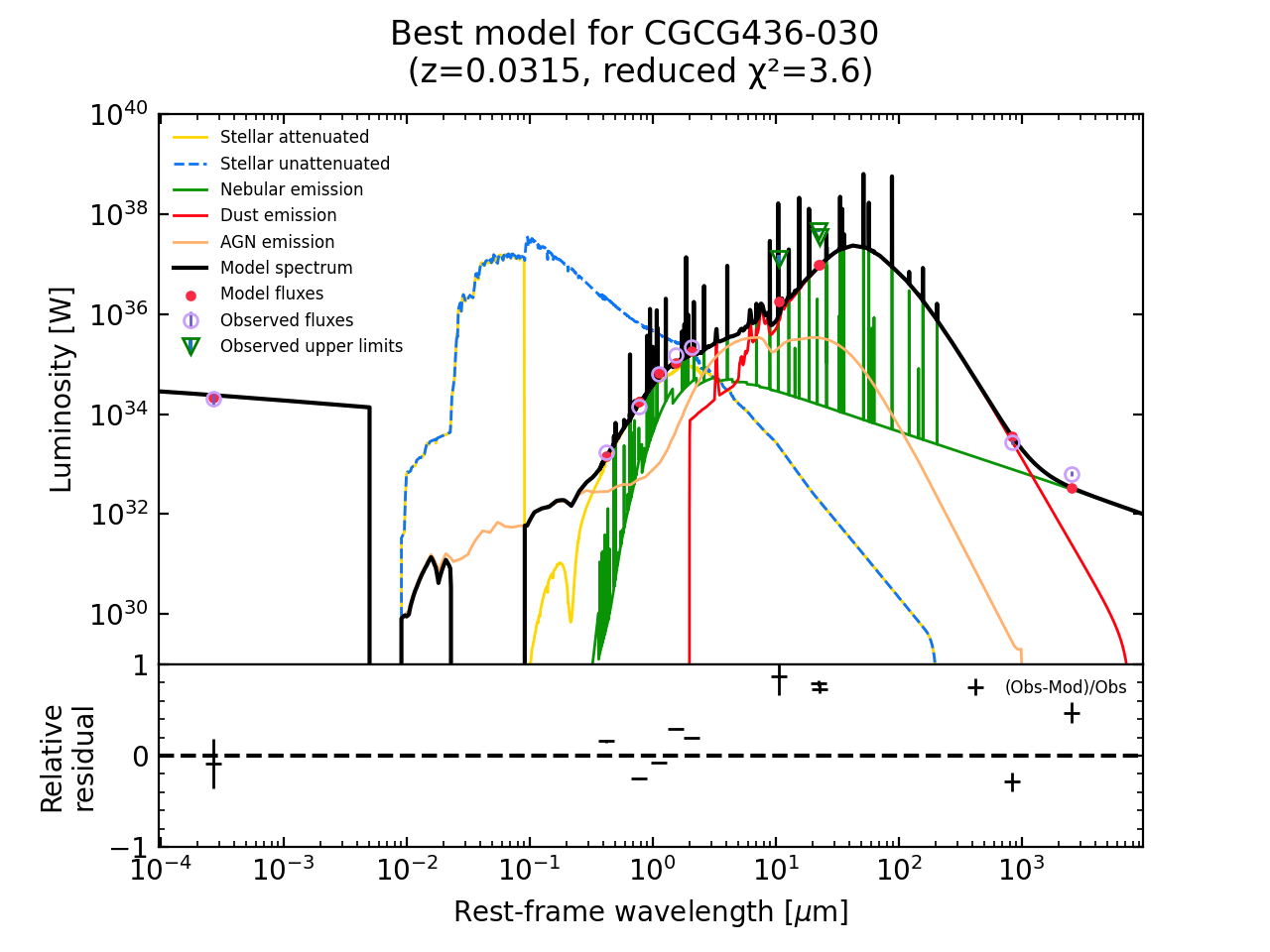}
        \includegraphics[width=0.45\textwidth]{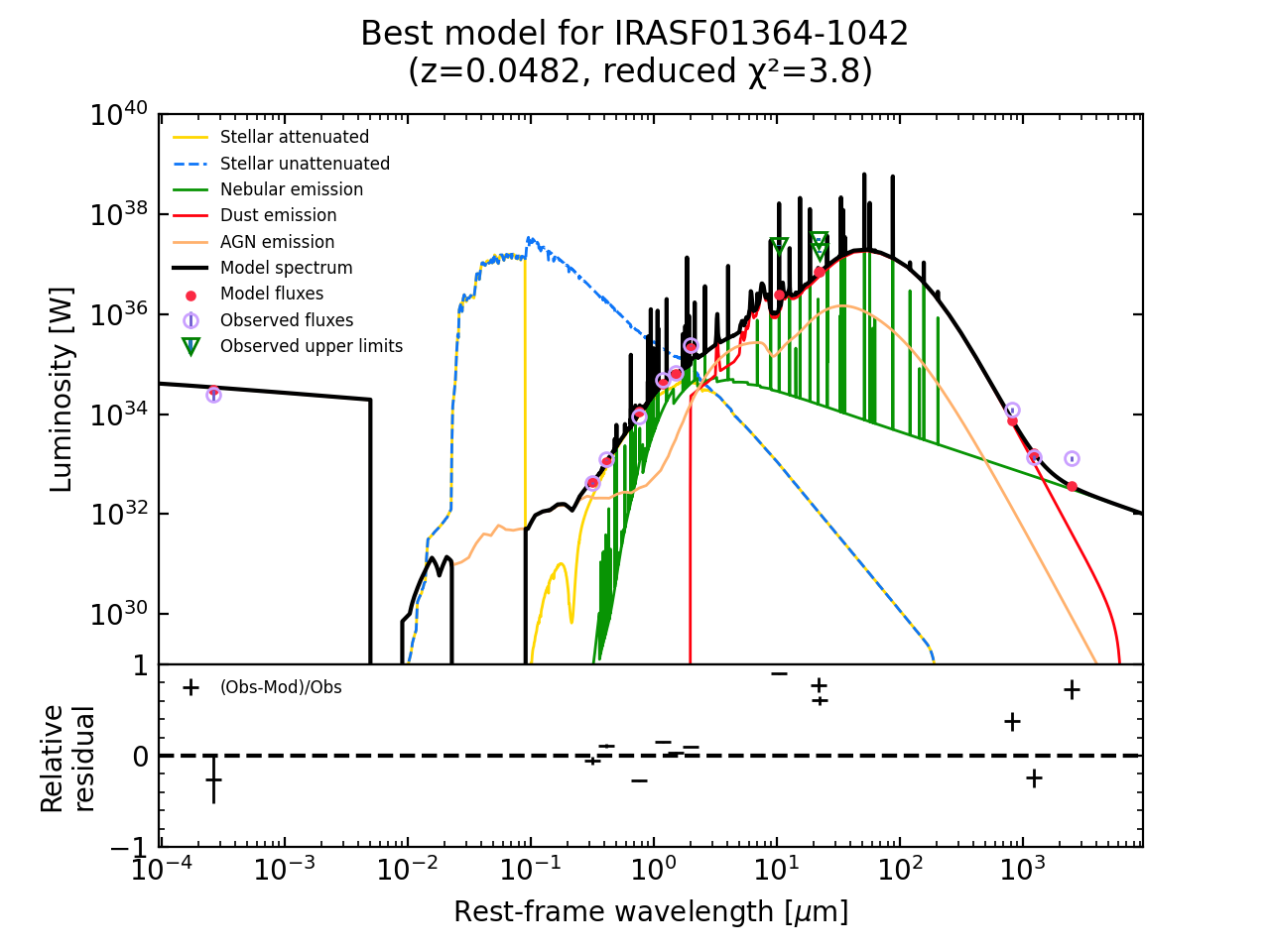}
        \includegraphics[width=0.45\textwidth]{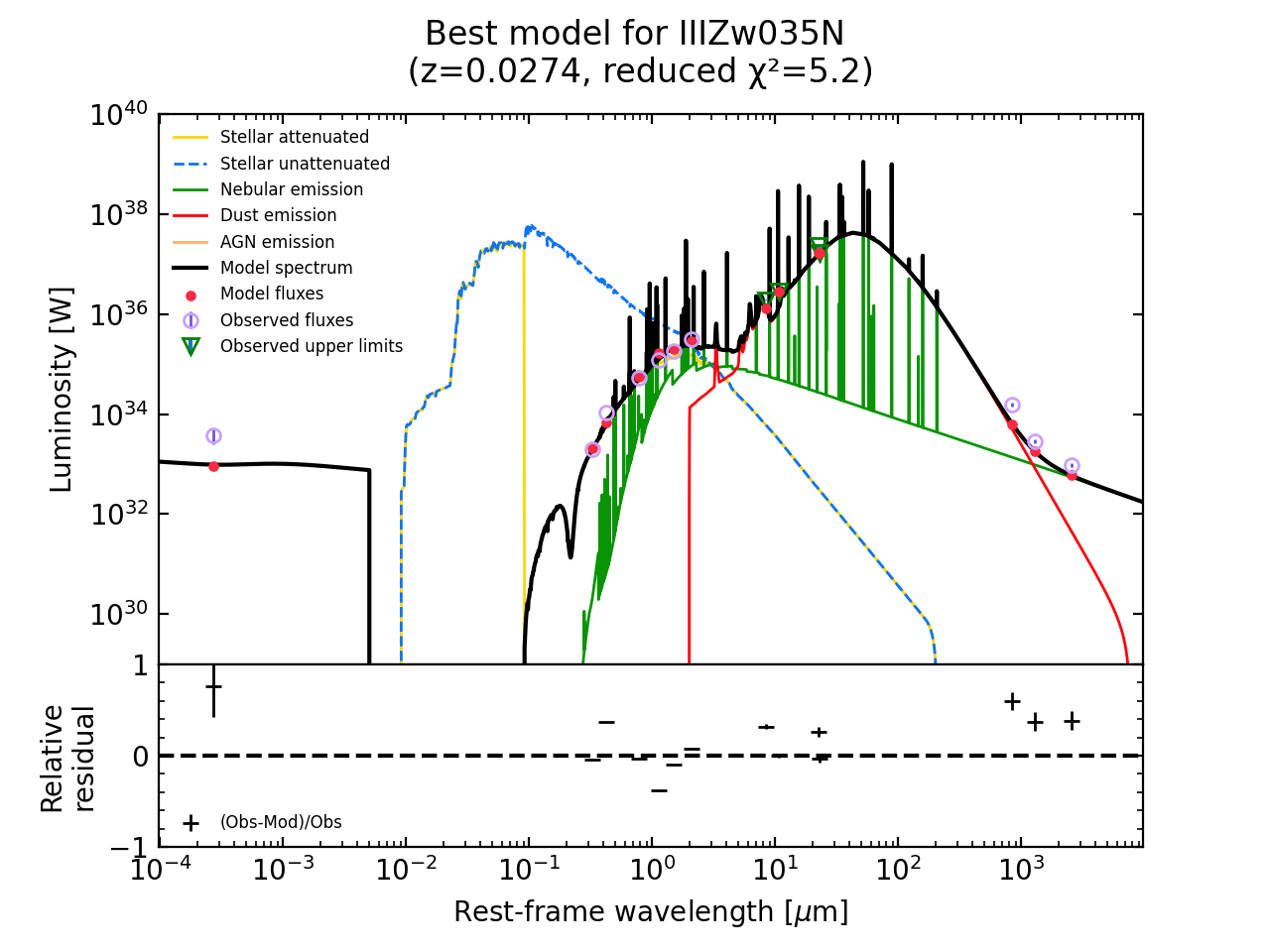}
        \includegraphics[width=0.45\textwidth]{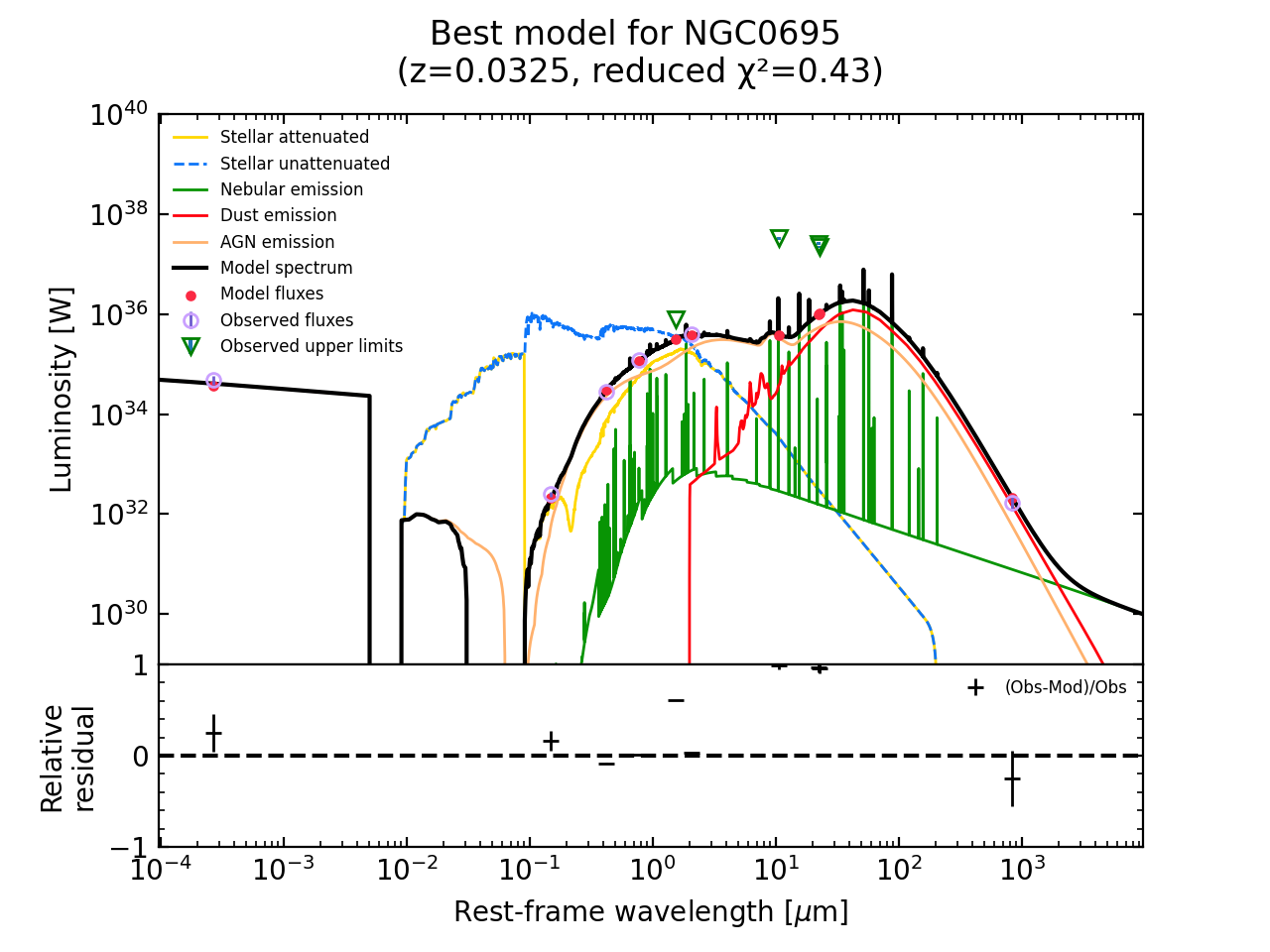}
	\caption{SED fits of all the nuclei in our sample. The observed fluxes are shown in open purple circles, while the upper limits are plotted as green downward-facing triangles. The model spectrum (black solid line) encompassing the individual components (stellar attenuated: yellow solid; stellar unattenuated: blue dashed; nebular emission: green solid; dust emission: red solid; AGN emission: orange solid) and the model fluxes (red circles) are labeled. The bottom plot within each panel shows the residuals at the wavelengths of the observed data points. The reduced $\chi^2$ is labeled in the title of each subplot.}
	\label{fig:allseds}
\end{figure*}

\addtocounter{figure}{-1}

\begin{figure*}[htbp]
	\centering
        \includegraphics[width=0.45\textwidth]{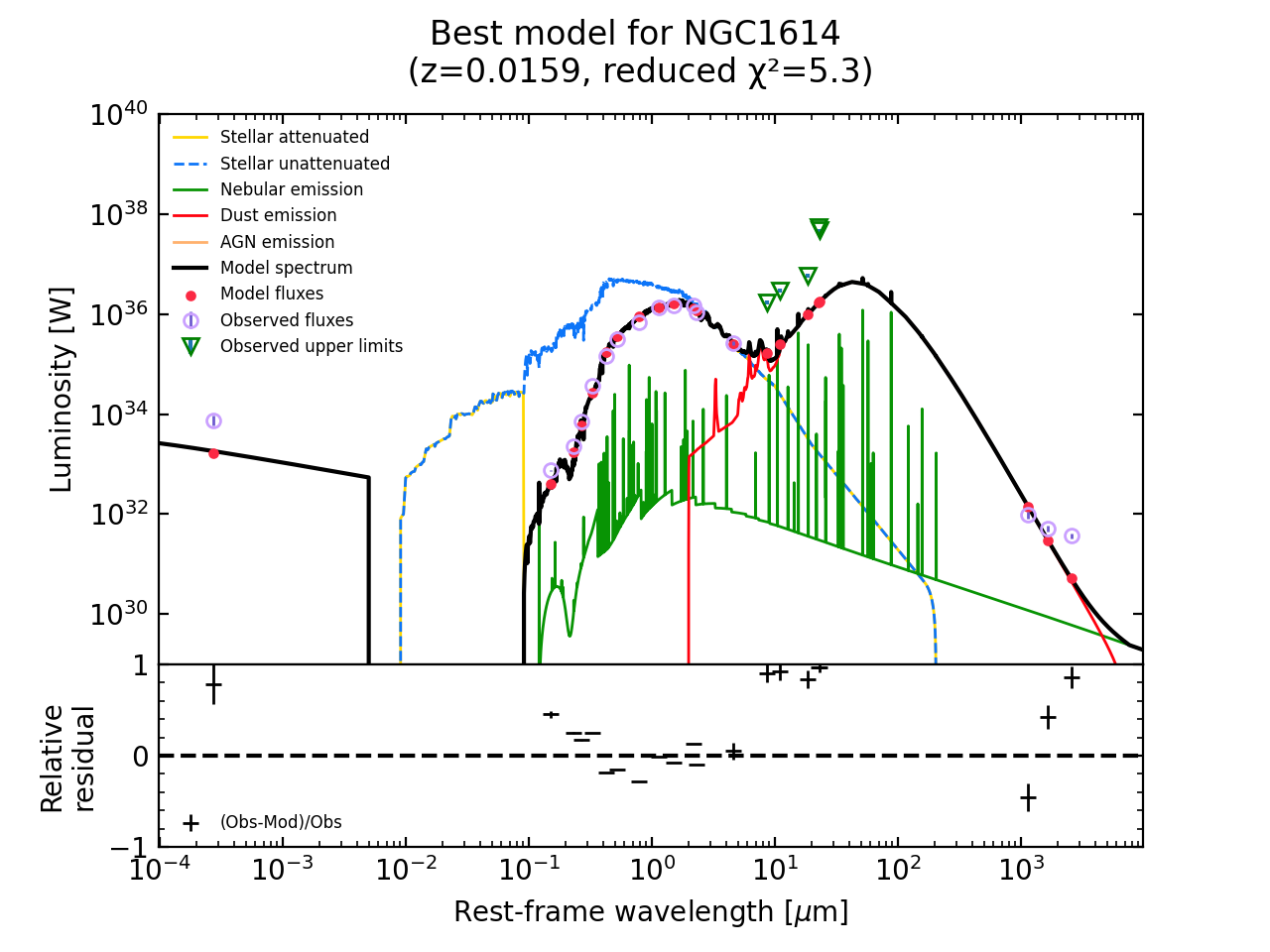}
        \includegraphics[width=0.45\textwidth]{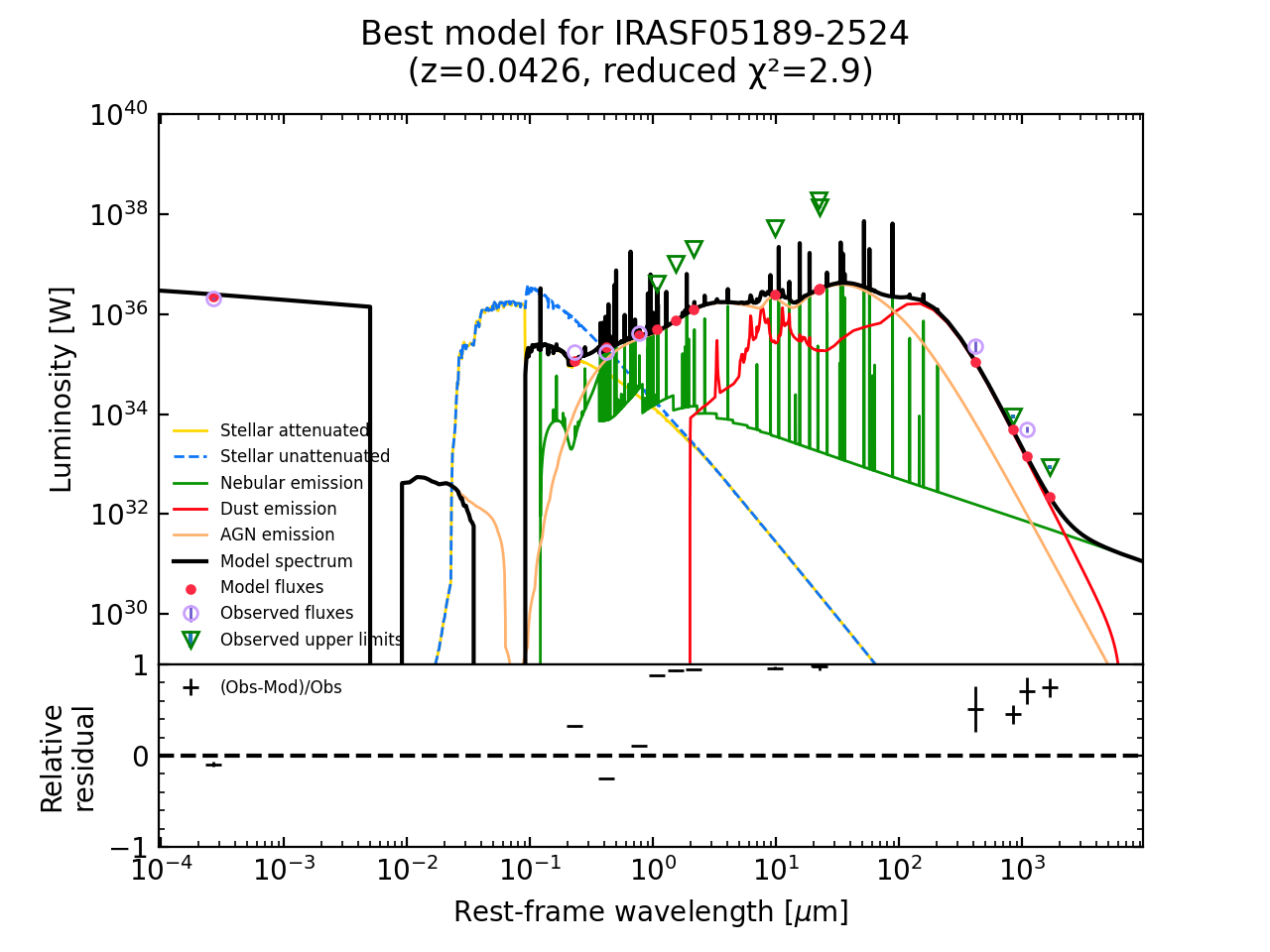}
        \includegraphics[width=0.45\textwidth]{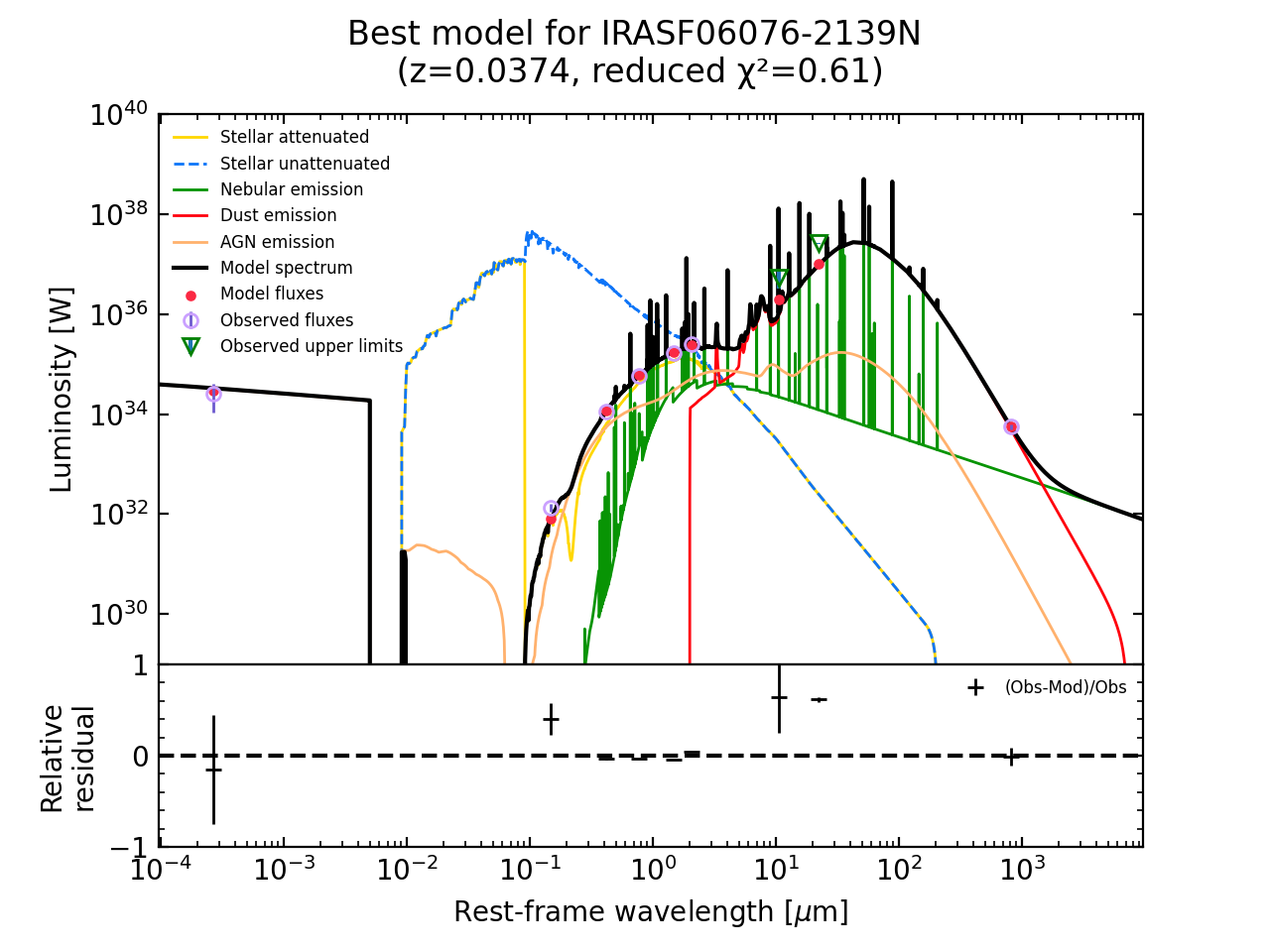}
        \includegraphics[width=0.45\textwidth]{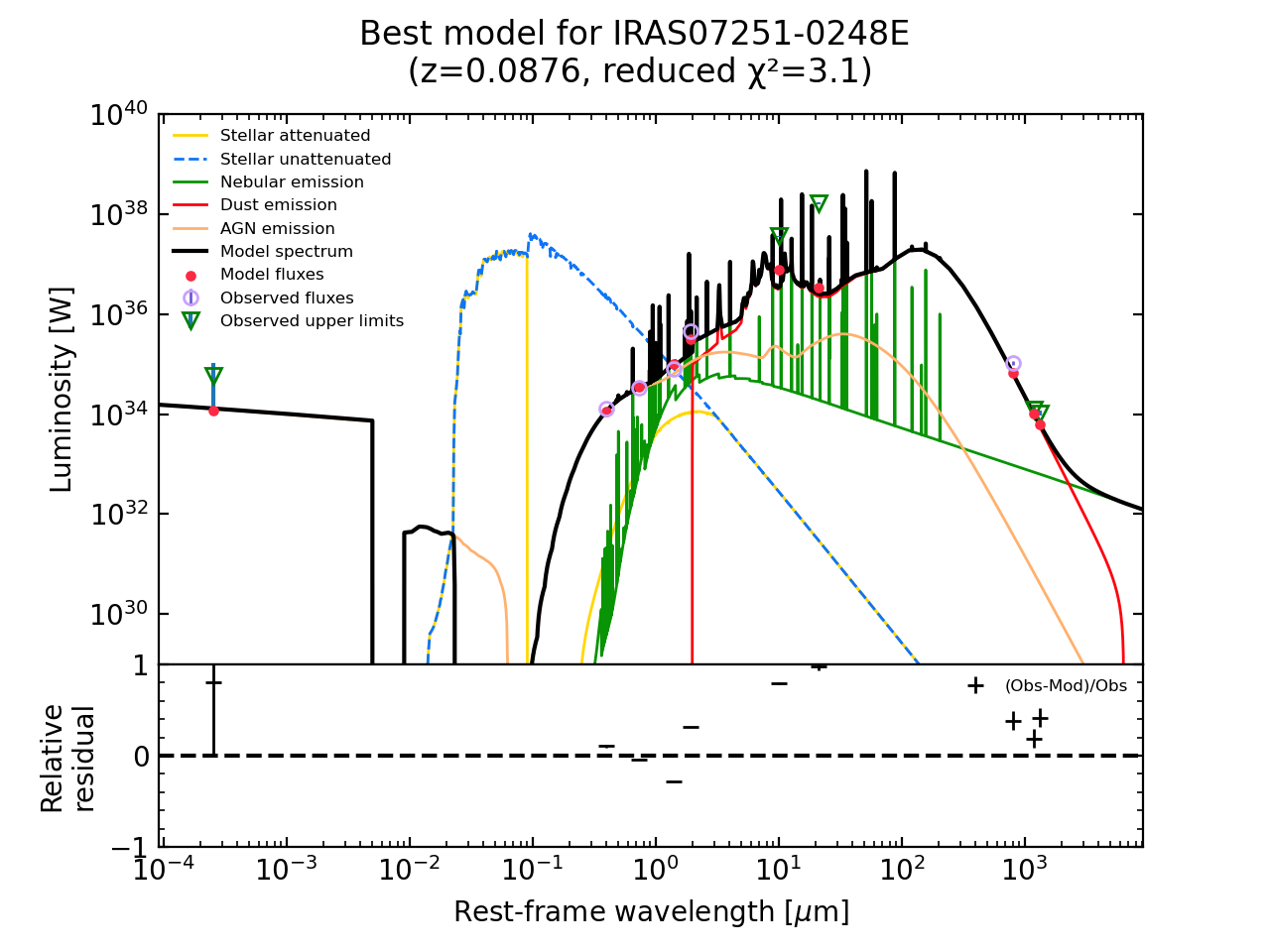}
        \includegraphics[width=0.45\textwidth]{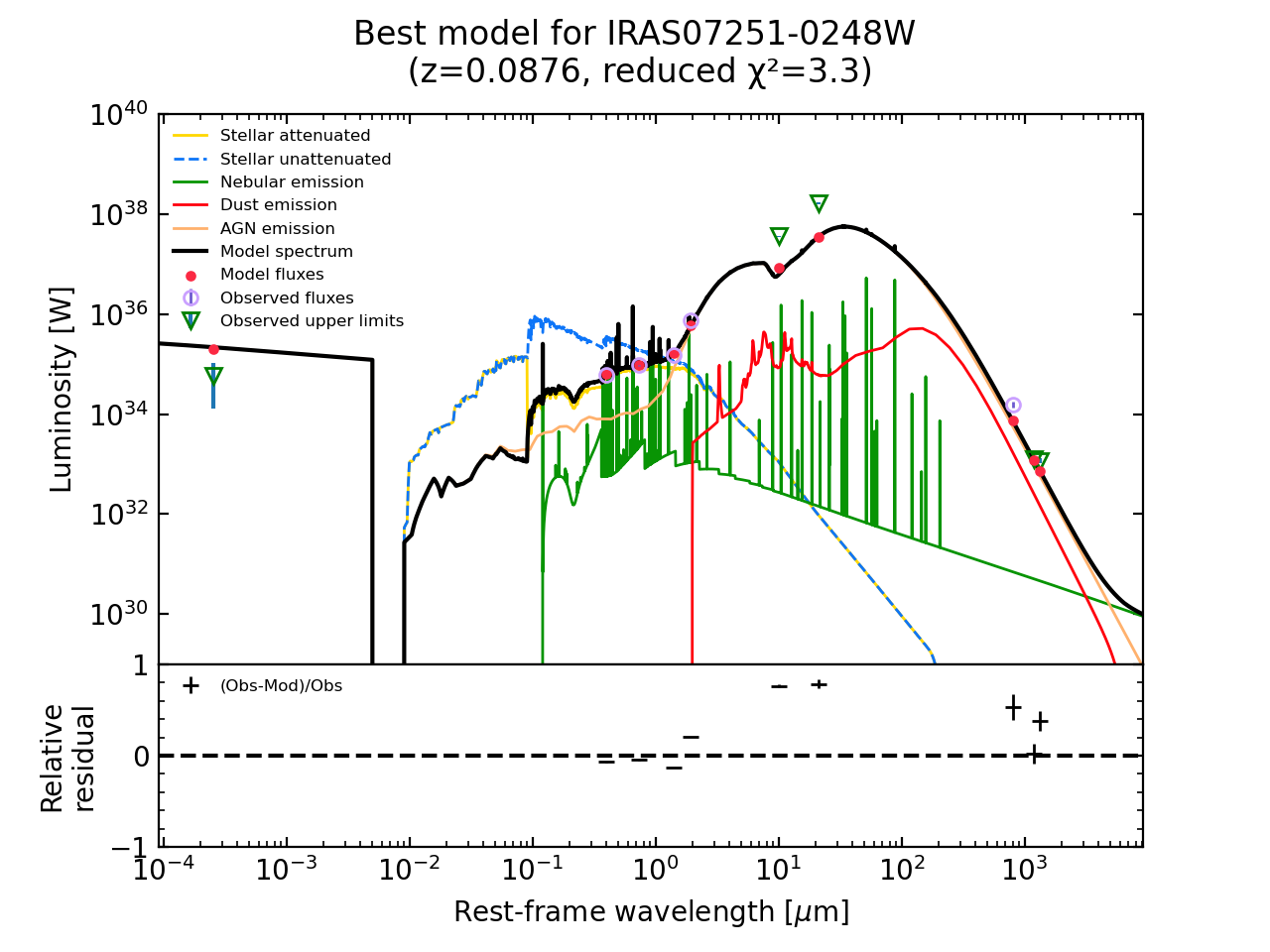}
        \includegraphics[width=0.45\textwidth]{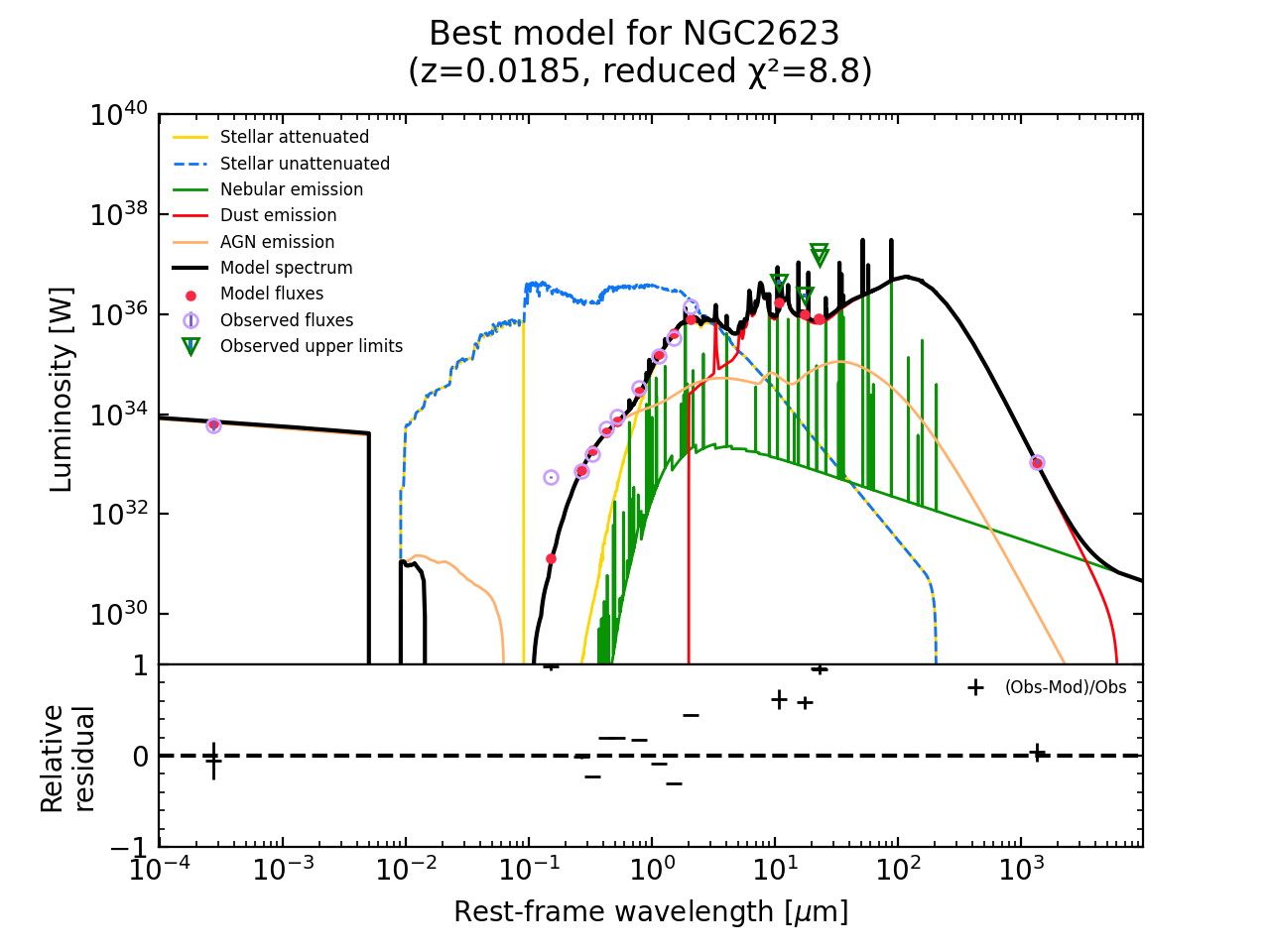}
        \includegraphics[width=0.45\textwidth]{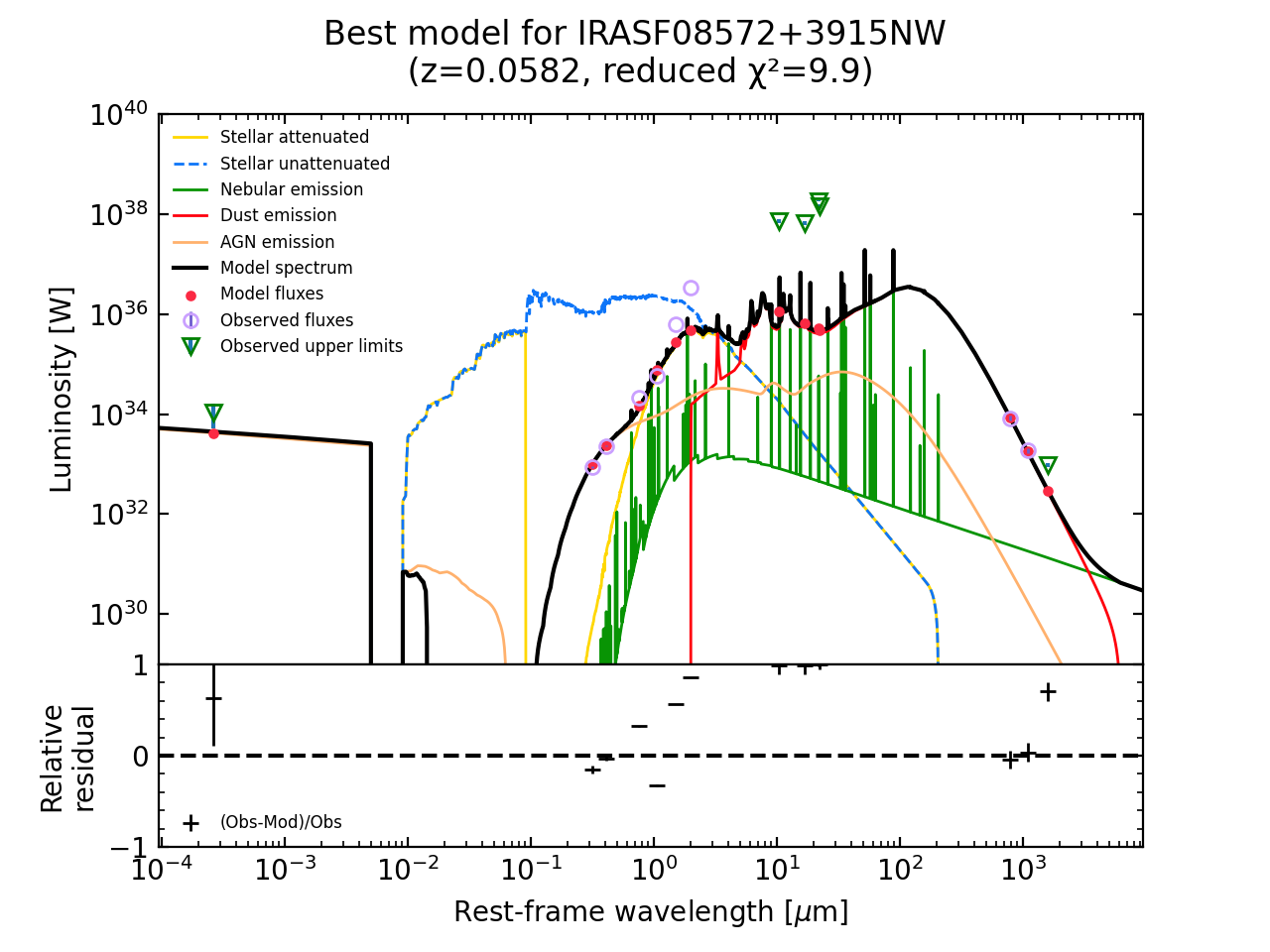}
        \includegraphics[width=0.45\textwidth]{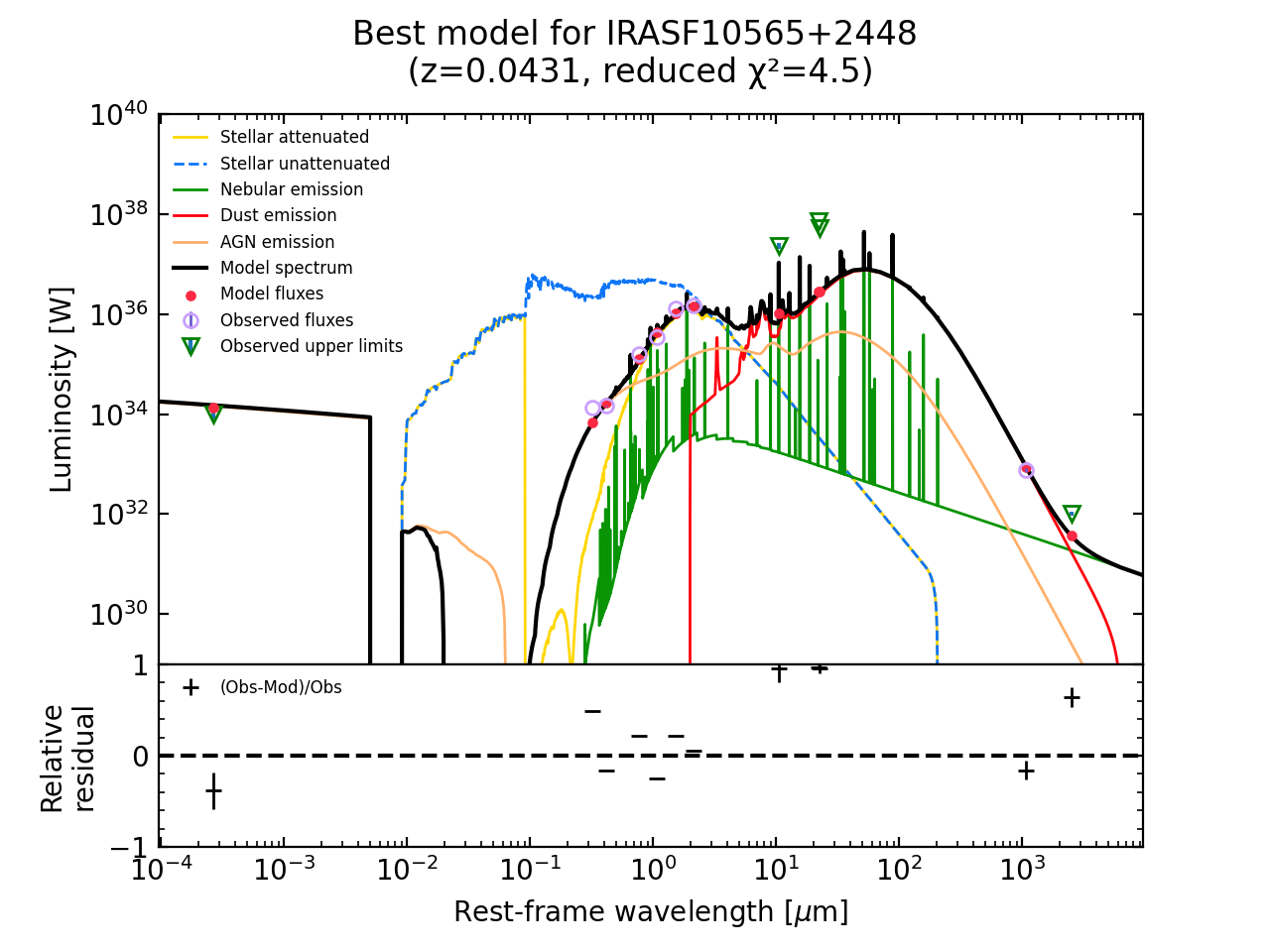}
	\caption{Continued}
	\label{fig:allseds2}
\end{figure*}

\addtocounter{figure}{-1}

\begin{figure*}[htbp]
	\centering
        \includegraphics[width=0.45\textwidth]{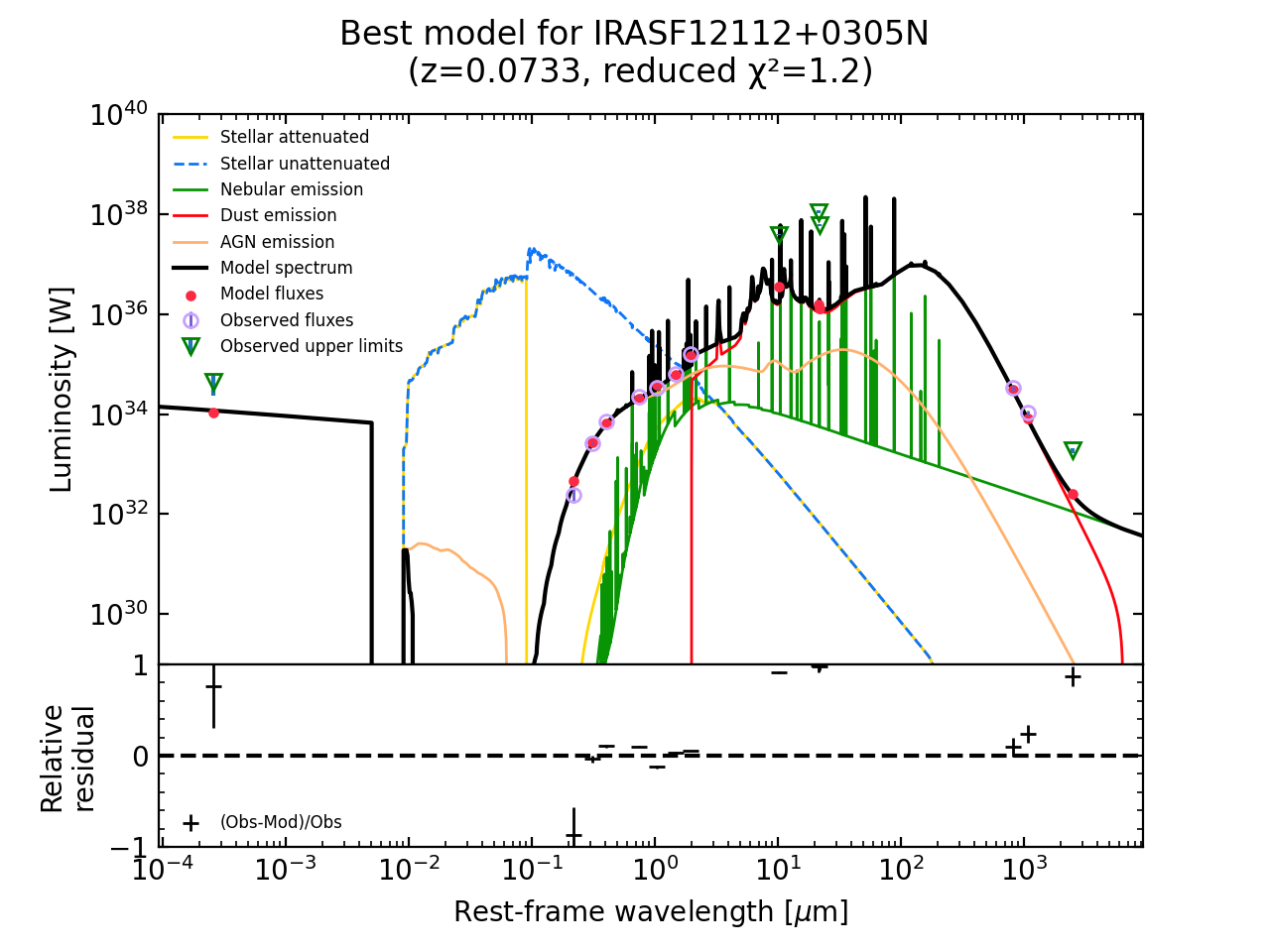}
        \includegraphics[width=0.45\textwidth]{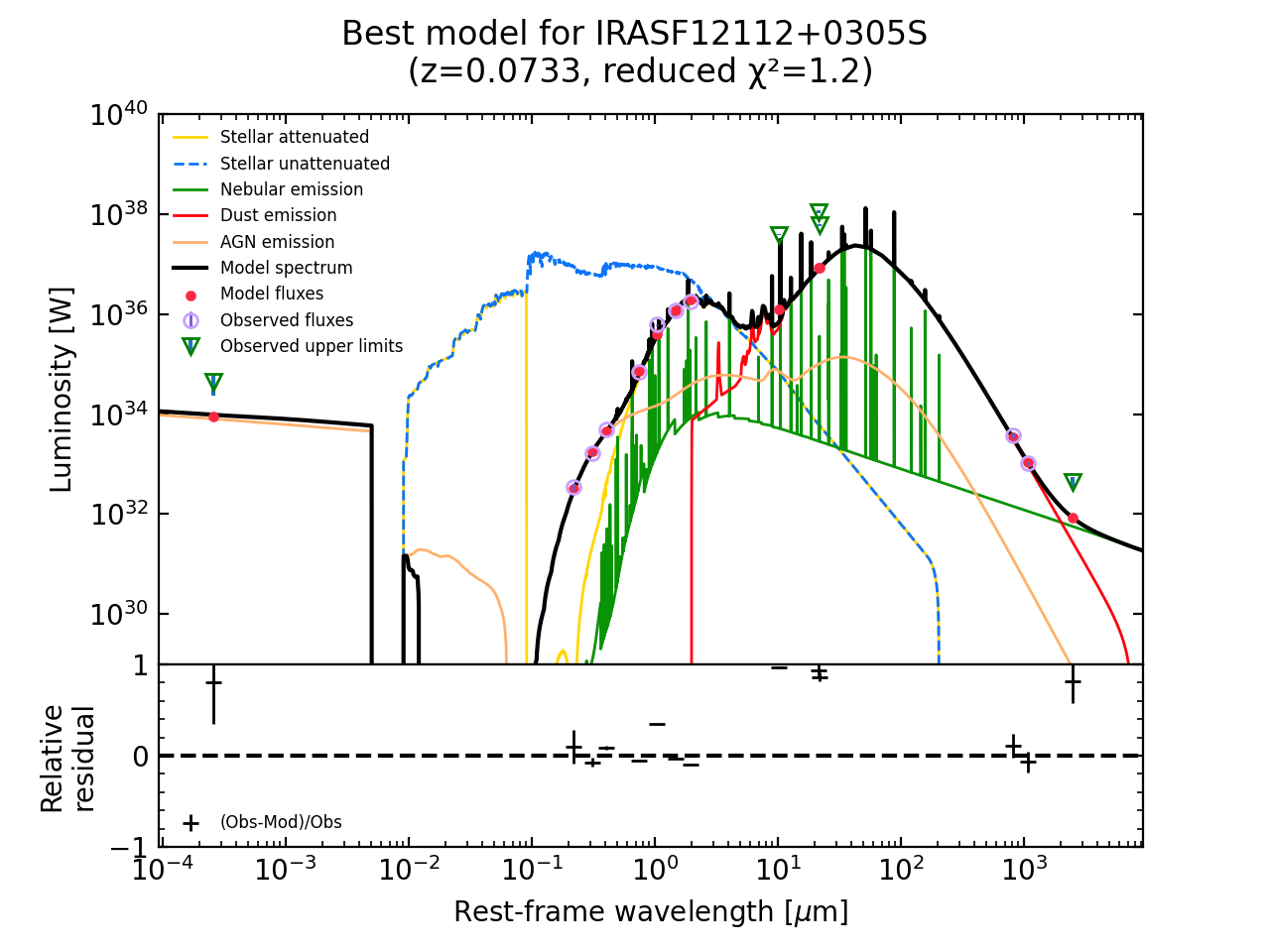}
        \includegraphics[width=0.45\textwidth]{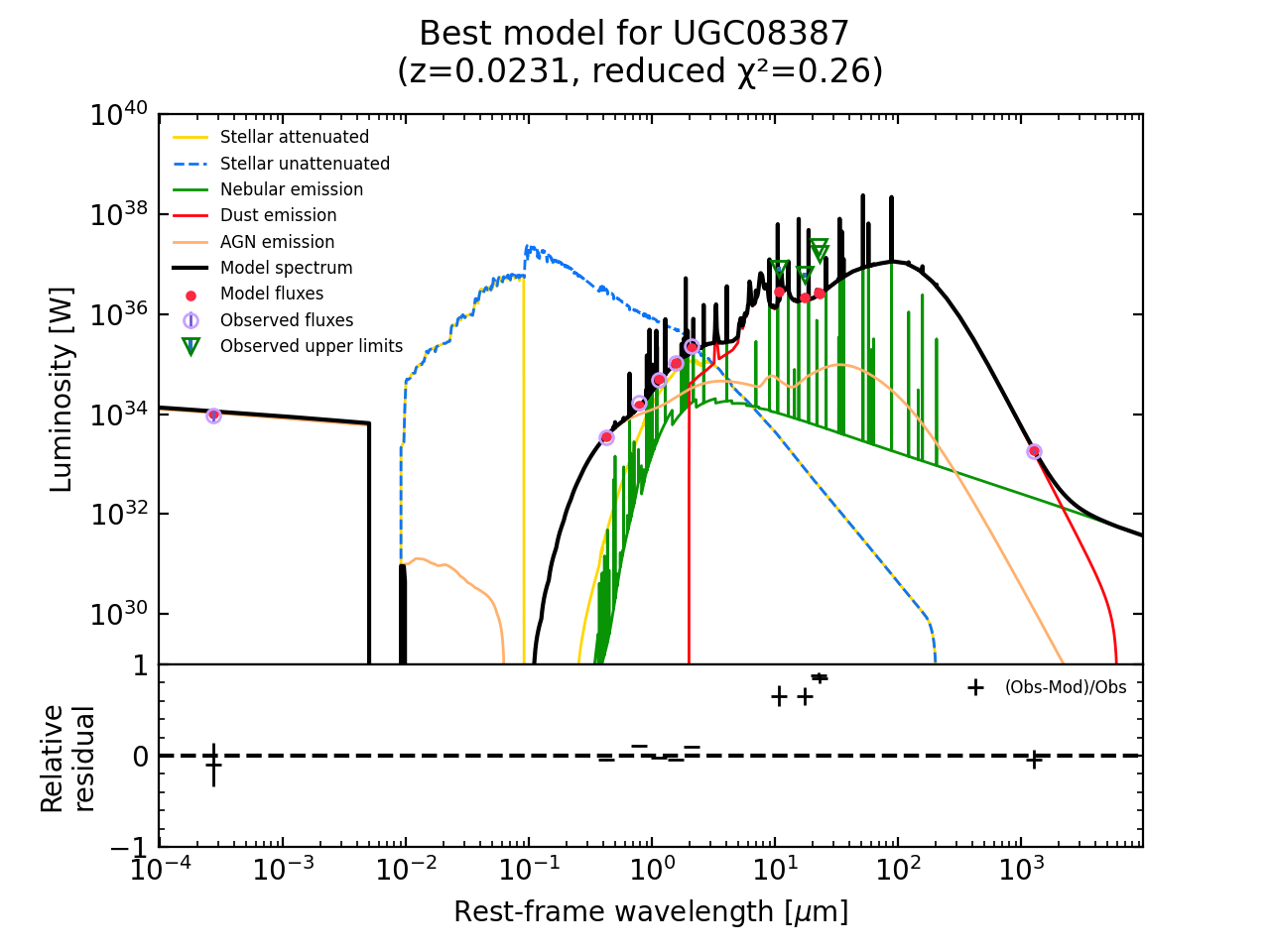}
        \includegraphics[width=0.45\textwidth]{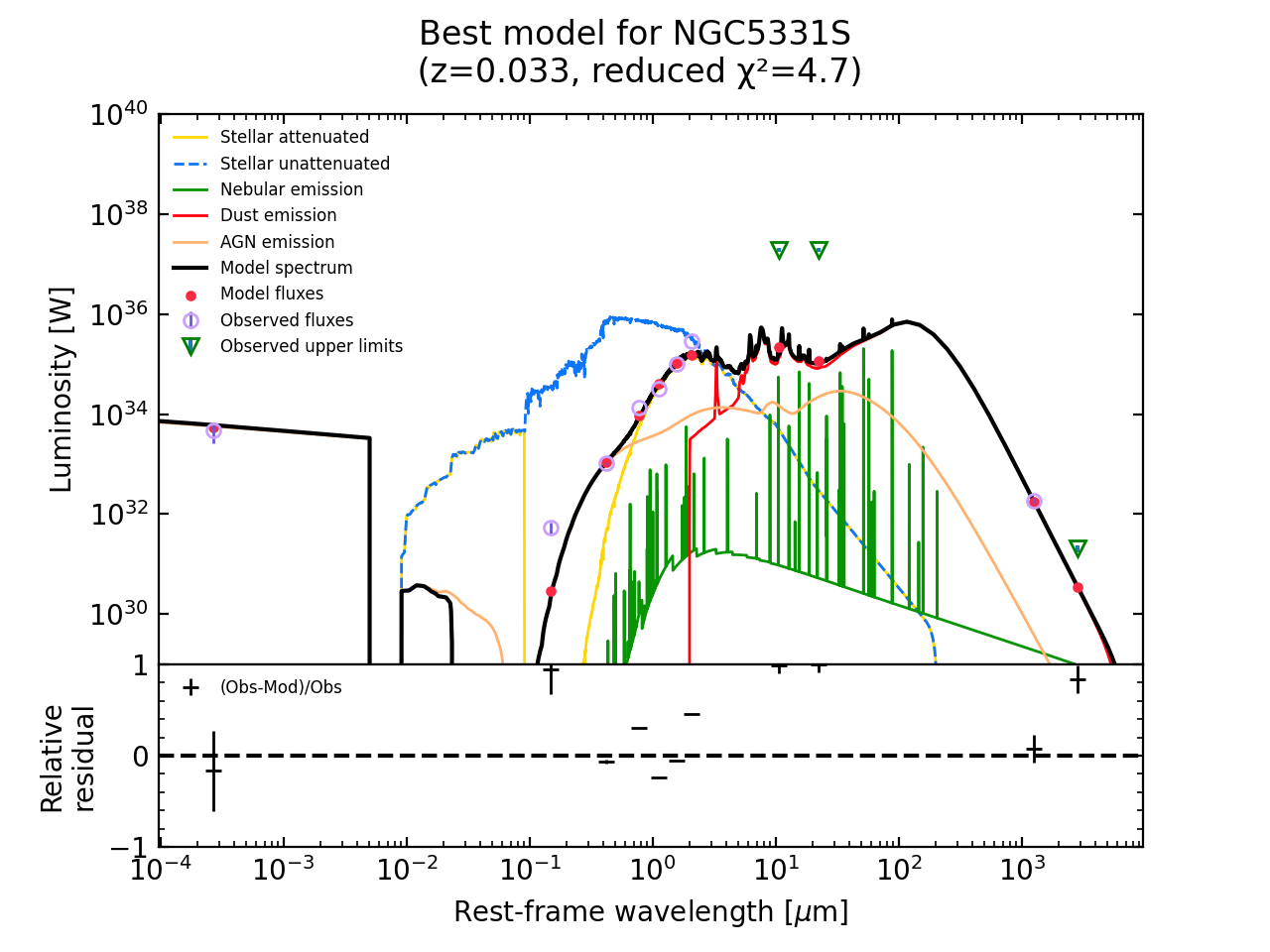}
        \includegraphics[width=0.45\textwidth]{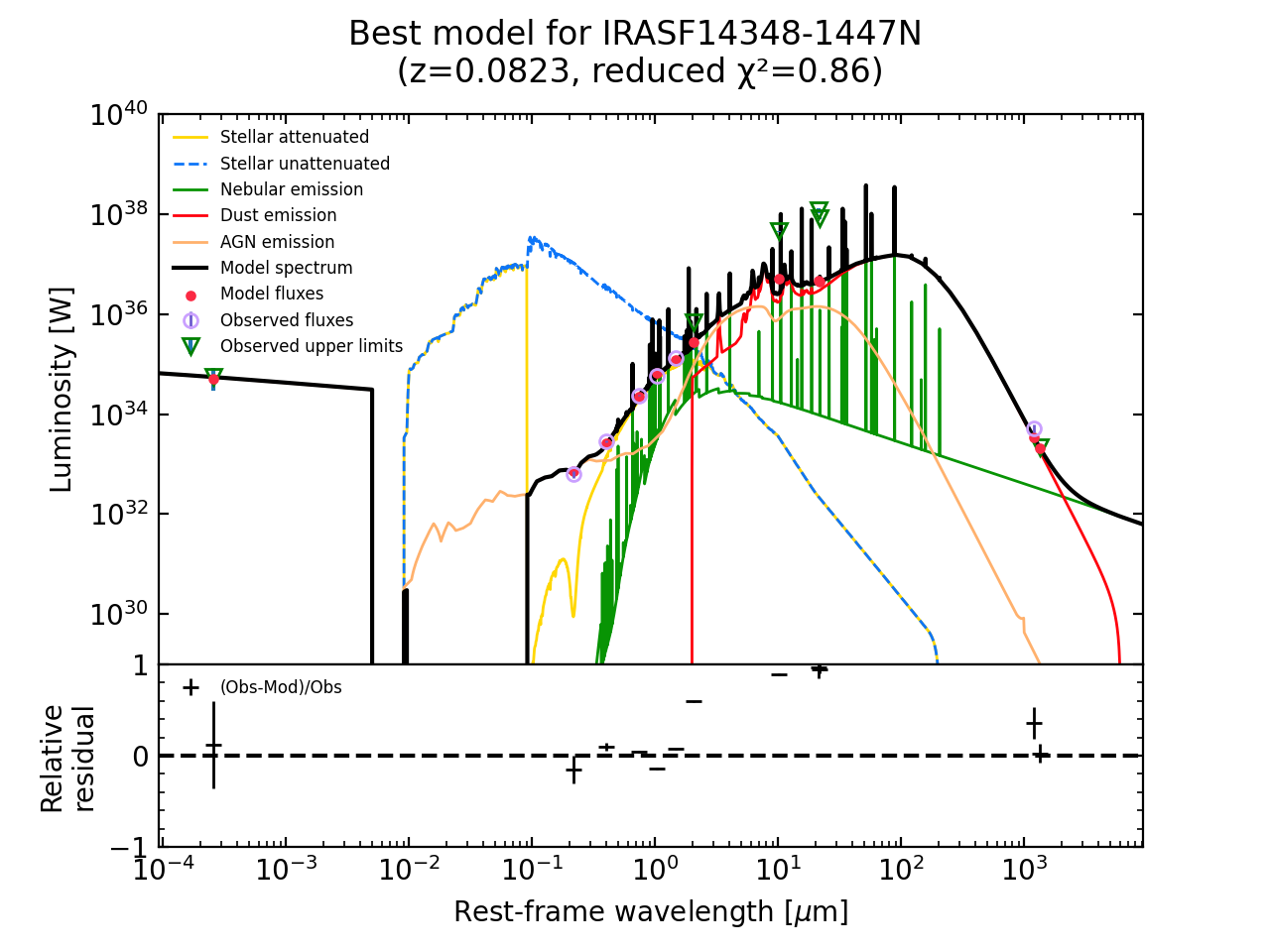}
        \includegraphics[width=0.45\textwidth]{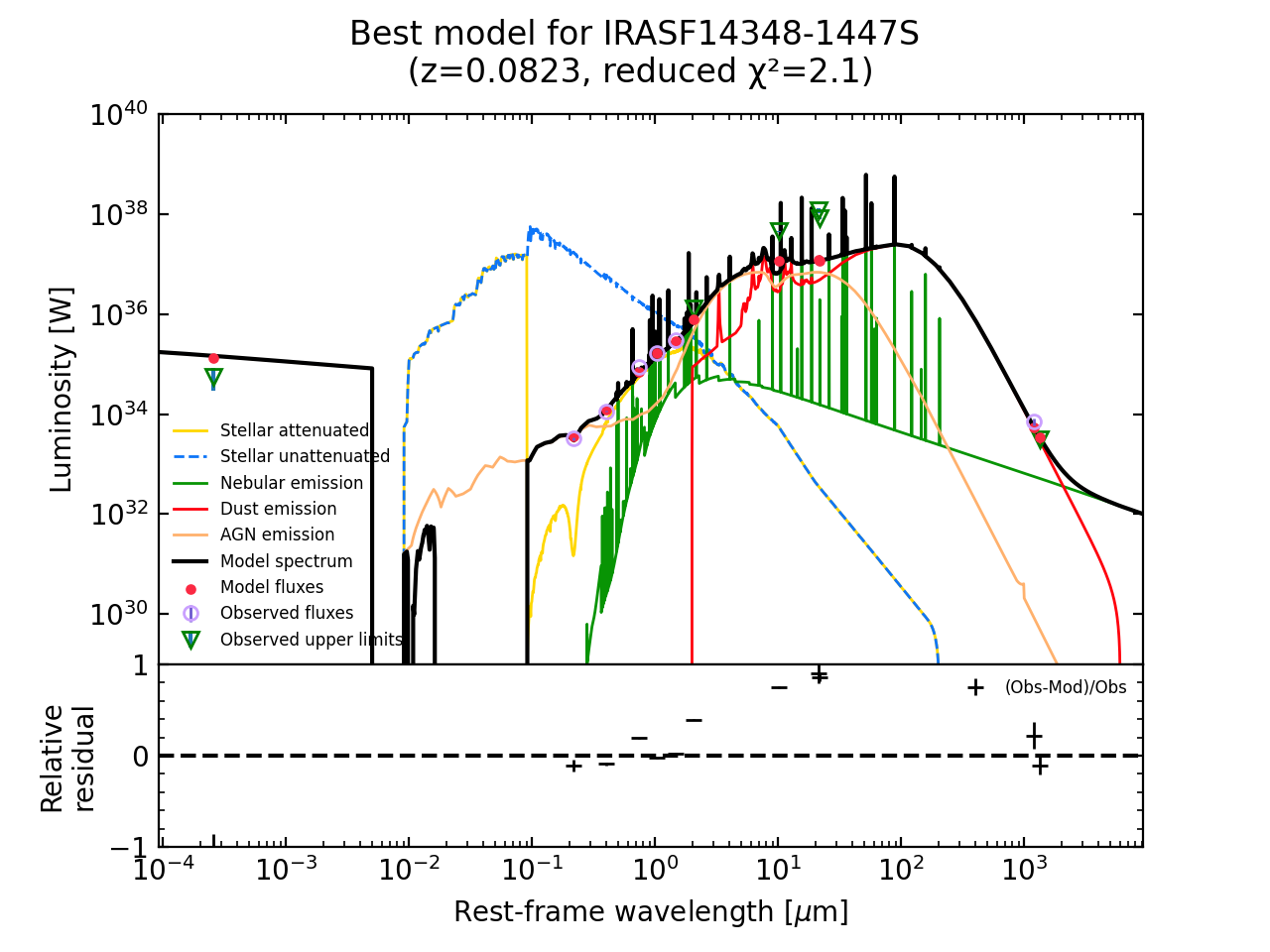}
        \includegraphics[width=0.45\textwidth]{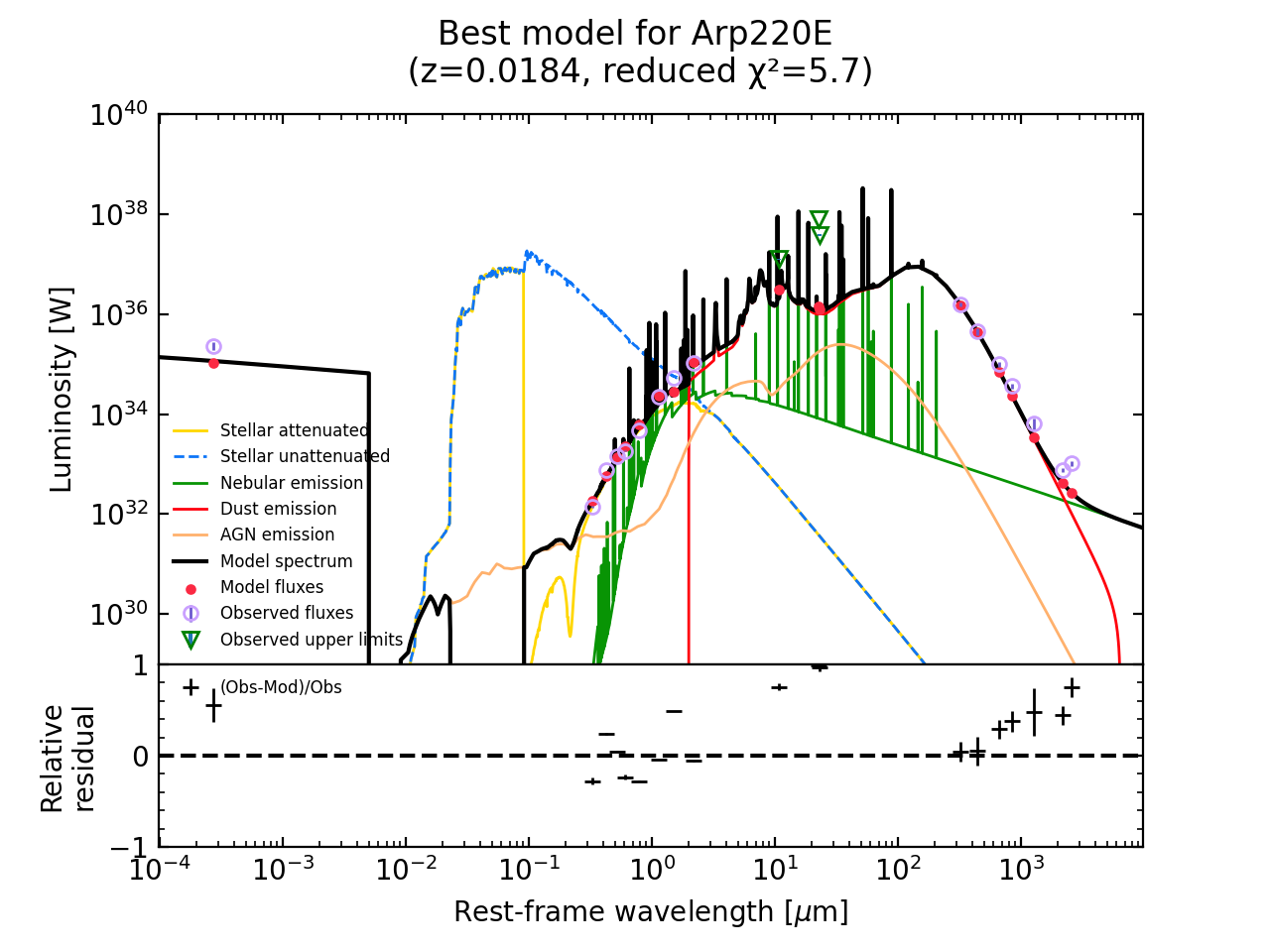}
        \includegraphics[width=0.45\textwidth]{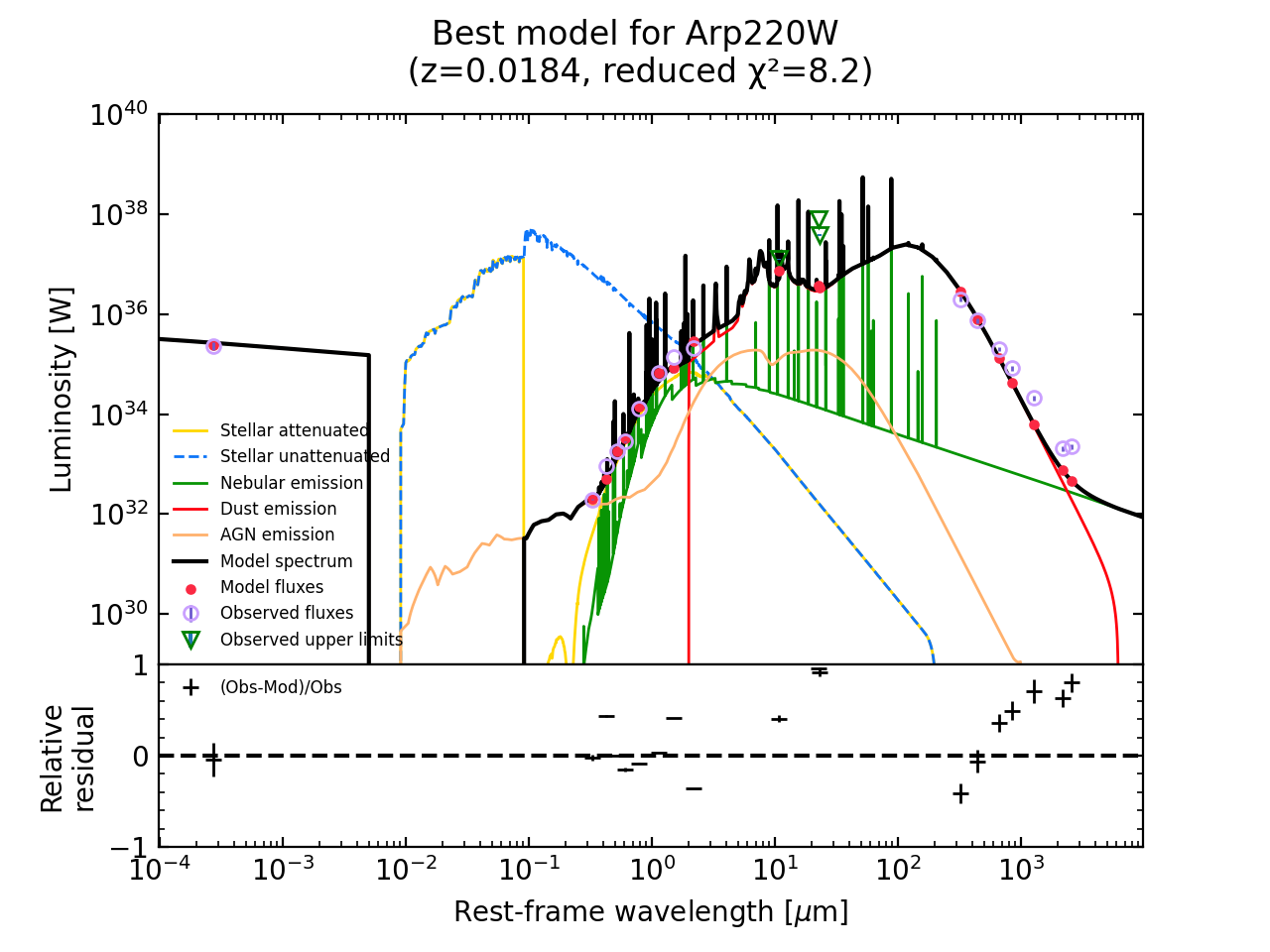}
	\caption{Continued}
	\label{fig:allseds3}
\end{figure*}

\addtocounter{figure}{-1}

\begin{figure*}[htbp]
	\centering
        \includegraphics[width=0.45\textwidth]{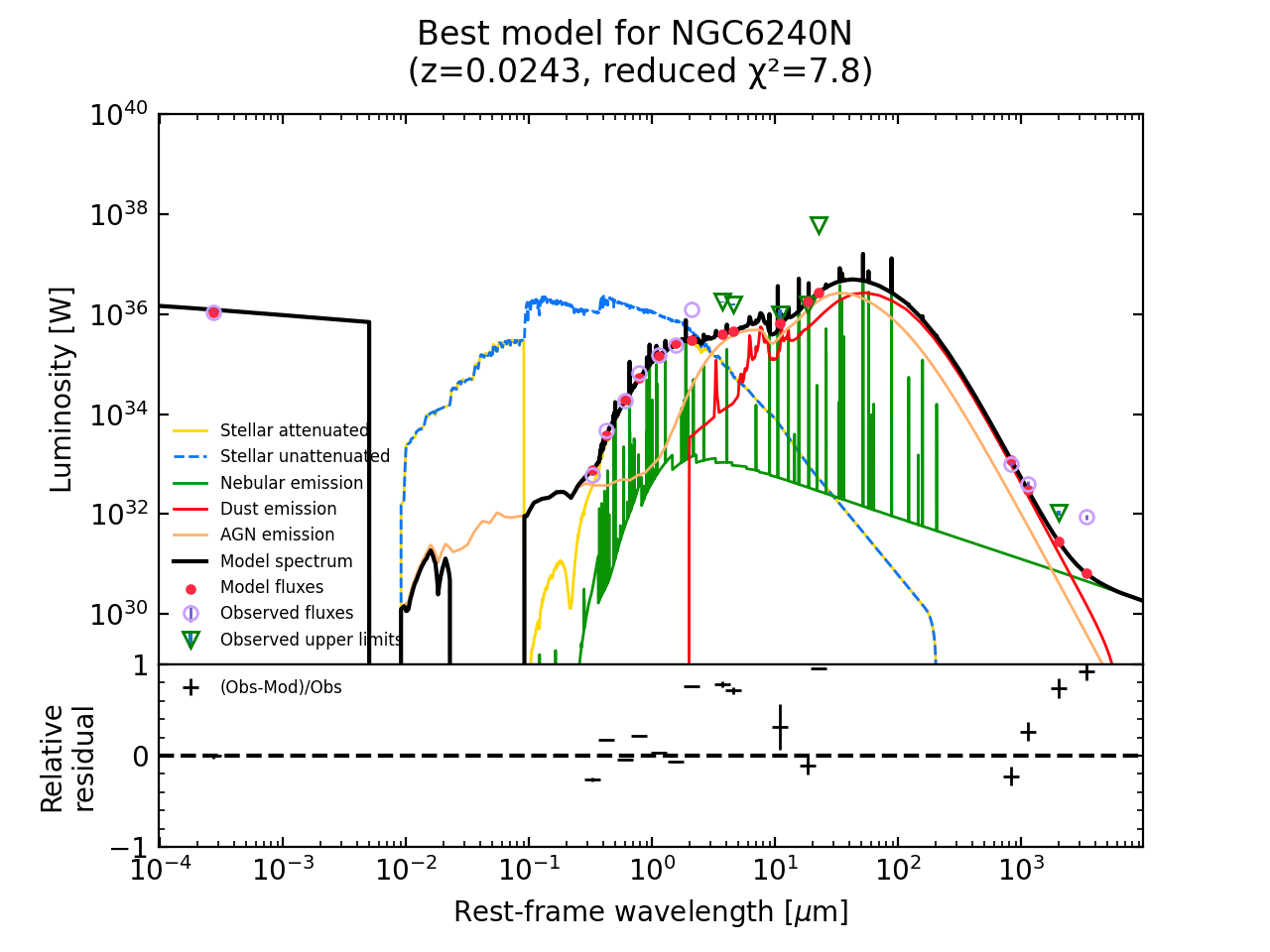}
        \includegraphics[width=0.45\textwidth]{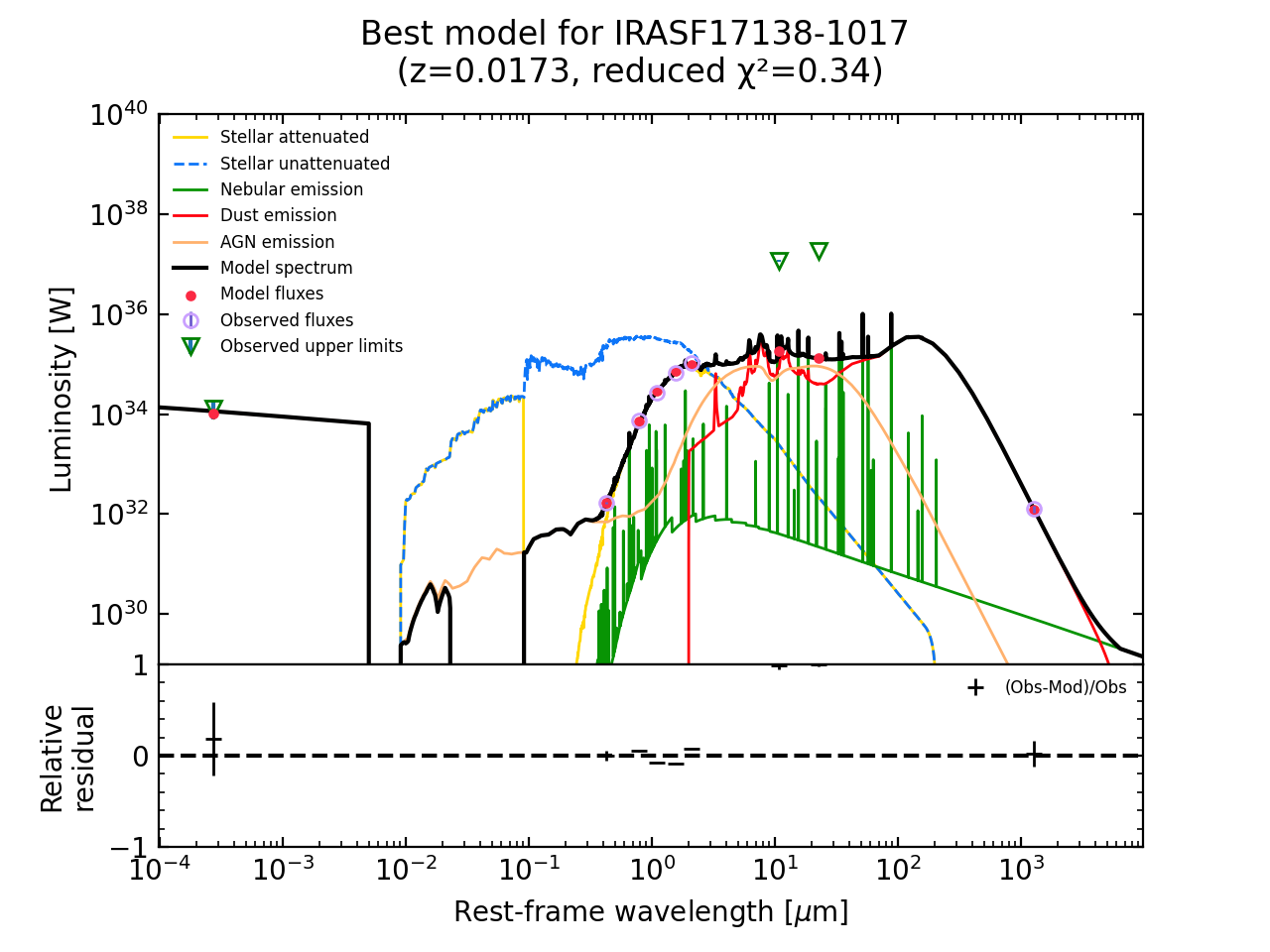}
        \includegraphics[width=0.45\textwidth]{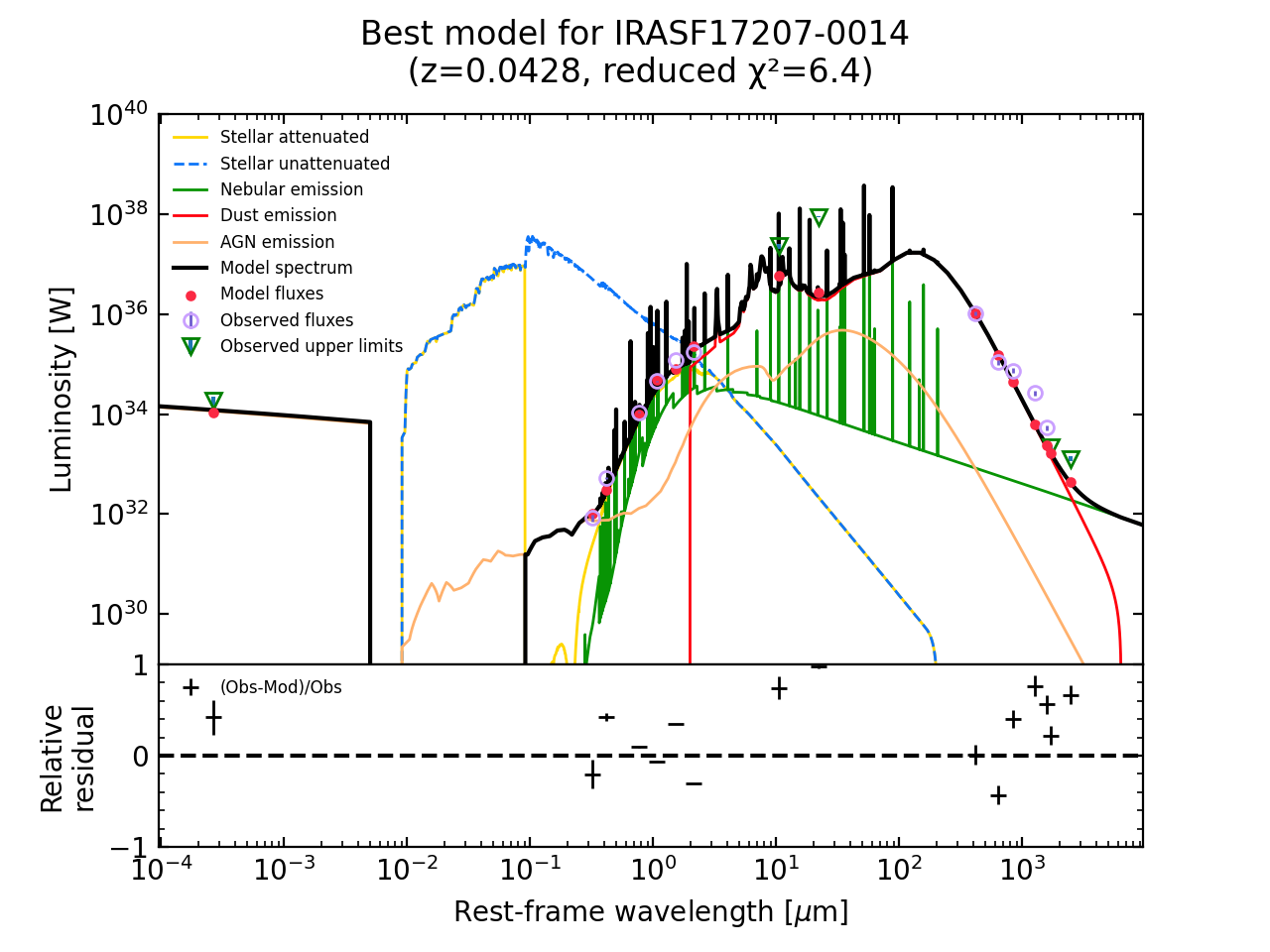}
        \includegraphics[width=0.45\textwidth]{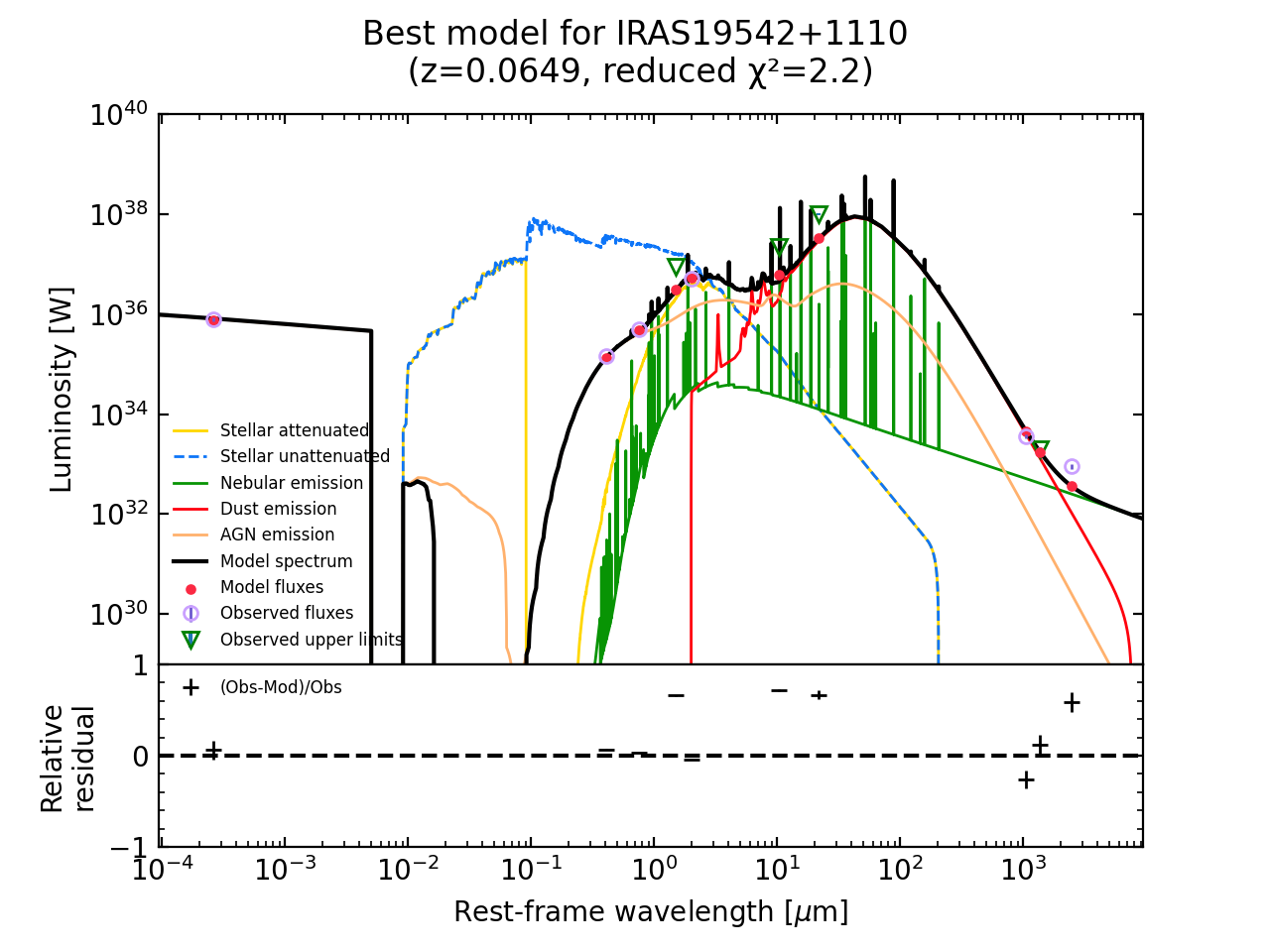}
        \includegraphics[width=0.45\textwidth]{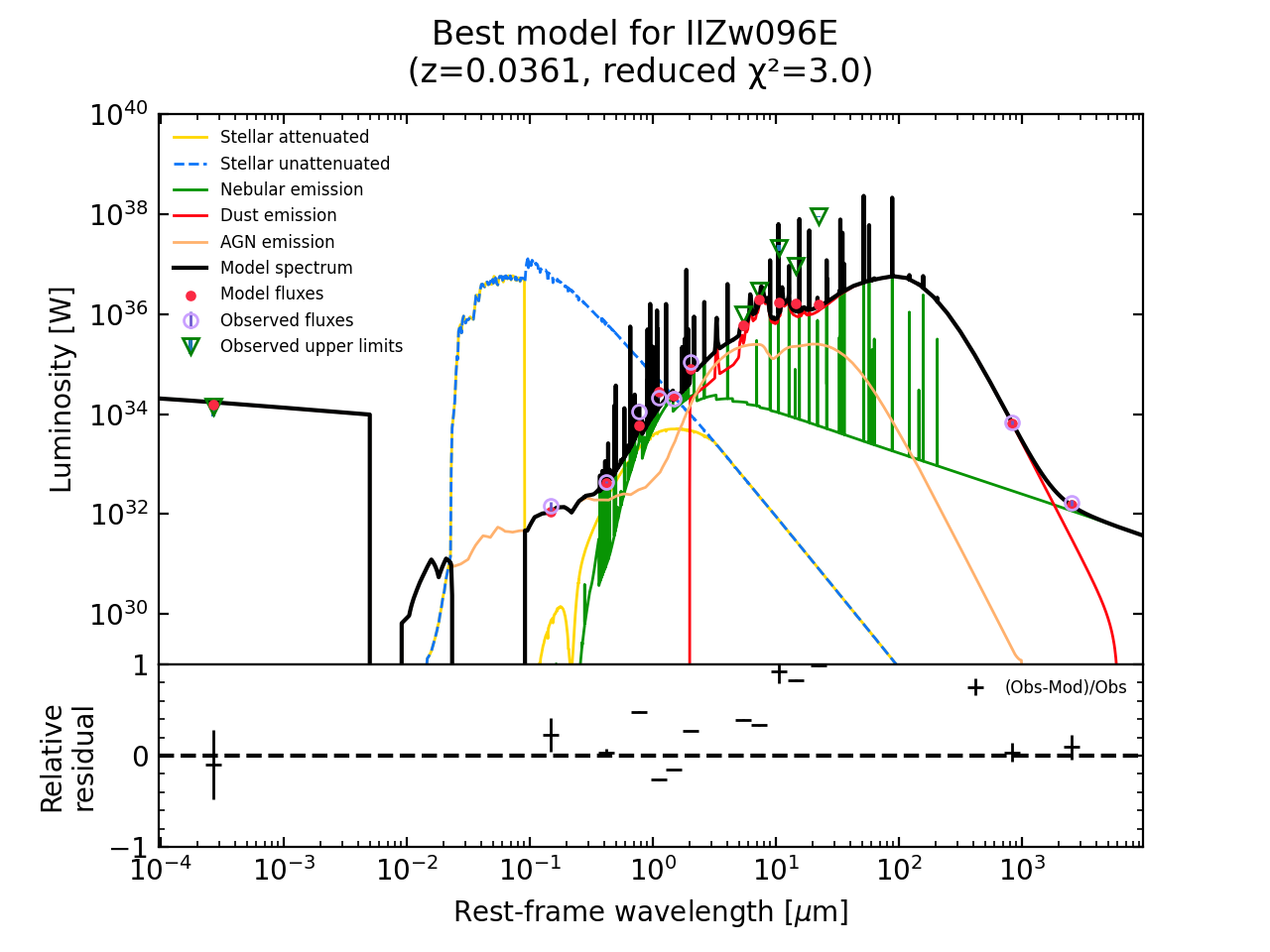}
        \includegraphics[width=0.45\textwidth]{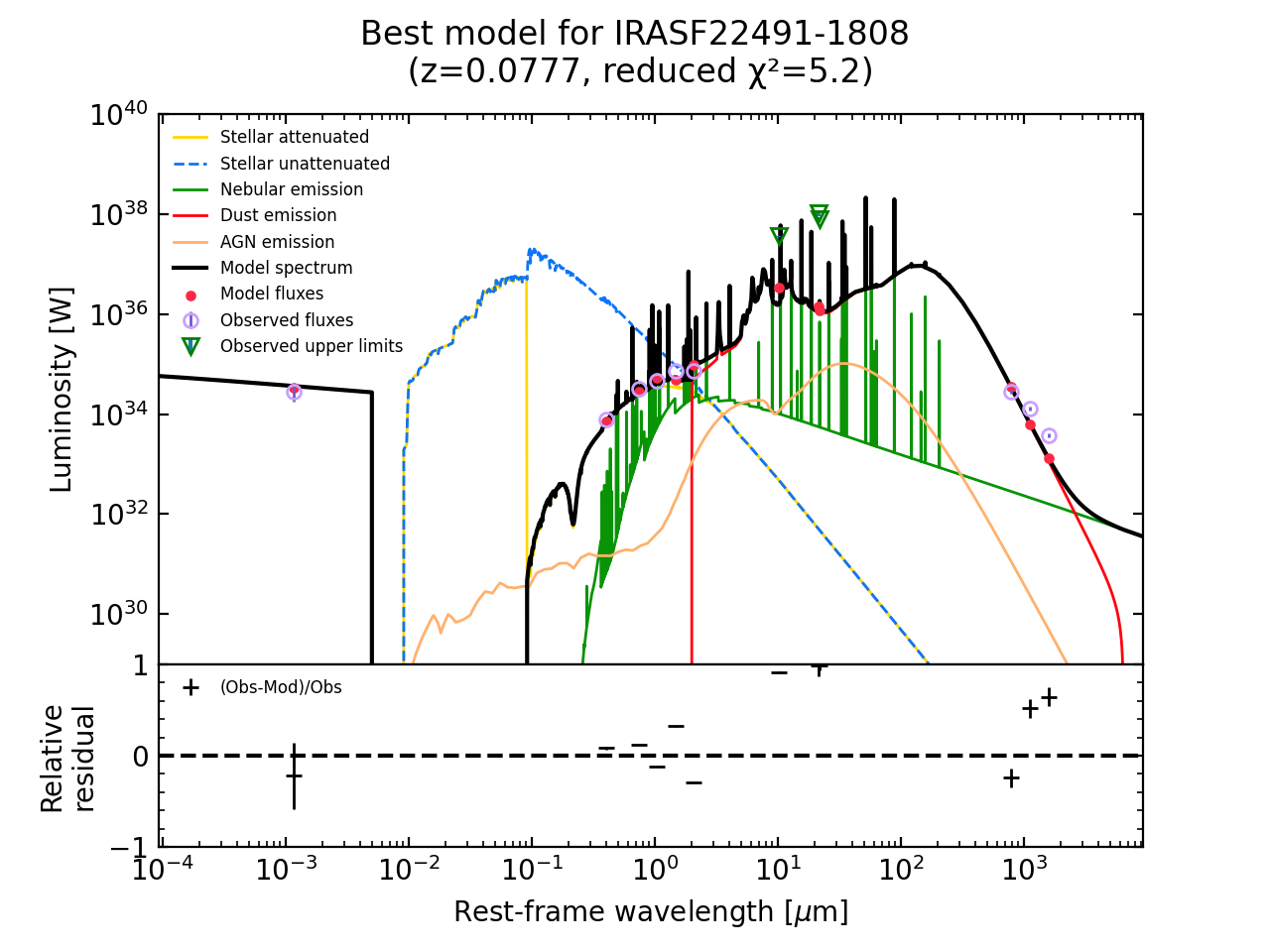}
        \includegraphics[width=0.45\textwidth]{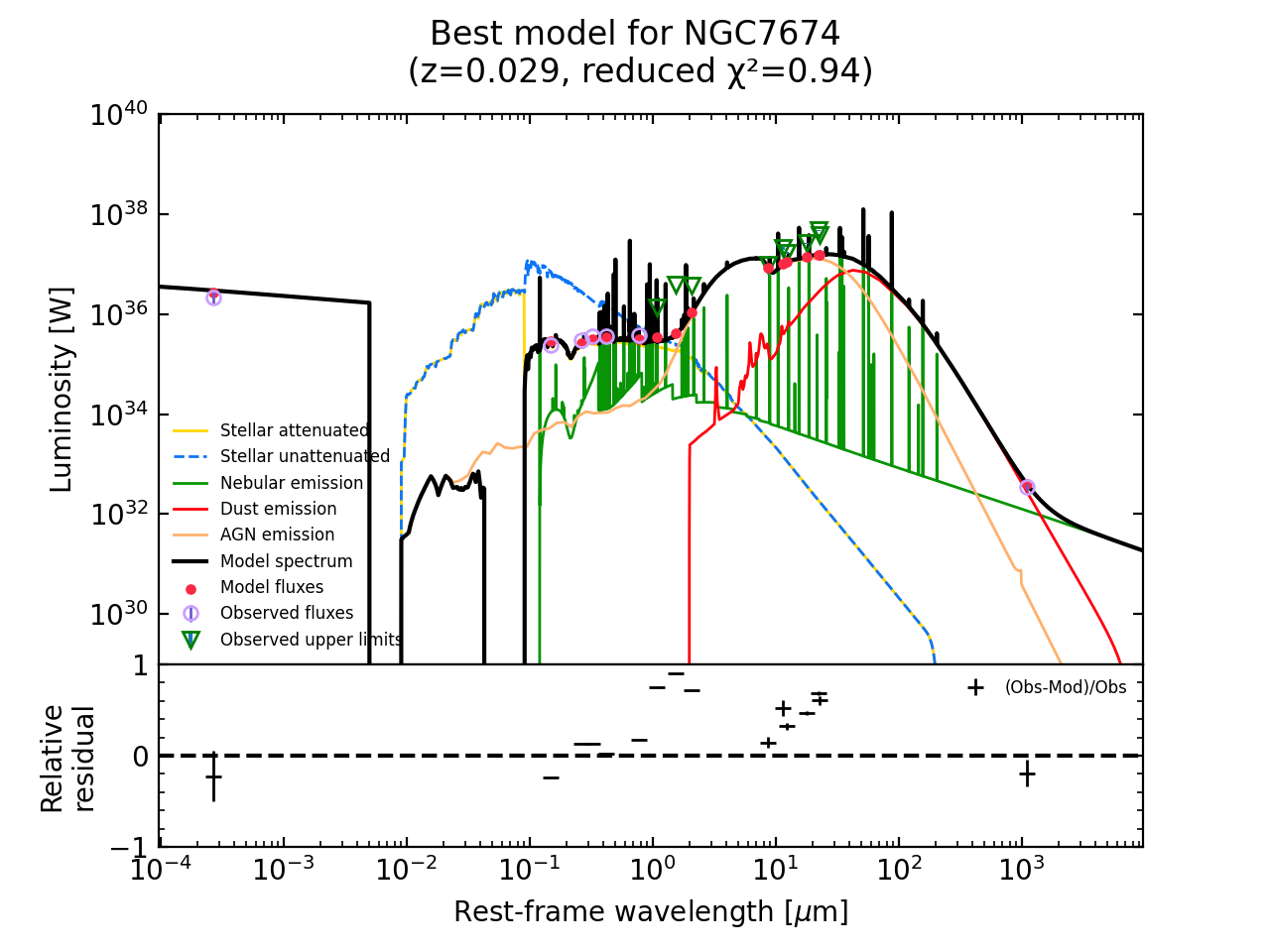}
        
	\caption{Continued}
	\label{fig:allseds4}
\end{figure*}

\section{DISCUSSION} \label{sec:discussion}
\subsection{Nuclear SEDs of (U)LIRGs} \label{sec:median seds}

SEDs are highly efficient tools for studying various ongoing astrophysical processes in galaxies, such as star formation, dust attenuation, and AGN activity. In previous work, \cite{2012U} analyzed radio through X-ray SEDs for 64 local (U)LIRGs with log($L_{\rm IR}/L_{\odot})=11.14–12.57$, and computed the total infrared luminosity, dust temperature, dust mass, stellar mass, and global SFR for each source. The SEDs for all objects show a broad, thermal stellar peak ($\sim0.3-2\rm\mu m$) and a dominant far-infrared ($\sim40-200\rm\mu m$) thermal dust peak, where $\nu f_{\nu}(60\,{\rm\mu m})/\nu f_{\nu}(V)$ varies from $\sim$2 to 30 with increasing $L_{\rm IR}$. \cite{2021Paspaliaris} replaced \emph{IRAS} far-infrared photometry with measurements from \emph{Herschel} and modeled the SEDs of 67 local (U)LIRGs using the CIGALE code. They derived physical properties, constrained the emission from photodissociation regions, and compared these to those of early- and late-type galaxies. Nowhere is the effect of merging more apparent than in the SFR, where in late-stage mergers the stars are formed at a median rate of up to $99\,M_{\odot}\,\rm yr^{-1}$, compared to $26\,M_{\odot}\,\rm yr^{-1}$ in isolated (U)LIRGs. More recently, \cite{2023Yamada} performed a global X-ray to radio SED decomposition on a sample of 57 local (U)LIRGs that had been observed by \emph{NuSTAR} and/or \emph{Swift}/BAT in the GOALS sample. The SFRs increase with the merger stage and are correlated with radio luminosity. The comparison between the SFRs and bolometric AGN luminosities suggests that starbursts precede the emergence of AGN, with their overall growth rates following the coevolutionary relation between the host galaxies and SMBHs.

It is worth mentioning that in this work, they employed the infrared CLUMPY model \edit1{\citep{2008NenkovaA,2008NenkovaB}}, which allowed the multi-wavelength study with the consistent X-ray torus model (XCLUMPY). Meanwhile, they demonstrated that the results derived from the SED modeling using either the CLUMPY or SKIRTOR model were not significantly different. Therefore, we continued to use the SKIRTOR model.

Amid all regions of the electromagnetic spectrum, the mid-infrared part of the SEDs in (U)LIRGs can be a mixture of emissions from stars, dust grains, and AGN, if present. Thus, mid-infrared observations (e.g. $5-25\,\rm\mu m$) are critical for the decomposition of the SEDs. Since only \edit1{8} galaxies in our sample have available mid-infrared imaging data from the VLT, Gemini and Subaru telescopes \cite[][see Table \ref{tab:photometry2}]{2014Asmus}, we are limited in how accurately we can determine the contributions of the individual components to the bolometric luminosity until more high-resolution mid-infrared data (e.g. from \emph{JWST}) are available. Here we examine how the nuclear SEDs of (U)LIRGs vary as a function of merger stage and infrared luminosity. 

\begin{figure*}[htbp]
	
	\centering
	\includegraphics[width=\textwidth]{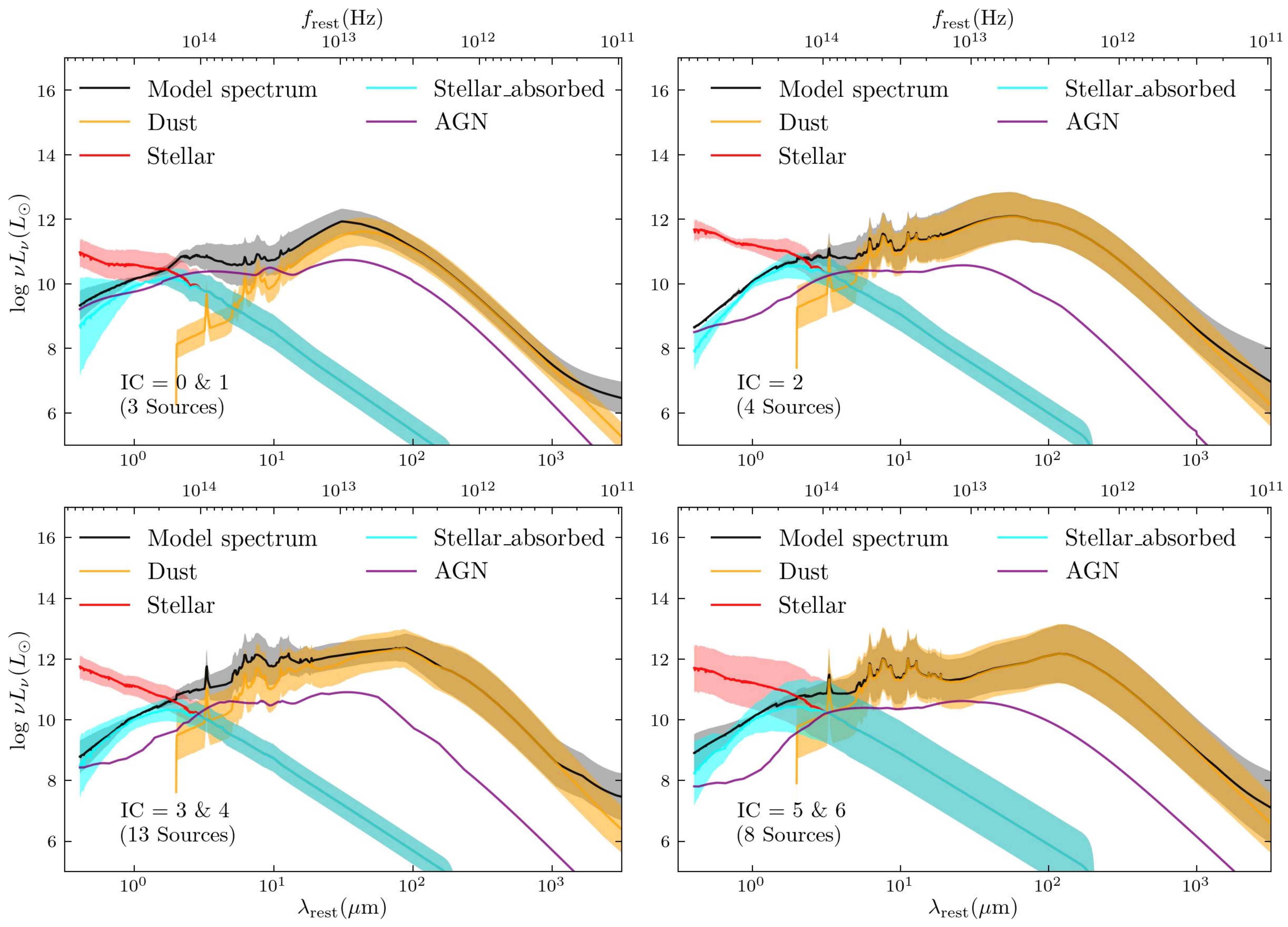}
	\caption{Median model SEDs (black solid) of the nuclear regions of local (U)LIRGs in our sample as categorized by their merger stages. The Interaction Class (IC) denotes the following bins: non- and pre-merging (0 \& 1), early-stage (2), late-stage (3 \& 4), and post-merger (5 \& 6). The components from SED-fitting are shown separately: dust (dark yellow solid), unattenuated stellar (red solid), attenuated stellar (cyan solid), and AGN (purple solid). The shaded region within each color indicates the $\pm 1 \sigma$ uncertainty for each component. As the AGN component may vary substantially, we are unable to present its uncertainty.}
	\label{fig:MedianSEDMS}
	
\end{figure*}

In Figure \ref{fig:MedianSEDMS}, we show the median nuclear SEDs as stacked by interaction classes (IC, see \S \ref{sec:sample} for more details) corresponding to the following bins: non- and pre-merging (IC 0 \& 1), early-stage \edit1{(IC 2), late-stage (IC 3 \& 4)}, and post-merger (IC 5 \& 6). The total SEDs as well as the individual components do vary as a function of the galaxy interaction class. 
One thing to note is an apparent low AGN contribution to the median SED among the early-stage \edit1{(IC 2)} mergers \edit1{(average $f_{\mathrm{AGN}}$=0.05)}, followed by an enhancement among the late-stage \edit1{mergers (average $f_{\mathrm{AGN}}$=0.16) and post-merger} sources \edit1{(average $f_{\mathrm{AGN}}$=0.14)}. Only three galaxies in our sample are classified as stage 0 \& 1\edit1{: the face-on spiral NGC 0695,} the face-on type 2 Seyfert NGC 7674 and the compact ULIRG IRAS 19542+1110, which exhibit less dust in the nuclei. Consequently, galaxies in IC 0 \& 1 \edit1{(average $f_{\mathrm{AGN}}$=0.44)} \textit{appear} to showcase stronger AGN contribution compared to \edit1{IC 2} mergers. \edit1{Despite the small sample size,} our resolved SEDs show that the AGN fraction for \edit1{post-merger (U)LIRGs (IC 5 \& 6) might be lower than that of} non- and pre-mergers. This phenomenon can be attributed partially to the buildup of heavy nuclear obscuration during the interaction, up to $A_V$ of $\sim$25$-$40 mag \citep{2019U}. Deeply buried AGN may be expected in \edit1{late-stage mergers and post-merger systems} where the SMBH is actively accreting gas behind the thick dust screen~\citep{2008Imanishi,2010Imanishi,2012Lee}. 
  
Meanwhile, the nuclear regions of galaxies in stages 0 \& 1 show obviously reduced dust emission compared to \edit1{early-stage mergers (IC 2), late-stage mergers (IC 3 \& 4) and post-merger (U)LIRGs (IC 5 \& 6)}. This is likely due to the less perturbed spiral morphology of both NGC 0695 and NGC 7674, leading to reduced obscuration in their nuclear regions. More recent hydrodynamic galaxy simulations coupled with dust radiative transfer have uncovered obscured signatures of merger-triggered AGN fueling~\citep{2018Blecha}. The obscuration phase coincides with the peak of black hole fueling, such that signatures of merger-triggered fueling could thus be missed among \edit1{both late-stage mergers and post-merger systems} from optical selections but recovered with infrared signatures~\citep{2019Pfeifle,2022Ub,2023Barrows}. 
At this peak, both AGN activity and intense starbursts take place in a heavily obscured environment with interaction-induced inflows, causing peak gas densities near the SMBH to increase by orders of magnitude~\citep{2022Sivasankaran}. While this scenario predicts a high AGN contribution to the SEDs of the sources in IC 5 \& 6 that is not apparent in our results, our archival data-driven sample selection may be limited in sample size in the last bin; a larger sample that extends to higher-luminosity ULIRGs could further support our claim. It should also be noted that degeneracies between model parameters can undermine the search for obscured AGN. \edit1{While the mid- and far-infrared emission originates from dust primarily heated by star formation and/or buried AGN, AGN can dominate the warm dust emission at mid-infrared wavelengths ($\sim3-30\,\mu m$) by heating the dust in the torus and surrounding environment. In contrast, its contribution to the cold dust emission at far-infrared wavelengths ($\sim30-1000\,\mu m$) can be insignificant \citep{2013Diaz-Santos}. However, in this study, the lack of data points to constrain the infrared SED shape makes it challenging to break the degeneracy between stellar population and AGN model parameters, potentially leading to inaccurate estimates of the AGN contribution.}

Apart from the AGN emission, the larger dispersion of dust emission in stage 5 \& 6 (Figure \ref{fig:MedianSEDMS}) may imply that the nuclear regions of \edit1{post-merger} galaxies in this bin span a wide range of dust features. This is possibly due to the complex nature of the obscured nuclei in these \edit1{post-merger} systems. Deep infrared observations that can distinguish the mid-infrared emission of AGN from that of the warm dust heated by starbursts will be invaluable for finding obscured AGN in ULIRGs in stages 5 \& 6 \citep{2018Hickox,2021Satyapal, 2022Sajina,2022Inami,2022Evans,2023Rich}. 

\begin{figure}[htbp]
	\centering
	\includegraphics[width=0.5\textwidth]{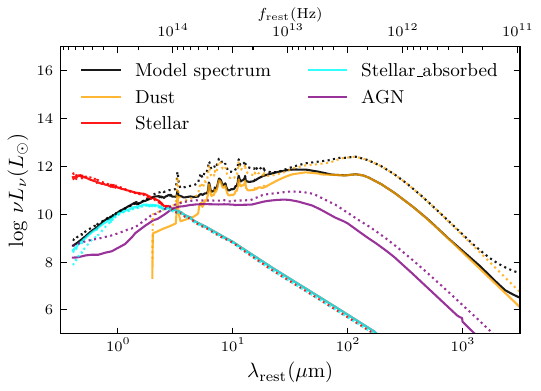}
	\caption{Median model SEDs of the nuclear regions of local LIRGs (solid lines) and ULIRGs (dotted lines) in our sample.  The \edit1{legends are} the same as those described in Figure \ref{fig:MedianSEDMS}, while the shaded regions have been omitted for clarity. 
 }
	\label{fig:MedianSED_LIR}
	
\end{figure}

In addition to merger stage, we compare the median SEDs of the nuclear regions in LIRGs and ULIRGs in Figure \ref{fig:MedianSED_LIR}. The nuclear SEDs of ULIRGs show a higher far-infrared dust hump along with slightly higher level of AGN emission. Our result is consistent with past findings that the contribution of AGN to the infrared luminosity increases with the total infrared luminosity~\citep{1995Veilleux,2007Desai,2011Petric}. Meanwhile, ULIRGs in our sample also exhibit a higher level of mid-infrared emission, which may also be related to the significant contribution from the warm dust in the AGN torus to the mid-infrared continuum \citep{2010Diaz,2011Petric}. The observed higher compactness of the mid-infrared continuum in ULIRGs may also contribute to the mid-infrared excess evident in their nuclear regions \citep{2010Diaz}. A larger sample with much tighter constraints on mid-infrared emission will enable a more precise assessment of the role of AGN as an energy source in (U)LIRGs.

To further interpret our results based on spatially-resolved SEDs of the 100$-$150 pc regions around SMBHs, we consider the similarities and discrepancies between our work and those on global SEDs of overlapping targets analyzed by~\cite{2021Paspaliaris} and~\cite{2023Yamada}. The AGN fraction identified in our sample is relatively similar to those found by \cite{2023Yamada}, except for three sources (II Zw 096, IRAS F10565+2448, and IRAS F12112+0305) that show no AGN signature according to their work. 
The absence of AGN in IRAS F10565+2448 and IRAS F12112+0305 is further corroborated by \cite{2021Paspaliaris}. The discrepancy could be due to the dilution of the AGN signal in the spatially-integrated SEDs. Moreover, among the overlapping targets between our work and that of \cite{2021Paspaliaris}, our sample features a greater number of AGN hosts. This can be attributed to the incorporation of X-ray data in our SED fitting, which proves to be crucial for the accurate identification of AGN. We also compared the AGN luminosity reported by \cite{2023Yamada} with our results. For galaxies present in both samples, we found that our estimated AGN luminosity was $0.5-2$ orders of magnitude lower than the results from \cite{2023Yamada}. Specifically, we refitted the SED of IRAS F05189-2524 from \S \ref{sec:modeling} without treating the mid-infrared data points as upper limits. The new result agrees with the global AGN luminosity reported \edit1{by \cite{2023Yamada}}, thus providing an upper limit to the AGN emission from the $\sim100$ pc nuclear region of this source. This reflects the crucial role of high-resolution mid-infrared observations in distinguishing hot dust emission from the AGN torus from other dust emission, thereby constraining AGN luminosity and other parameters.

\subsection{Relationship between X-ray and Infrared Emission} \label{sec:xray vs ir}

The X-ray emission of starburst galaxies mainly comes from high-mass X-ray binaries (HMXB), supernova remnants, young stars, and hot gas heated by supernova explosions, all of which bear a close relationship with recent starbursts. Therefore, in the absence of AGN, hard X-ray emission originating from the HMXB is often assumed to be a good indicator of the SFR, and a correlation between hard X-ray luminosity and infrared luminosity is observed in nearby star-forming galaxies \cite[e.g.][]{2003Ranalli,2003Grimm,2004Persic}. This relation appears to extend to objects with even higher SFR (up to $100\,M_{\odot}\,\rm yr^{-1}$) like (U)LIRGs and at high redshifts \cite[e.g.][]{2005Hornschemeier,2007Persic,2008Lehmer}. Previous studies of large samples of (U)LIRGs \citep{2011Iwasawa,2017Ricci,2018Torres,2021Ricci} have highlighted the power of X-ray observation in detecting dust-embedded AGN and estimating its contribution to the total luminosity.
Analyzing \emph{Chandra} observations of a large subset of the GOALS (U)LIRGs, \citet[\emph{Chandra}-GOALS I, or CG1 hearafter]{2011Iwasawa} and \citet[\emph{Chandra}-GOALS II, or CG2 hereafter]{2018Torres} both observed lower X-ray luminosities than predicted by the correlation between infrared and X-ray luminosities in nearby star-forming galaxies. They suggested that the X-ray faintness could be explained by several factors: the presence of a self-absorbing starburst or an obscured AGN, both of which would contribute to the IR emission but escape X-ray detection; or extremely young star-forming regions, where HMXBs no longer dominate the hard X-ray emission.

Building on past work, we investigate the relationship between X-ray and infrared emission in the central 100$-$150 pc regions of (U)LIRGs, minimizing the contamination from circumnuclear stars and dust except for those along the line of sight. We obtain the soft (SX) and hard X-ray (HX) luminosities of the nuclei following the steps described in \S \ref{sec:chandra}. The luminosity distributions in the soft and hard X-ray bands peak at log\,$L_{\rm SX} \approx$ 40.5$-$41\,$\rm erg\,s^{-1}$ and log\,$L_{\rm HX} \approx$ 41$-$41.5\,$\rm erg\,s^{-1}$, respectively. The infrared counterpart for the nuclear regions is more difficult to measure directly given that current mid- and far-infrared instruments lack the high-angular resolution capability to resolve the galactic nuclei at similar spatial scales. Thus, we derive the nuclear infrared luminosities \edit1{for all sources in our sample by using the formulae from \cite{1996Sanders}, which was also used to obtain the total infrared luminosities of GOALS (U)LIRGs \citep{2003Sanders}, including those in the CG1 and CG2 samples. Specifically, the total infrared luminosity can be expressed as:}

\begin{equation}
    L_{\mathrm{IR}}=4\pi D_{\mathrm{L}}^{2}F_{\mathrm{IR}}[L_{\odot}]
\end{equation}
\edit1{where $F_{\mathrm{IR}}=1.8\times10^{-14}\times(13.48f_{\mathrm{12\mu m}}[\mathrm{Jy}]+5.16f_{\mathrm{25\mu m}}[\mathrm{Jy}]+2.58f_{\mathrm{60\mu m}}[\mathrm{Jy}]+f_{\mathrm{100\mu m}}[\mathrm{Jy}])[\mathrm{W\,m^{-2}}]$. The differences between the nuclear infrared luminosities calculated using the above formulae and the dust luminosities provided by X-CIGALE based on the best-fit results are mostly $\sim10\%$. Meanwhile, the ratios of these nuclear infrared luminosities to the global values of the corresponding galaxies from \cite{2009Armus} range from less than $1\%$ to more than $30\%$. This range also reflects the significant uncertainty in infrared luminosity estimates caused by the lack of mid- and far-infrared data.}

\begin{figure*}[htbp]

    \centering
	\includegraphics[width=\textwidth]{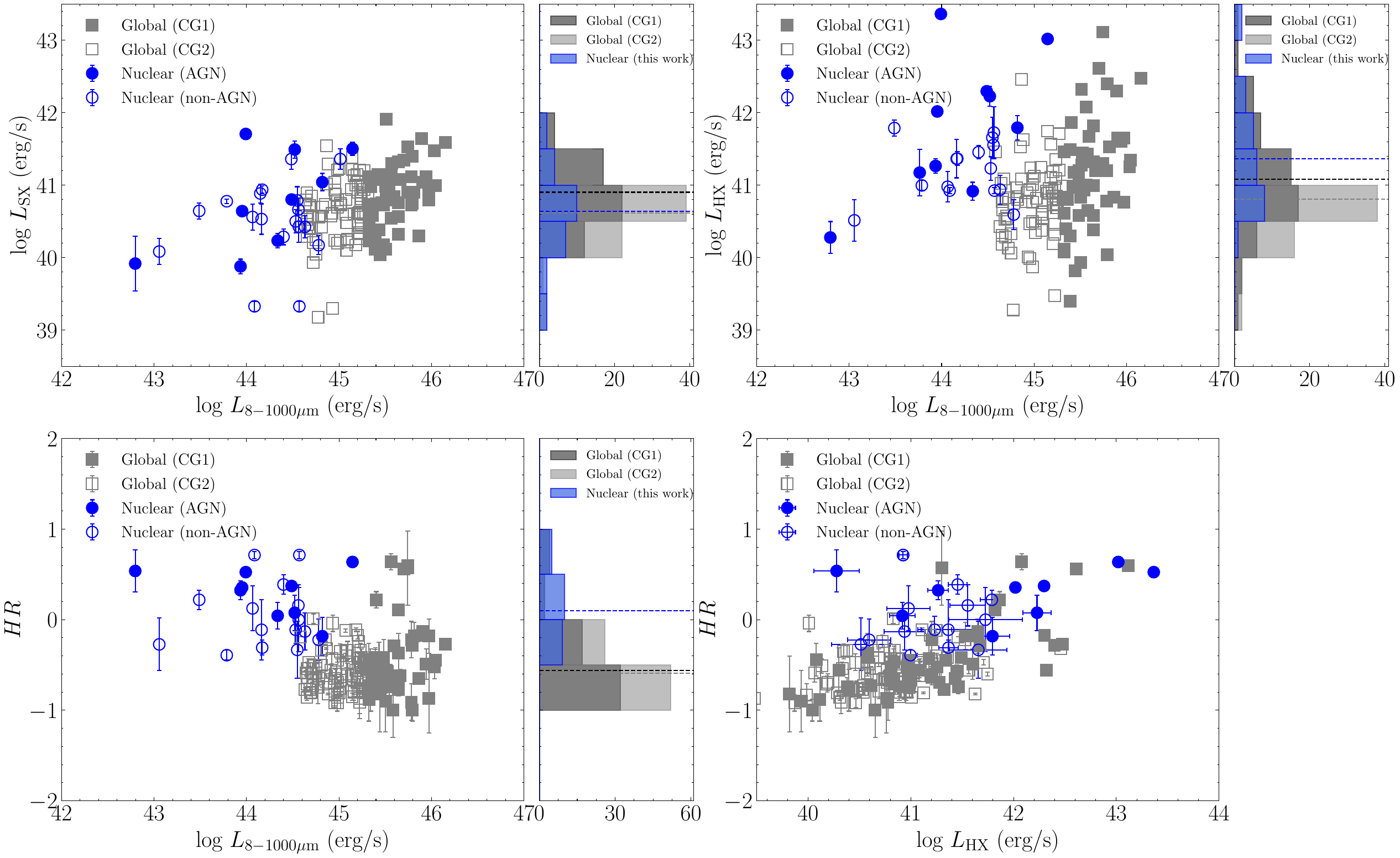}
	\caption{Top panels: soft X-ray (0.5$–$2 keV) versus infrared luminosity (left) and hard X-ray (2$–$7 keV) versus infrared luminosity (right) as extracted from the nuclear regions (blue circles). X-ray selected AGN are marked with filled circles. Global values from CG1 (solid grey squares) and CG2 (open grey squares) are shown for comparison. Side histograms show the distributions of the corresponding X-ray luminosities for CG1 (dark grey), CG2 (light grey), and nuclear values (blue) along with their median values (dashed lines), respectively. Bottom panels: X-ray color \emph{HR} versus infrared luminosity (left) and hard X-ray luminosity (right). Colored symbols are the same as those in the upper panels. Side histograms show the distributions of $HR$ for CG1 (dark grey), CG2 (light grey), and nuclear values (blue) along with their median values (dashed lines), respectively.}
	\label{fig:xray_all}
    
\end{figure*}

We present the relationship between the nuclear infrared luminosities and the soft and hard X-ray luminosities, respectively, in Figure \ref{fig:xray_all} (upper panels), where AGN identified in previous studies are marked with solid circles, while non-AGN are shown as open circles. \edit1{As for AGN classification, we adopt the classification from CG1 \citep{2011Iwasawa} and CG2 \citep{2018Torres}, which was performed using both X-ray and mid-infrared criteria. The X-ray color, also called the “hardness ratio”, is defined as $HR=(H-S)/(H+S)$, where $H$ and $S$ are net count rate in the 2$-$7 keV and 0.5$-$2 keV bands, respectively~\citep{2011Iwasawa}. Therefore, it characterizes the relative intensity of the hard and soft X-ray emission. Since strong emission above 2 keV could indicate the presence of an obscured AGN with a neutral hydrogen column density $N_{\rm H}\approx 10^{22-24}\,\rm cm^{-2}$, even when considering the soft excess produced by scattered radiation into the line of sight \citep[e.g.][]{2007Ueda,2012Brightman} and the reflection of the primary X-ray continuum caused by circumnuclear material \citep[e.g.][]{2008Ajello}, \emph{HR} can serve as a proxy for AGN up to $z\sim 3-4$ \citep{2021Peca}. The AGN selection threshold was set at $HR>-0.3$ \citep{2011Iwasawa}. Additionally, the detection of a strong Fe K$\alpha$ line at 6.4 keV, as introduced in \S \ref{sec:chandra}, also serves as a criterion for AGN selection. Meanwhile, the mid-infrared selection relies on the detection of the [Ne V] 14.32 $\mathrm{\mu m}$ line, which traces high-ionization gas that cannot be produced by young stars \citep[e.g.][]{2011Petric}.}

Several points are of note: first, the nuclear regions for AGN and non-AGN feature similar $L_\mathrm{SX}$, whereas the $L_\mathrm{HX}$ for AGN are larger than for non-AGN at the same infrared luminosity. This is because the AGN contribute to the Fe K$\alpha$ emission at 6.4 keV~\citep{2004page} in addition to featuring hard radiation, whereas the SX emission may contain a mix of starburst origin. In fact, the spread in global $L_\mathrm{HX}$ at a given  $L_\mathrm{IR}$ appears to be larger than that in global $L_\mathrm{SX}$, whereas the two are more consistent in the nuclear regions. This trend is because the infrared emission in the nucleus is dominated by dust and AGN emission with relatively little contribution from star formation. 

Second, when we compare our results with \edit1{CG1 and CG2}, we note that our nuclear and the global data points span the same $L_\mathrm{SX}$ range (log\,$L_{\rm SX} \approx$ 39.2$-$43.3~$\rm erg\,s^{-1}$), while the nuclear $L_\mathrm{HX}$ values (log\,$L_{\rm HX} \approx$ 40.3$-$43.3~$\rm erg\,s^{-1}$) overlap with those of the more X-ray luminous sources in both \edit1{CG1 and CG2}. This indicates that the hard X-ray emission tends to be more nucleated, and the soft X-ray emission is more evenly distribute across the galaxy. Compared with the global $L_\mathrm{IR}$ values, the nuclear $L_\mathrm{IR}$ are \edit1{overall} smaller, suggesting that while much of the infrared luminosities may be dominated by the center, the merger-induced, widespread star formation throughout the host galaxy still drives the infrared luminosities beyond the central 100$-$150\,pc region of the galaxy~\citep{2020Larson,2023Linden}.

The lower panels of Figure \ref{fig:xray_all} show the nuclear hardness ratios as a function of the infrared (left) and hard X-ray luminosities (right). We note that the nuclear regions exhibit higher \emph{HR} values than the global data points, regardless of whether an AGN is detected. The hard X-ray excess seen in the nuclear region can be interpreted primarily as the concentration of hard X-ray emission rather than lack of soft X-ray emission in the center of the system. While nuclear \emph{HR} is elevated relative to those of the LIRGs from \edit1{CG2}, the lack of the Fe K$\alpha$ indicator makes it challenging to distinguish AGN from non-AGN based on \emph{HR} alone~\citep{2011Iwasawa}.

We also note that several galaxies from \edit1{CG1} have overall \edit1{$HR>0.5$, comparable to those} of the nuclear regions. \edit1{This can be explained by the fact that,} for the obscured AGN, the X-ray hardness does not depend on the aperture used for measurement, since their soft X-ray emission is absorbed and their hard X-ray emission is point-like. For other galaxies, as the aperture size increases, more and more soft X-ray emission from stars is included, thereby decreasing the hardness of its observed X-ray radiation field. An alternative explanation could be a temperature difference in the X-ray emitting gas, with higher temperatures in the center near the starburst, gradually cooling off as the X-ray bubble expands.

\subsection{Dual Nuclei Systems}
\label{sec:duals}

There are five systems in our sample, Arp 220, NGC 6240, IRAS 07251-0248, IRAS F12112+0305 and IRAS F14348-1447, that feature resolved dual nuclei $d_\mathrm{nuc} >$ 100\,pc apart for which we can generate separate SEDs \edit1{in \S \ref{sec:modeling}.} This approach allows us to start tracking the state of the SMBH pairs in merging galaxies as we build up a larger sample with similarly highly resolved data. \edit1{We plot the SEDs for the dual nuclei of each galaxy together in Figure \ref{fig:dual_nuclei_1}. For clarity, we remove the nebular emission lines, leaving only the continuum.} Based on matching the projected separation distance between the nuclei to N-body merger simulations by \cite{2014Stickley}, we estimate the interaction timescale for four of the galaxy systems here. For all four galaxies, the maximum time for transitioning from their current state to final nuclei coalescence is $\sim$340 Myrs. The minimum time required increases with nuclear separation, ranging from 10 to 50 Myrs. \cite{2022Sobolenko} more precisely constrained the upper limit for the merging time of the double SMBH system in NGC 6240, estimating that it would take $\sim70\,\rm Myr$ from now until final coalescence. The nuclear separation of Arp 220 is smaller than the resolution of the simulation \citep{2014Stickley}, making it challenging to directly estimate the merging time. Nevertheless, due to its small nuclear separation compared to NGC 6240 and a morphology that indicates late-stage merging, 70 Myr can also serve as an upper limit for its merging time. According to the simulation, all five galaxies are likely in the phase between the second and third passages of the nuclei of two progenitor galaxies. During this period, significant accretion onto SMBHs may occur, corresponding to a bolometric luminosity greater than $10^{44}\,\rm erg\,s^{-1}$. SMBHs are likely to be active alone rather than as a pair at this luminosity threshold, as discussed in more detail by \cite{2012VanWassenhove}. Here, we discuss these dual-nuclei systems in more detail.

\paragraph{Arp 220}
It is the closest ULIRG, and its nuclear region has been the subject of many high-resolution studies. Gas and dust disks around both nuclei, with sizes of $\sim100\,\rm pc$, have been observed \citep{2014Wilson,2015Rangwala,2015Scoville,2017Scoville,2017Sakamoto}. \emph{HST} observation shows that 50\% of the Pa$\beta$ emission is emitted by the nuclear disks \citep{2023Chandar}. \cite{2015Loreto} estimate extremely high molecular gas surface densities of $\rm \Sigma_{IR}\sim4.2^{+1.6}_{-0.7}\times10^{13}$ (east) and $\rm \Sigma_{IR}\sim9.7^{+3.7}_{-2.4}\times10^{13}$ $L_{\odot}\,\rm kpc^{-2}$ (west), and star formation rate surface densities of $\rm \Sigma_{SFR}\sim10^{3.7\pm0.1}$ (east) and $\rm \Sigma_{SFR}\sim10^{4.1\pm0.1}$ $M_{\odot}\,\rm yr^{-1}\,kpc^{-2}$ (west). MUSE observation also reveals a disturbed kpc-scale disk in its nuclear region, along with a multi-phase galactic-scale outflow emerging from its eastern nucleus \citep{2020Perna}. Collimated, bipolar outflow of molecular gas with speed of up to 840 $\rm km\,s^{-1}$ has also been observed by ALMA, but whether it is driven by starburst or AGN is still unclear \citep{2018Loreto,2021SakamotoA,2021SakamotoB}. Moreover, the inferred gas column density ($N_{\rm H}\sim10^{25-26}\,\rm cm^{-2}$) suggests that any possible AGN in the western nuclear region can be highly Compton thick \citep{2014Wilson,2021SakamotoA}. The two nuclei are so close that their X-ray emission cannot be resolved by \emph{Chandra}. Both nuclei exhibit similar SED shapes, with the western nucleus being brighter than the eastern one across all wavelengths (Figure \ref{fig:dual_nuclei_1}, panel A). Additionally, both show excess emission in the submillimeter bands. Due to heavy obscuration, the two nuclei are most distinct in the ALMA images but are difficult to discern at other wavelengths. High-resolution mid-infrared observations from \emph{JWST} would be crucial for determining the nature of its nuclei.

\paragraph{NGC 6240}
A well-known late-stage merger hosting dual X-ray AGN. \emph{Chandra} observation resolved the X-ray emission into two nuclei separated by $\sim$ 1\,kpc that host CT AGN through the detection of reflection-dominated spectra \citep{2003Komossa}. High-resolution ground-based infrared AO observations not only resolved the two SMBHs~\citep{2005Max,2007Max} but also measured their masses dynamically~\citep{2011Medling,2015Medling}. As we can see in Figure \ref{fig:dual_nuclei_1} (panel B), the southern nucleus is much more prominent across most of the electromagnetic spectrum even though both nuclei suffer from dust extinction and appear similarly dim in the UV regime. Both nuclei exhibit similarly shaped SEDs, which could be a result of their proximity to each other and their sharing of the circumnuclear environments (where cold molecular gas has been detected to reside in between the nuclei)~\citep{2014Wang,2019Medling,2020Treister}. 

\paragraph{IRAS 07251$-$0248}
Although it has been classified as an IC 5 system with a single or obscured nucleus~\citep{2013Kim}, its dual nuclei are well resolved in the \emph{HST}, Keck, and ALMA images, with a separation of $\sim2\,\rm kpc$. The eastern nucleus is dustier than its western counterpart, which is reflected in both the downturn in UV emission and the stronger submillimeter emission (Figure \ref{fig:dual_nuclei_1}, panel C). Its strong submillimeter photometric values drive the overall SED fitting. \cite{2015Kirkpatrick} argued that the far-infrared/submillimeter emission should be dominated by cold dust, which is associated with star formation. Additionally, IRAS 07251$-$0248 exhibits disk-like motions as traced by H$\alpha$ 
in the nuclear region, but with a kinematic center offset from either of its two nuclei~\citep{2022Perna}. We note that the dual nuclei are not resolved by \emph{Chandra}, where the X-ray emission is only detected in the soft $0.5-2\,\rm keV$ band. It is thus unclear whether either component in this galaxy system hosts an AGN based on existing data.

\paragraph{IRAS F12112+0305}
In this case of another late-stage merger featuring large-scale tidal tails and dual SMBHs with a separation of $d_\mathrm{nuc} \sim$ 4\,kpc, the X-ray emission from both nuclei are comparable. However, the northern nucleus is significantly brighter in the submillimeter while the southern nucleus is more luminous in the \edit1{optical/near-infrared, presumably because it has more unattenuated stellar emission (see also Figure \ref{fig:allseds3}).} Figure \ref{fig:dual_nuclei_2} (panel D) shows an SED with a clear stellar hump in the optical and infrared for the southern nucleus, while the northern nucleus is dominated by dust. \edit1{Based on mid-infrared spectroscopy, this ULIRG is considered to be starburst-dominated~\citep{1998Genzel}, while its optical spectrum hosts signatures of a LINER~\citep{1999Veilleux}.} However, it is possible that if both of these nuclei contain AGN, then the northern one is likely to appear extincted and may evade optical spectroscopic searches. The detection of bright vibrationally excited HCN emission indicates that its northeastern nucleus is a compact obscured nucleus, which can be optically thick up to millimeter wavelengths~\citep{2021Falstad}. We note that the longest-wavelength submillimeter data points are not all well-fitted; more high-resolution radio data may help improve the fit.

\paragraph{IRAS F14348$-$1447}
It consists of two galaxies that are in the pre-coalescent stage of a merger, with a projected nuclear separation of $\sim$6 kpc. \cite{2018Pereira-Santaella} found nuclear molecular gas disks around both nuclei, while the CO($2-1$) emission is dominated by the southern nucleus. Molecular gas outflows with different temperatures have also been observed \citep{2013Veilleux,2017Emonts,2018Pereira-Santaella}. The hot $\rm H_{2}$ gas outflow from the southern nucleus has a velocity of $\sim900\,\rm km\,s^{-1}$ and a mass outflow rate of $\sim85\,M_{\odot}\,\rm yr^{-1}$, and is likely driven by a buried AGN \citep{2017Emonts}.  Both mid-infrared and X-ray observations have classified the southern nucleus of IRAS F14348-1447 as an AGN \citep{2016Alonso-Herrero,2011Iwasawa}. Although the two nuclei appear not to be in a common envelope, they exhibit similar SED shapes, including a subtle optical stellar peak. The southern nucleus, which is believed to host the AGN, appears brighter in all bands. However, the \emph{Chandra} observation only reveals faint diffuse emission, while \emph{NuSTAR} cannot detect this source at all \citep{2021Ricci}.

\begin{figure*}[htbp]
	
	\centering
	\includegraphics[width=\textwidth]{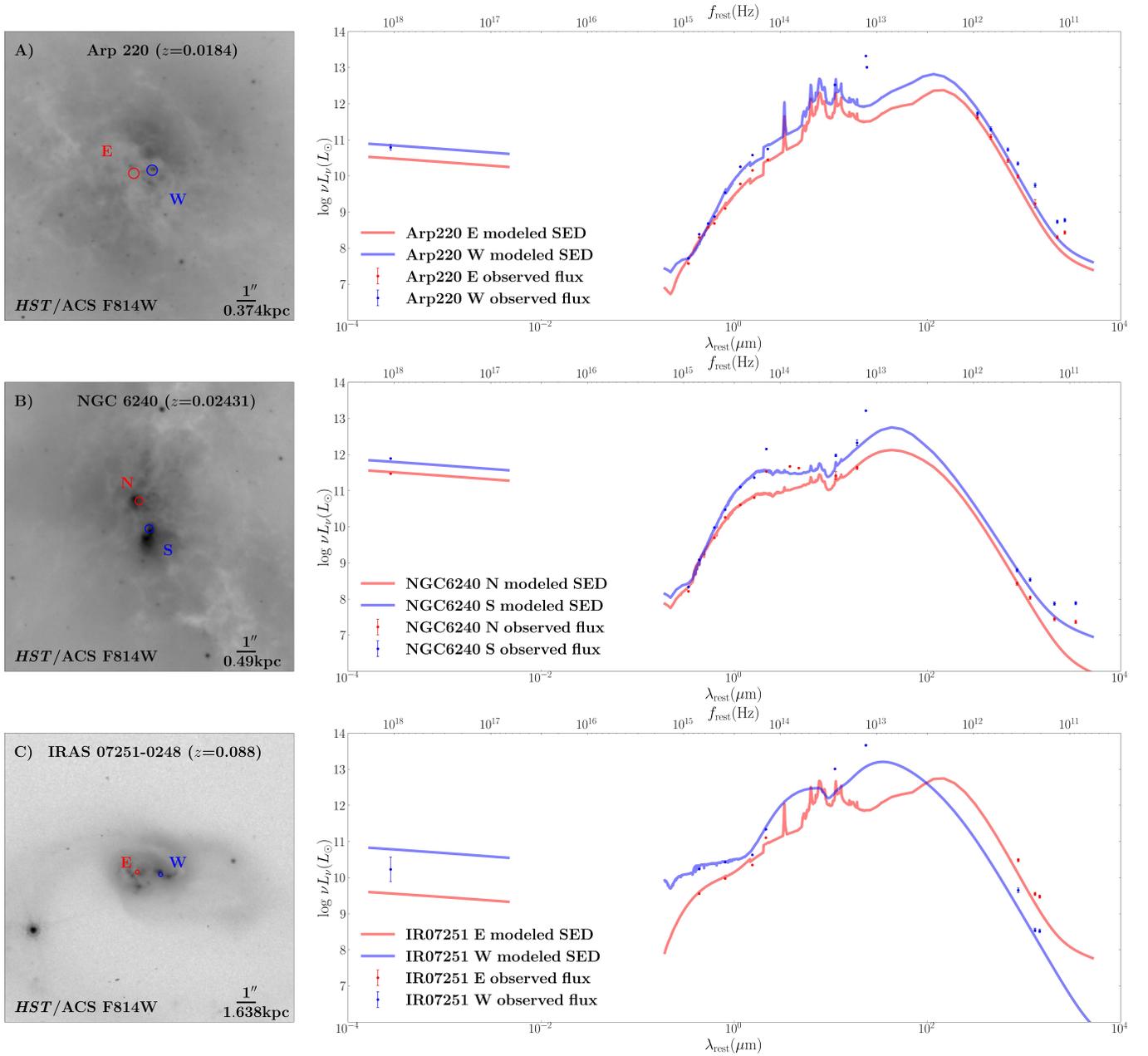}
	\caption{Resolved nuclear SEDs for five dual nuclei systems (from top to bottom: Arp 220E/W, NGC 6240N/S, IRAS 07251$-$0248E/W, IRAS F12112+0305N/S and IRAS F14348$-$1447N/S). The left panel shows the \emph{HST/ACS} F814W images, overlaid with the nuclear apertures of the resolved SMBH pair within each system. The right panel shows the observed flux and modeled SEDs for the different nuclei marked with the same colors. }
	\label{fig:dual_nuclei_1}
	
\end{figure*}

\addtocounter{figure}{-1}

\begin{figure*}[htbp]
	
	\centering
	\includegraphics[width=\textwidth]{dual_nuclei_part2.pdf}
	\caption{Continued}
	\label{fig:dual_nuclei_2}
	
\end{figure*}

Although the dual nuclei in the five (U)LIRGs mentioned above all have separations $<10\,\rm kpc$, a phase hosting the peak of SMBH growth as demonstrated by both simulations and observations~\citep{2012VanWassenhove,2021Stemo}, it remains controversial whether or not some of these sources contain dual or offset AGN. 
In some cases, the differences between the stellar and dust components shown in the SEDs suggest that AGN triggering might depend on the conditions of their immediate vicinity. According to simulations of galaxy mergers, a SMBH binary is expected to be embedded in a gaseous circumnuclear disk (CND) formed by infalling gas \citep{2007Mayer,2010Mayer}. CNDs with radii of $<1\,\rm kpc$ have also been observed in nearby (U)LIRGs \citep{2014Medling}. Compared to the SFR on galactic scales, the small-scale ($<100\,\rm pc$) SFR in the CND shows a stronger correlation with the accretion rate of the SMBH throughout the entire galaxy merger process, as both have similar variability timescales in simulations \citep{2015Volonteri}. This correlation is attributed to the fact that both small-scale SFR and the accretion rate of the SMBH are influenced by local merger dynamics. \edit2{A recent study, using a stellar mass-limited sample at $z<1$ from the Kilo-Degree Survey (KiDS) and selecting AGN based on MIR, X-ray, and SED-fitting criteria, reported a clear excess of AGN by a factor of 2–3 in mergers compared to non-mergers for MIR-selected AGN, along with a moderate excess for X-ray and SED AGN \citep{2024LaMarca}. The merger fraction ($f_{\rm merger}$) among the MIR AGN hosts (45–55\%) was found to be significantly higher than in the MIR non-AGN controls (20–30\%), as well as in the X-ray and SED AGN populations. Additionally, it was found that, across all redshift bins and AGN types, mergers exhibited an excess of AGN with high $f_{\rm AGN}$ values compared to non-mergers, although a strong correlation between $f_{\rm merger}$ and $f_{\rm AGN}$ appeared only for sources with $f_{\rm AGN}>0.8$. These results suggest that mergers are strongly linked to the presence of dust-obscured and the most powerful AGN, while less powerful AGN may be primarily triggered by secular processes or stochastically by mergers, as supported by previous studies across various wavelengths and redshifts \citep{2008Urrutia, 2012Treister, 2015Glikman, 2018Goulding, 2019Ellison, 2020Gao, 2022Pierce}. This is further consistent with our finding that the triggering of AGN during mergers is not solely determined by the merger stage, but remains susceptible to other contributing factors.} Moreover, the feedback triggered by the more massive SMBH, which is already active, could also heat up the surrounding gas reservoir, inhibiting gas accretion onto the less massive SMBH \citep{2016Steinborn,2017SouzaLima}. Measurements of the gas temperature distribution within the CND by the next-generation Extremely Large Telescopes can help us better understand the role of feedback in the formation process of dual AGN.

\section{CONCLUSION} \label{sec:conclusion}

We present the resolved, nuclear (100$-$150\,pc) SEDs for a sample of 23 merging (U)LIRGs, including five galaxies that host dual nuclei at $\lesssim$ \edit1{6}\,kpc separation. Leveraging high-resolution X-ray through submilliter archival data from space- and ground-based facilities, we amassed multi-band aperture photometry for the 28 nuclei in the merging galaxy systems based on coordinates localized by ALMA observations. We performed multi-component SED fitting using X-CIGALE and binned the resulting SEDs by merger stages and infrared luminosities. Our findings are summarized as follows:

\begin{itemize}
    \item We find that the contributions from the various components (stellar, dust, and AGN) to the nuclear SEDs vary with merger stage. Of note is that the AGN appears to have a lower contribution to the total IR luminosity in stages 2, possibly due to the buildup of dust in the nucleus, followed by an enhancement among the later stages. This phenomenon is consistent with the scenario in which deeply buried AGN are expected in \edit1{late-stage mergers and post-merger systems}, where the SMBH is actively accreting gas behind thick dust screens. \edit1{Meanwhile, compared to LIRGs, ULIRGs exhibit a higher level of AGN emission as well as enhanced dust emission in the mid- to far-infrared within the nuclear region. Nevertheless, we would like to caution that due to the lack of high-resolution mid- and far-infrared data, the physical quantities underlying this conclusion may have significant uncertainties, as demonstrated by the mock analysis.} 
    \item Comparing the X-ray and infrared emission for global versus nuclear values, we find differences in trends associated with soft and hard X-ray emission. Specifically, the nuclear regions for AGN and non-AGN feature similar $L_{\rm SX}$, whereas the $L_{\rm HX}$ for AGN are larger than for non-AGN because of the AGN contribution to the Fe K$\alpha$ line at 6.4 keV. The spread in global $L_{\rm HX}$ at a given $L_{\rm IR}$ appears larger than that in global $L_{\rm SX}$, suggesting that the infrared emission in the nucleus is dominated by dust and AGN with relatively little stellar contribution. Meanwhile, the $HR$ values of the nuclear regions are overall higher than those measured for entire galaxies, regardless of the presence of an AGN.
    \item Our resolved SED approach is particularly useful in dissecting the nature of individual SMBHs in dual systems in order to address the timescale question of dual AGN detections in large surveys. Our pilot sample contains five well-resolved dual nuclei systems where the projected separation of the nuclei ranges from 0.5 to 6 kpc and that contain SMBH pairs that may be quiescent, low-luminosity or obscured, or X-ray bright. Our SED fitting suggests that \edit1{the individual SMBHs in the pair can be in different states as their host galaxies move along the merger sequence}, but clearly a much larger sample of SMBH pairs resolved in mergers with high-resolution multi-wavelength data will be needed for the statistical inference of the dual AGN timescale.
\end{itemize}

Additional high-resolution imaging data for a larger sample of galaxies, particularly in the mid-infrared e.g. from \emph{JWST}, will help constrain the dust and AGN contribution to the central regions and thus, the true intrinsic obscured AGN fraction among (U)LIRGs. This may be the key to explaining the apparent deficit of dual closely-separated AGN in mergers as progenitors of gravitational wave emitters hosting the coalescence of supermassive black holes. The next-generation far-infrared space telescopes with high sensitivity and spatial resolution will sample the wavelength range where obscured star formation is the main energy source of dust emission. This capability will help in accurately distinguishing the two heating mechanisms in small-scale regions. Furthermore, these telescopes will enable more reliable estimates of AGN and host galaxy properties (e.g. SFR, SMBH accretion rate) at high redshifts, which will significantly enhance our understanding of the co-evolution between galaxies and SMBHs across cosmic time.

\facilities{\emph{Chandra}, \emph{HST} (ACS/NICMOS/WFC3), Keck:II (NIRC2), ALMA, VLT, Gemini, Subaru, MAST, KOA, IPAC/NED, NASA/ADS}

\software{astropy~\citep{2013astropy},
CIAO data analysis package~\citep{2006CIAO},
Cosmology calculator~\citep{2006Wright},
DrizzlePac~\citep{2012DrizzlePac},
HEASOFT / XSPEC~\citep{1996Xspec},
Keck AO Imaging DRP~\citep{2022lu},
Python \texttt{photutils} package~\citep{2020photutils},
TinyTim~\citep{1993Krist},
X-CIGALE~\citep{2022Yang}
}

\section*{Acknowledgments}
We would like to thank the anonymous referee for a detailed report, which greatly helps improve the quality of this manuscript. We would like to thank Guang Yang, the author of X-CIGALE, for helpful discussions on SED modeling. We thank Bailey Liu and Ansh Vashisth for their contribution to gathering the imaging data. This research is supported primarily by the National Aeronautics and Space Administration (NASA) under Grant No. 80NSSC20K0450 issued through the Astrophysics Data Analysis Program (ADAP). 
T.G. acknowledges support from ARC Discovery Project DP210101945. V.U further acknowledges partial funding support from NASA grants \#80NSSC23K0750 and \#80NSSC25K7477, National Science Foundation (NSF) Astronomy and Astrophysics Research Grant \#2408820, as well as \emph{STScI} grants \#JWST-GO-01717.001-A, \#HST-AR-17063.005-A, and \#HST-GO-17285.001-A, which were provided by NASA through a grant from the Space Telescope Science Institute, which is operated by the Association of Universities for Research in Astronomy, Inc., under NASA contract NAS 5-03127.  V.U had partially performed work for this project at the Aspen Center for Physics, which is supported by National Science Foundation grant PHY-2210452. D.B.S. and C.A. acknowledge support from National Science Foundation (NSF) AST Grant \#1716994.

This research has made use of data obtained from the Chandra Data Archive and the Chandra Source Catalog, and software provided by the Chandra X-ray Center (CXC) in the application packages CIAO and Sherpa.
This research is based on observations made with the NASA/ESA Hubble Space Telescope obtained from the Space Telescope Science Institute, which is operated by the Association of Universities for Research in Astronomy, Inc., under NASA contract NAS 5–26555. These observations are associated with program(s) 10592, 11235, \& 13690.
This research has made use of the NASA/IPAC Extragalactic Database (NED), which is funded by the National Aeronautics and Space Administration and operated by the California Institute of Technology.

Some of the data presented herein were obtained at the W. M. Keck Observatory, which is operated as a scientific partnership among the California Institute of Technology, the University of California and the National Aeronautics and Space Administration. The Observatory was made possible by the generous financial support of the W. M. Keck Foundation.
This research has made use of the Keck Observatory Archive (KOA), which is operated by the W. M. Keck Observatory and the NASA Exoplanet Science Institute (NExScI), under contract with the National Aeronautics and Space Administration.
The authors wish to recognize and acknowledge the very significant cultural role and reverence that the summit of Maunakea has always had within the indigenous Hawaiian community.  We are most fortunate to have the opportunity to use observations conducted from this mountain.

This paper makes use of the following ALMA data: ADS/JAO.ALMA\#2012.1.00077.S, 2012.1.01022.S, 2013.1.01165.S, 2015.1.00102.S, 2015.1.00708.S, 2015.1.00804.S, 2015.1.01439.S, 2015.1.01448.S, 2016.1.00051.S, 2016.1.00140.S, 2016.1.00991.S, 
2016.1.01385.S, 2017.1.00022.S, 2017.1.00255.S, 2017.1.00395.S, 2017.1.00759.S, 2017.1.01085.S, 2017.1.01235.S, 2018.1.00279.S, 2018.1.00486.S, 2018.1.00699.S, 2018.1.01123.S, 2018.1.01344.S, 2019.1.00027.S, 2019.1.00811.S. ALMA is a partnership of ESO (representing its member states), NSF (USA) and NINS (Japan), together with NRC (Canada), MOST and ASIAA (Taiwan), and KASI (Republic of Korea), in cooperation with the Republic of Chile. The Joint ALMA Observatory is operated by ESO, AUI/NRAO and NAOJ. The National Radio Astronomy Observatory is a facility of the National Science Foundation operated under cooperative agreement by Associated Universities, Inc.

This research has made use of the SVO Filter Profile Service "Carlos Rodrigo", funded by MCIN/AEI/10.13039/501100011033/ through grant PID2023-146210NB-I00.

\bibliographystyle{aasjournal}
\bibliography{ms_nuclearSEDs}

\appendix
\renewcommand\thefigure{\thesection.\arabic{figure}}    
\setcounter{figure}{0}

\section{X-ray Spectra and Derived Properties}\label{sec:appendix_xray}

This section presents the X-ray spectra of the nuclear regions for all galaxies in our sample, as well as the X-ray properties obtained through spectral fitting.

\begin{figure*}[htbp]
	\centering
        \includegraphics[width=0.9\textwidth]{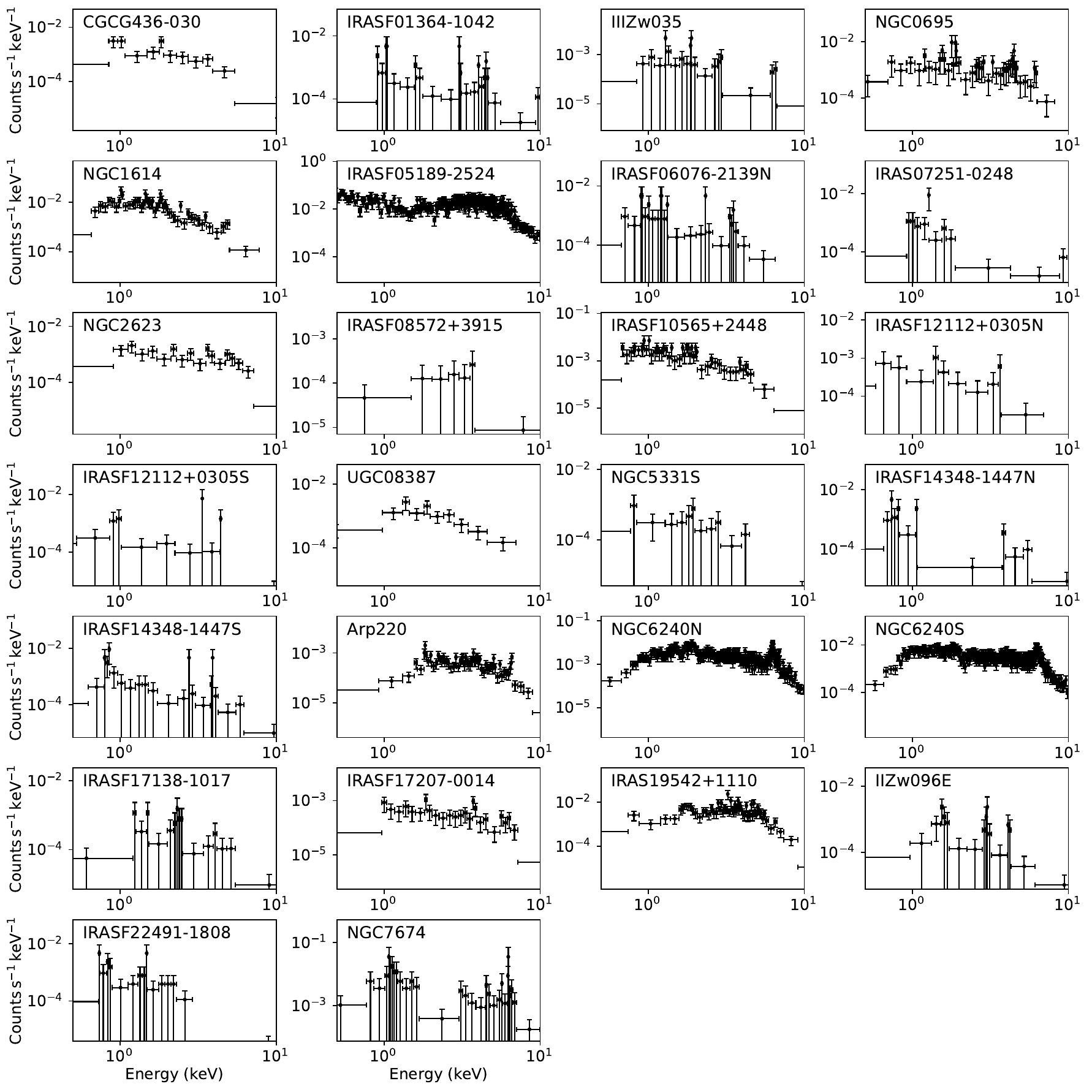}
	\caption{All 0.5–10 keV X-ray spectra extracted from the nuclear regions of galaxies in our sample.}
	\label{fig:xray-spectra}
\end{figure*}

\clearpage
\begin{deluxetable*}{lccccc}
\tablenum{A.1}
\tablecaption{Nuclear X-ray Spectral Properties for Each Galaxy\label{tab:x-ray_derived-properties}}
\tabletypesize{\footnotesize}
\tablewidth{0pt}
\tablehead{
\colhead{Galaxy} & \colhead{SX/HX Photon Counts} & \colhead{$kT$ (keV)} & \colhead{Metallicity ($Z_{\odot}$)} &
\colhead{$N_{\mathrm{H}} (10^{22}\,\mathrm{cm^{-2}})$} & \colhead{Photon Index}}
\startdata
CGCG 436-030 & 25/20 & $2.11^{+0.67}_{-1.04}$ & $0.43^{+0.39}_{-0.11}$ & $12.50^{+3.63}_{-1.91}$ & $1.76^{+1.03}_{-1.02}$ \\
IRAS F01364-1042 & 8/13 & 0.70 & $1.28^{+1.15}_{-0.92}$ & 10.00 & $1.25^{+0.94}_{-0.75}$ \\
III Zw 035N & 11/7 & 0.70 & $0.45^{+1.25}_{-0.30}$ & 10.00 & $2.60^{+1.43}_{-1.33}$ \\
NGC 0695 & 32/52 & $2.42^{+0.43}_{-0.99}$ & $0.55^{+1.13}_{-0.35}$ & $13.47^{+3.86}_{-2.62}$ & $2.34^{+0.46}_{-0.65}$ \\
NGC 1614 & 190/83 & $4.47^{+0.39}_{-0.66}$ & $0.57^{+0.68}_{-0.35}$ & $4.35^{+2.47}_{-2.20}$ & $3.13^{+0.55}_{-0.58}$ \\
IRAS F05189-2524 & 460/1479 & $1.97^{+0.02}_{-0.04}$ & $0.31^{+0.03}_{-0.01}$ & $^{(a)}12.59^{+1.21}_{-0.57}$ & $0.03^{+0.04}_{-0.02}$ \\
IRAS F06076-2139N & 13/10 & 0.70 & $1.17^{+1.23}_{-0.94}$ & $18.00^{+7.64}_{-5.49}$ & $5.87^{+1.45}_{-1.56}$ \\
IRAS 07251-0248 & 10/$\dots$ & 0.70 & $1.42^{+1.07}_{-0.98}$ & $^{(b)}\dots$ & $\dots$ \\
NGC 2623 & 27/58 & $2.50^{+0.35}_{-0.53}$ & $1.10^{+1.21}_{-0.74}$ & $^{(a)}7.08^{+4.94}_{-2.81}$ & $0.12^{+0.42}_{-0.41}$ \\
IRAS F08572+3915NW & $\dots$/4 & $\dots$ & $\dots$ & $^{(b)}\dots$ & $1.19^{+1.12}_{-0.84}$ \\
IRAS F10565+2448 & 89/47 & $2.66^{+0.24}_{-0.40}$ & $0.41^{+0.52}_{-0.23}$ & $^{(b)}\dots$ & $2.02^{+0.47}_{-0.48}$ \\
IRAS F12112+0305N & 5/4 & 0.70 & $1.25^{+1.13}_{-0.93}$ & $^{(b)}\dots$ & $1.50^{+0.96}_{-0.96}$ \\
IRAS F12112+0305S & 4/4 & 0.70 & $1.52^{+1.00}_{-1.00}$ & $^{(b)}\dots$ & $0.37^{+1.37}_{-0.97}$ \\
UGC 08387 & 22/21 & $2.52^{+0.33}_{-0.53}$ & $0.72^{+1.27}_{-0.48}$ & 10.00 & $3.86^{+1.01}_{-0.98}$ \\
NGC 5331S & 7/4 & 0.70 & $1.00^{+1.29}_{-0.75}$ & 10.00 & $3.97^{+2.32}_{-2.28}$ \\
IRAS F14348-1447N & 6/3 & 0.70 & $1.50^{+1.05}_{-0.97}$ & $^{(b)}\dots$ & $0.50^{+0.34}_{-0.35}$ \\
IRAS F14348-1447S & 13/9 & 0.70 & $1.29^{+1.15}_{-0.92}$ & $^{(b)}\dots$ & $0.89^{+0.59}_{-0.58}$ \\
Arp 220 & 44/282 & $2.71^{+0.22}_{-0.49}$ & $0.65^{+1.11}_{-0.43}$ & $^{(a)}>524.81$ & $3.13^{+0.55}_{-0.58}$ \\
NGC 6240N & 184/373 & $8.68^{+0.93}_{-1.40}$ & $2.43^{+0.42}_{-0.77}$ & $^{(a)}154.88^{+74.21}_{-31.85}$ & $0.24^{+0.07}_{-0.07}$ \\
NGC 6240S & 328/551 & $0.96^{+0.53}_{-0.07}$ & $0.27^{+0.20}_{-0.08}$ & $^{(a)}147.91^{+21.91}_{-19.09}$ & $-0.35^{+0.06}_{-0.06}$ \\
IRAS F17138-1017 & 3/10 & 0.70 & $0.47^{+1.67}_{-0.24}$ & $^{(b)}\dots$ & $1.74^{+0.82}_{-1.06}$ \\
IRAS F17207-0014 & 22/50 & $1.74^{+0.19}_{-0.32}$ & $1.19^{+1.57}_{-0.71}$ & $^{(b)}\dots$ & $0.38^{+0.40}_{-0.27}$ \\
IRAS 19542+1110 & 52/245 & $3.56^{+0.33}_{-0.80}$ & $0.61^{+1.18}_{-0.39}$ & $11.98^{+2.09}_{-1.39}$ & $2.31^{+0.34}_{-0.31}$ \\
II Zw 096E & 7/9 & 0.70 & $0.72^{+1.39}_{-0.54}$ & $^{(b)}\dots$ & $1.52^{+0.90}_{-1.11}$ \\
IRAS F22491-1808 & 11/2 & 0.70 & $2.08^{+0.93}_{-0.46}$ & 10.00 & $\dots$ \\
NGC 7674 & 12/16 & 0.70 & $0.81^{+1.41}_{-0.61}$ & $^{(a)}>301.99$ & $-0.40^{+0.69}_{-0.43}$ \\
\enddata
\tablecomments{Column 1: Galaxy name; Column 2: Photon counts in the soft (0.5$-$2 keV) and hard (2$-$10 keV) X-ray bands of the spectrum; Column 3: \textit{apec} model temperature; Column 4: \textit{apec} model metallicity; Column 5: Line-of-sight column densities of intrinsic obscuration; Column 6: Photon index of power-law model. $^{(a)}$Values from \cite{2021Ricci}. $^{(b)}$Intrinsic hard X-ray luminosity is available in \cite{2021Ricci}, but column density is not included.}

\end{deluxetable*}

\section{Aperture Photometry for the Sample}
\label{sec:appendix_apphot}

\setcounter{figure}{0}

We include here the representative multiwavelength data set with nuclear apertures used in our complete sample, complementing Figure \ref{fig:photometry}. 

\begin{figure*}[htbp]
	\centering
        \includegraphics[width=0.6\textwidth]{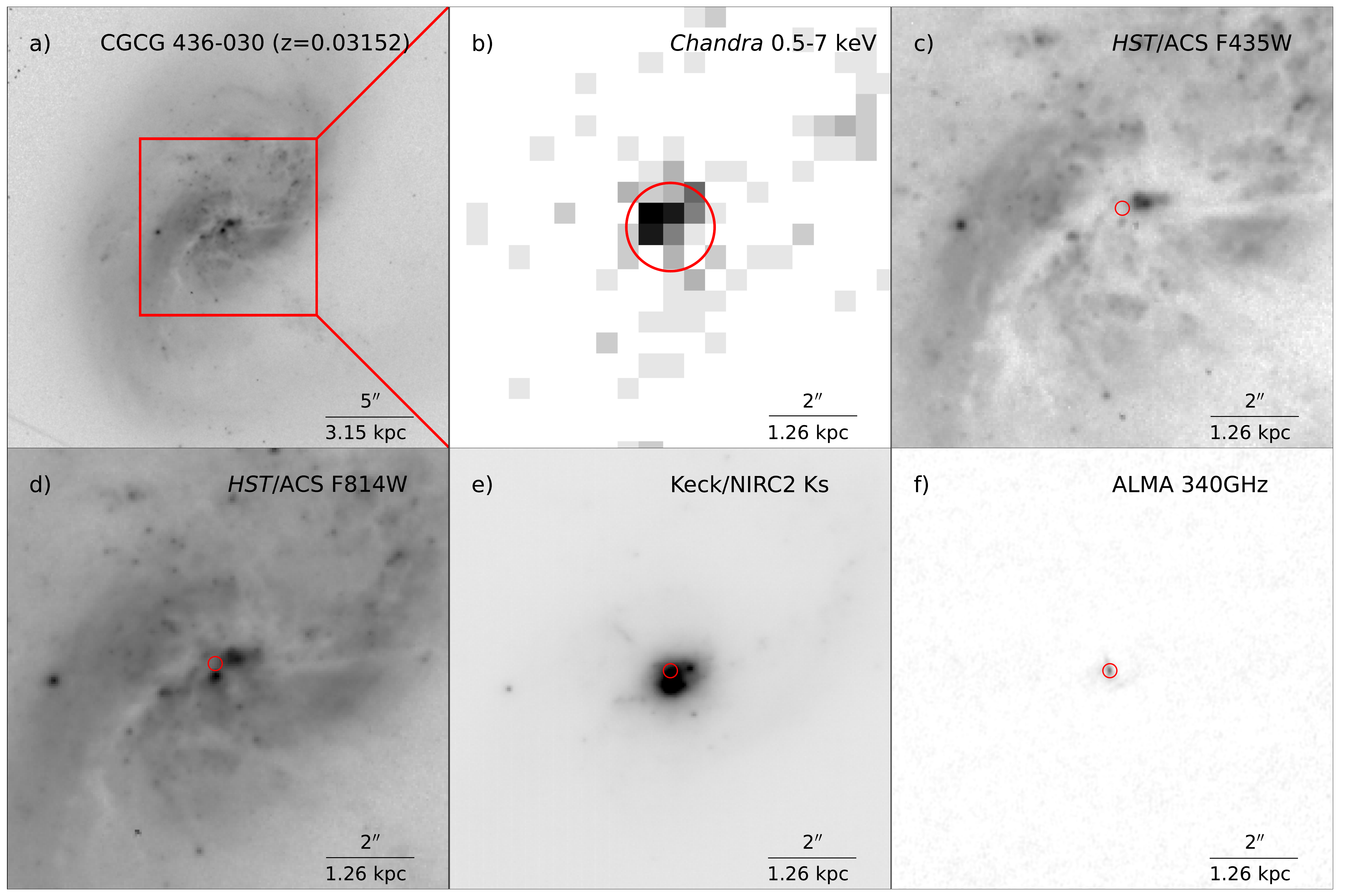}
        \includegraphics[width=0.6\textwidth]{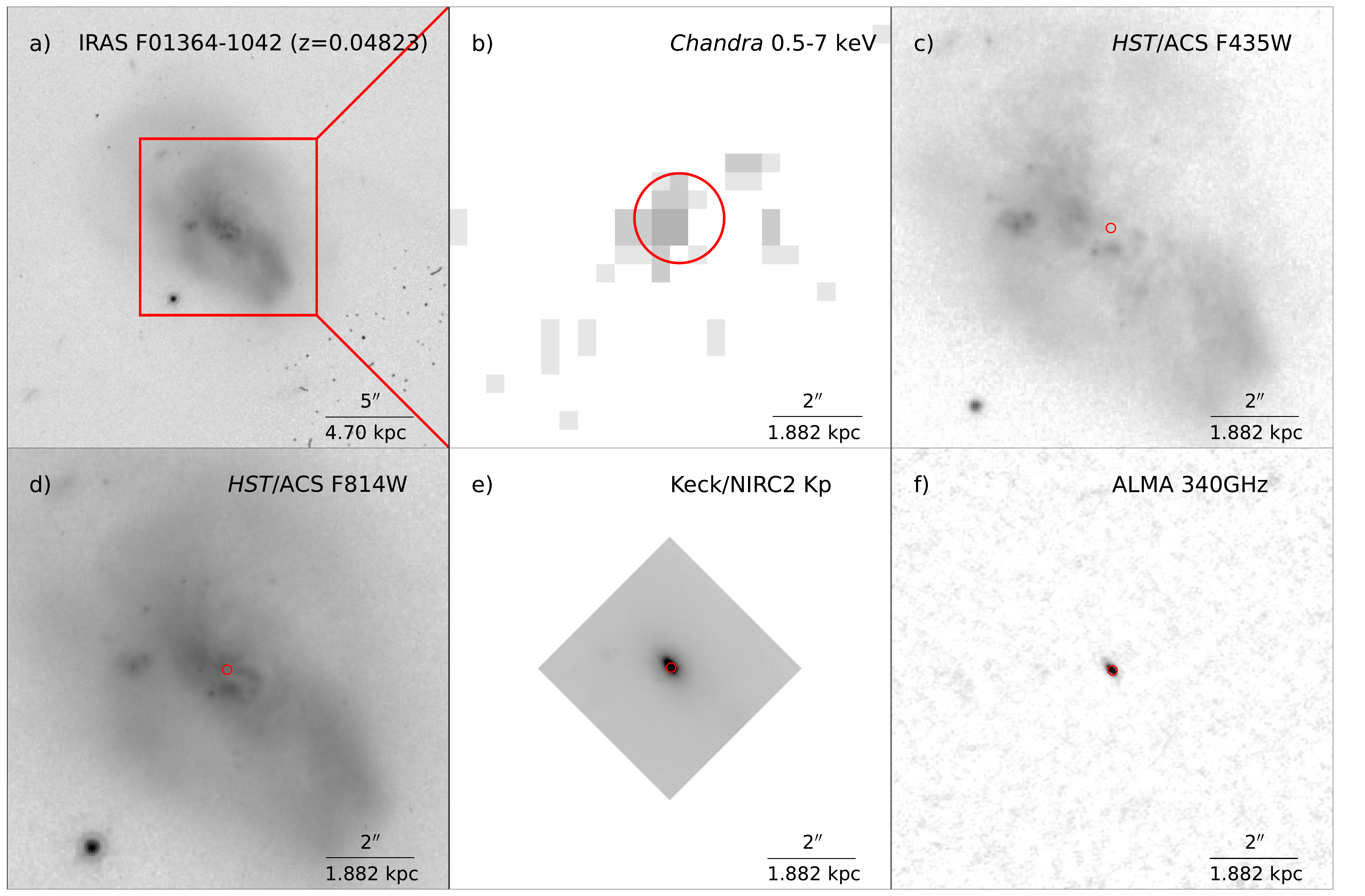}
	\caption{Apertures used for extracting nuclear photometry from a representative subset of our multiwavelength data set for our sample. See figure caption in Figure \ref{fig:photometry} for the detailed description.}
	\label{fig:append_photometry}

 \end{figure*}

\addtocounter{figure}{-1}

\begin{figure*}[htbp]
	\centering
        \includegraphics[width=0.6\textwidth]{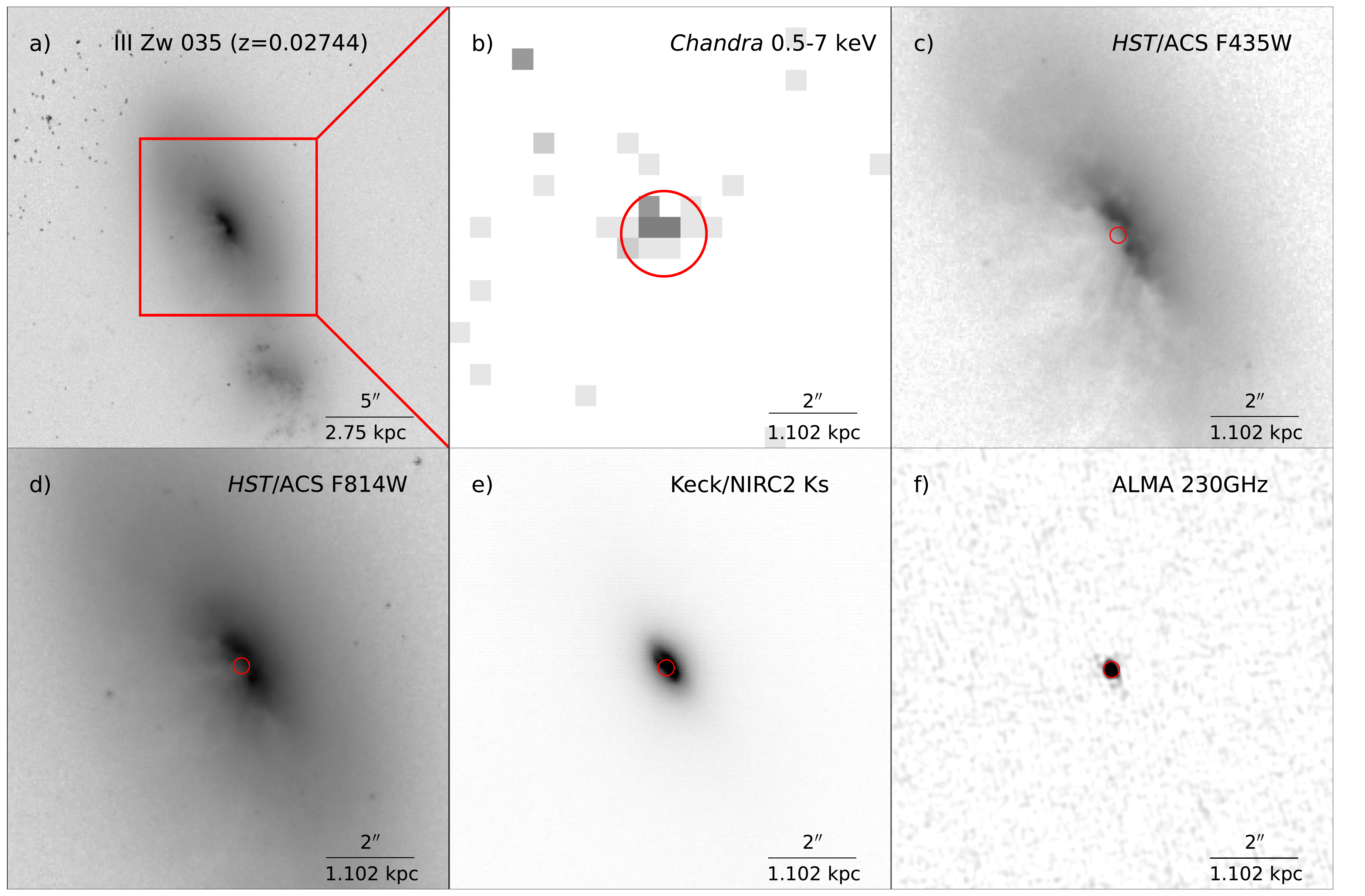}
        \includegraphics[width=0.6\textwidth]{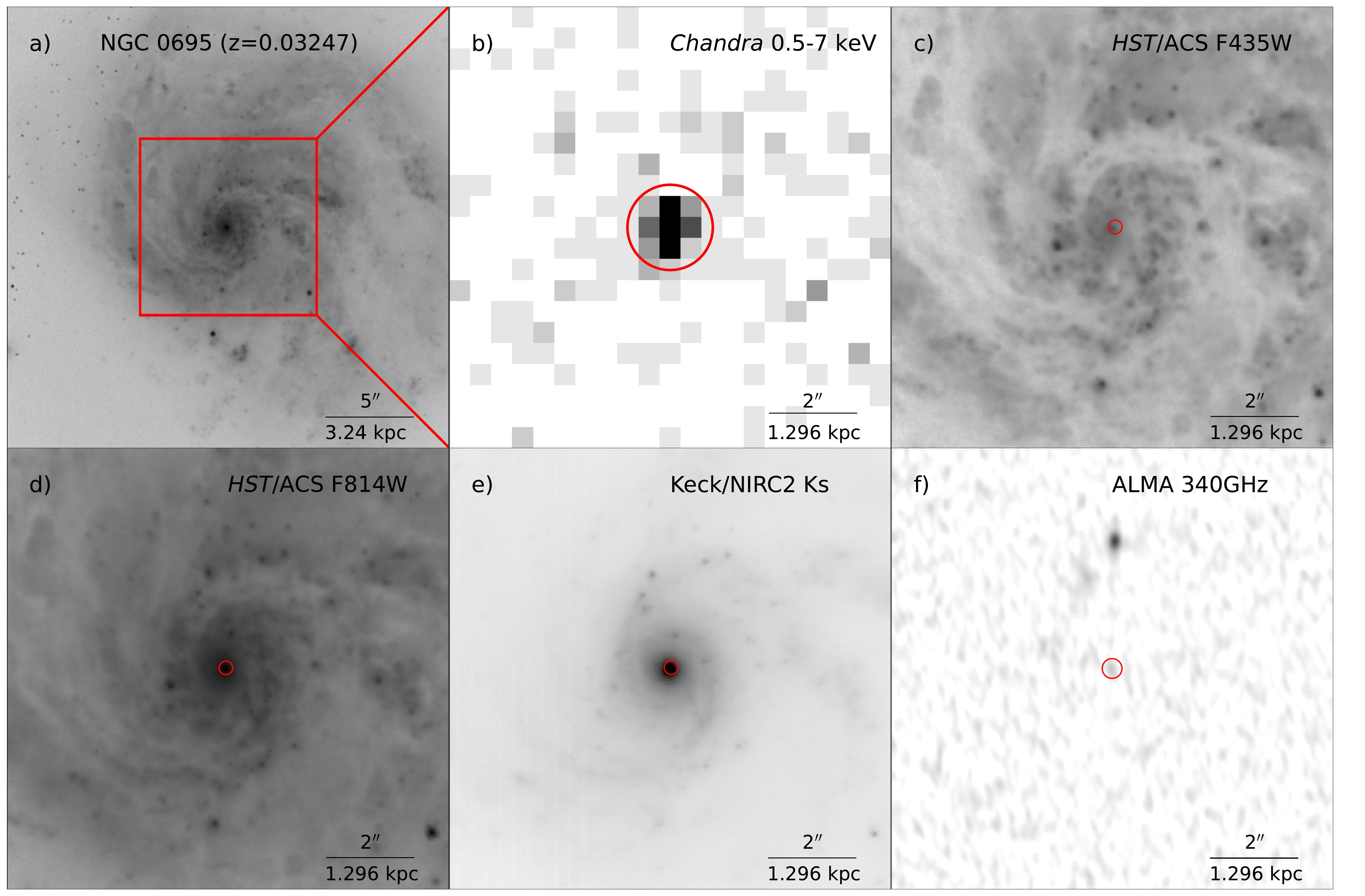}
        \includegraphics[width=0.6\textwidth]{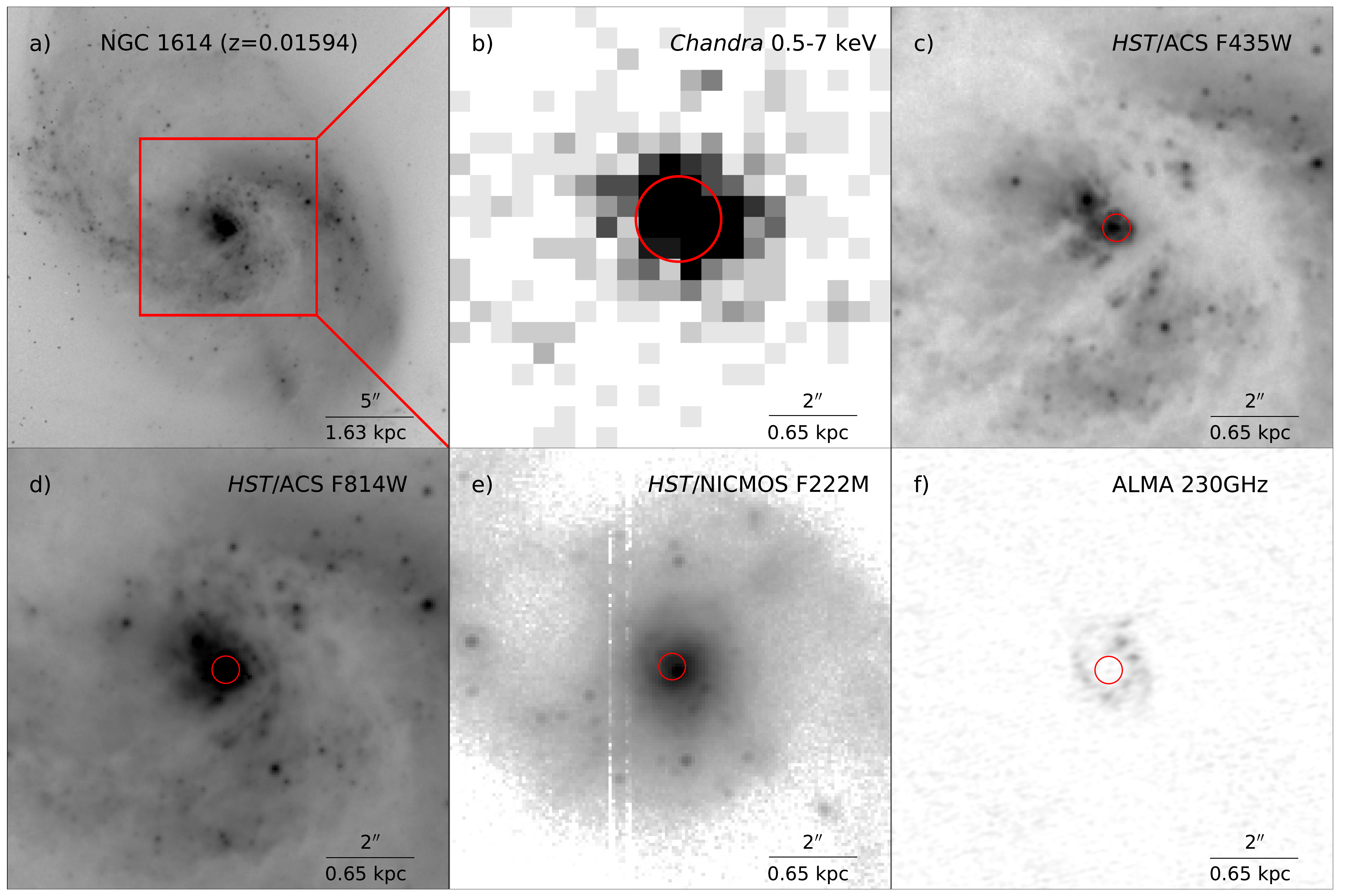}
        
	\caption{Continued}
	\label{fig:append_photometry2}
\end{figure*}

\addtocounter{figure}{-1}

\begin{figure*}[htbp]
	\centering
        \includegraphics[width=0.6\textwidth]{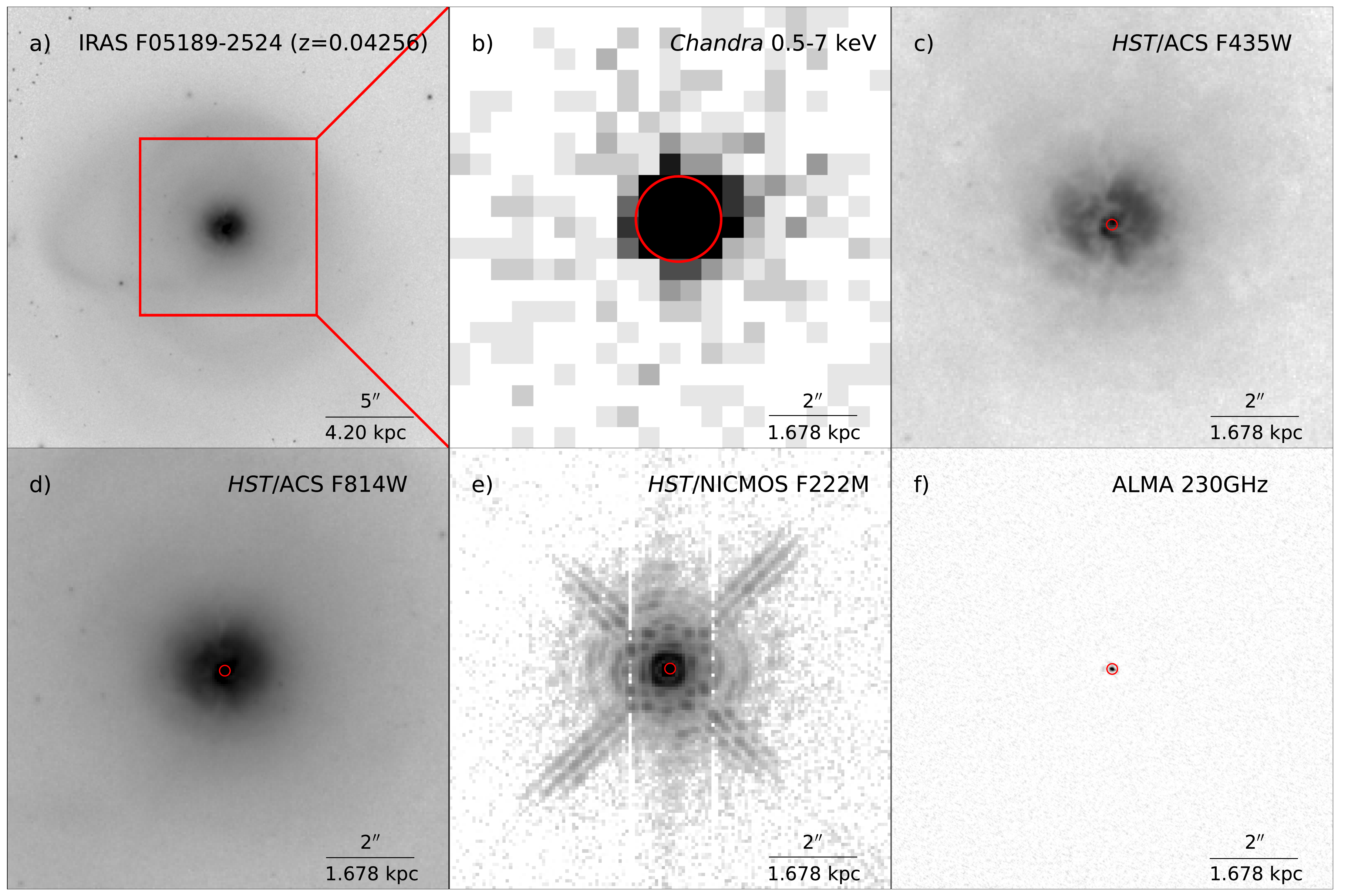}
        \includegraphics[width=0.6\textwidth]{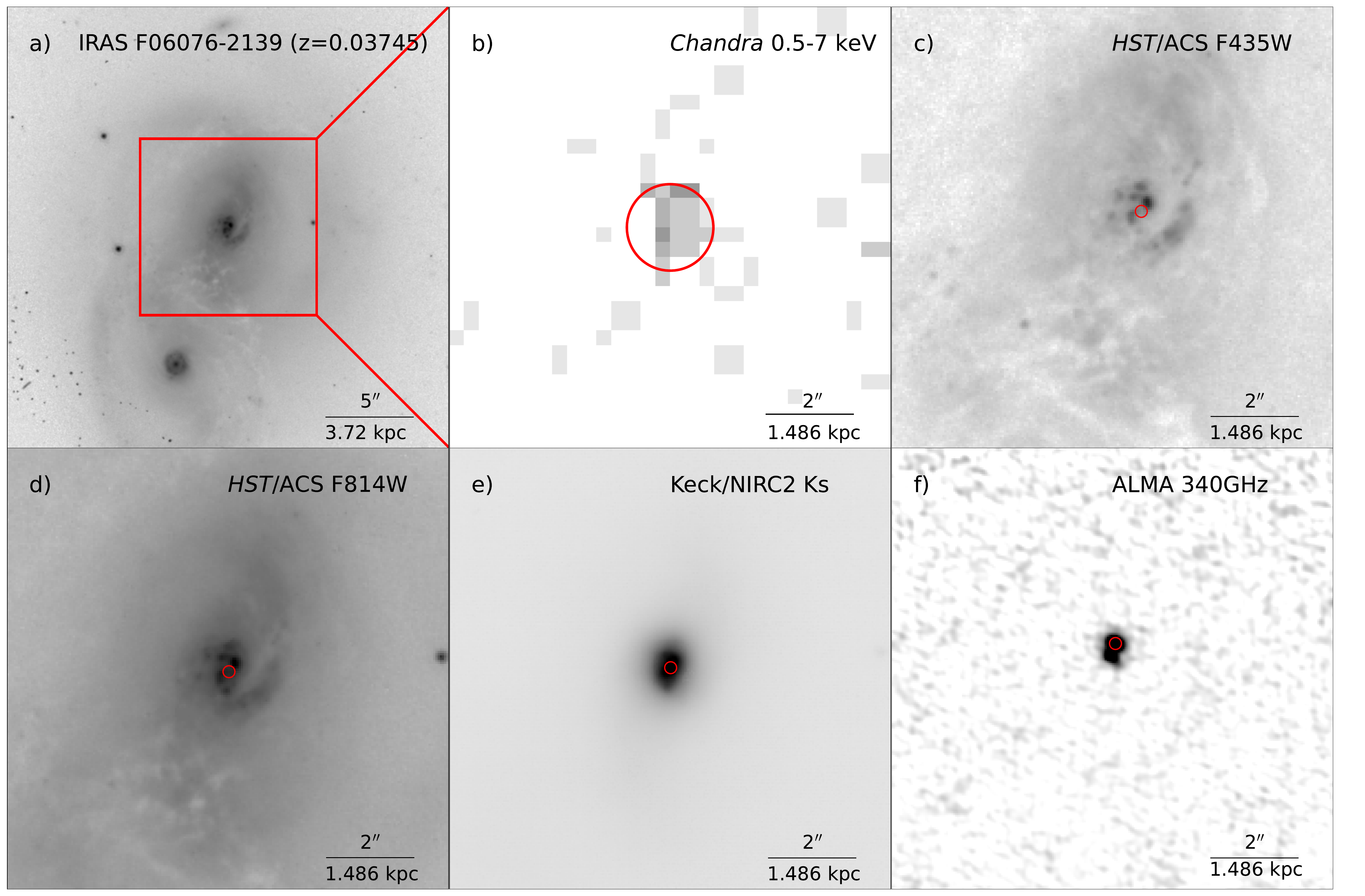}
        \includegraphics[width=0.6\textwidth]{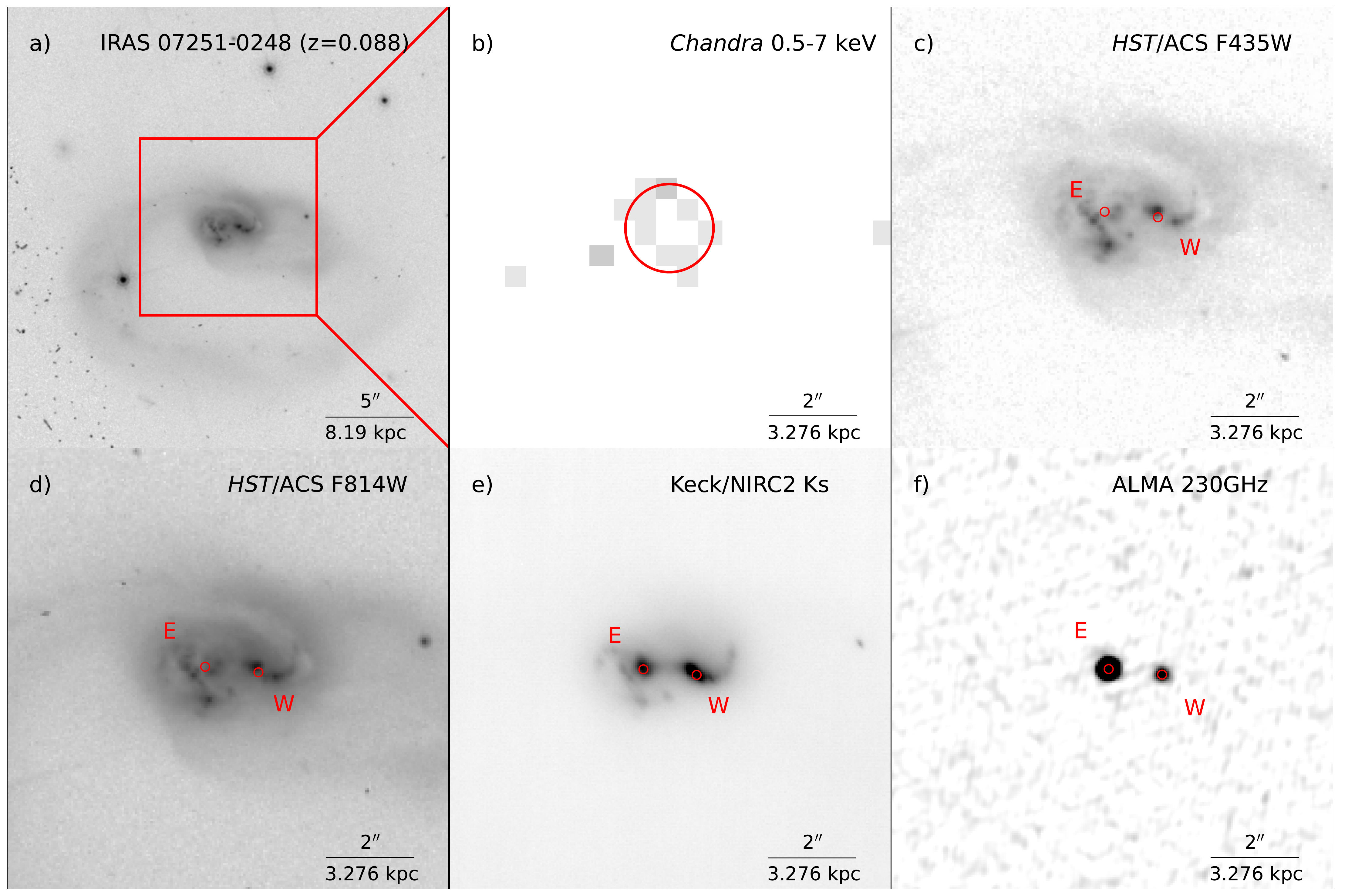}
	\caption{Continued}
	\label{fig:append_photometry3}
\end{figure*}

\addtocounter{figure}{-1}

\begin{figure*}[htbp]
	\centering
        \includegraphics[width=0.6\textwidth]{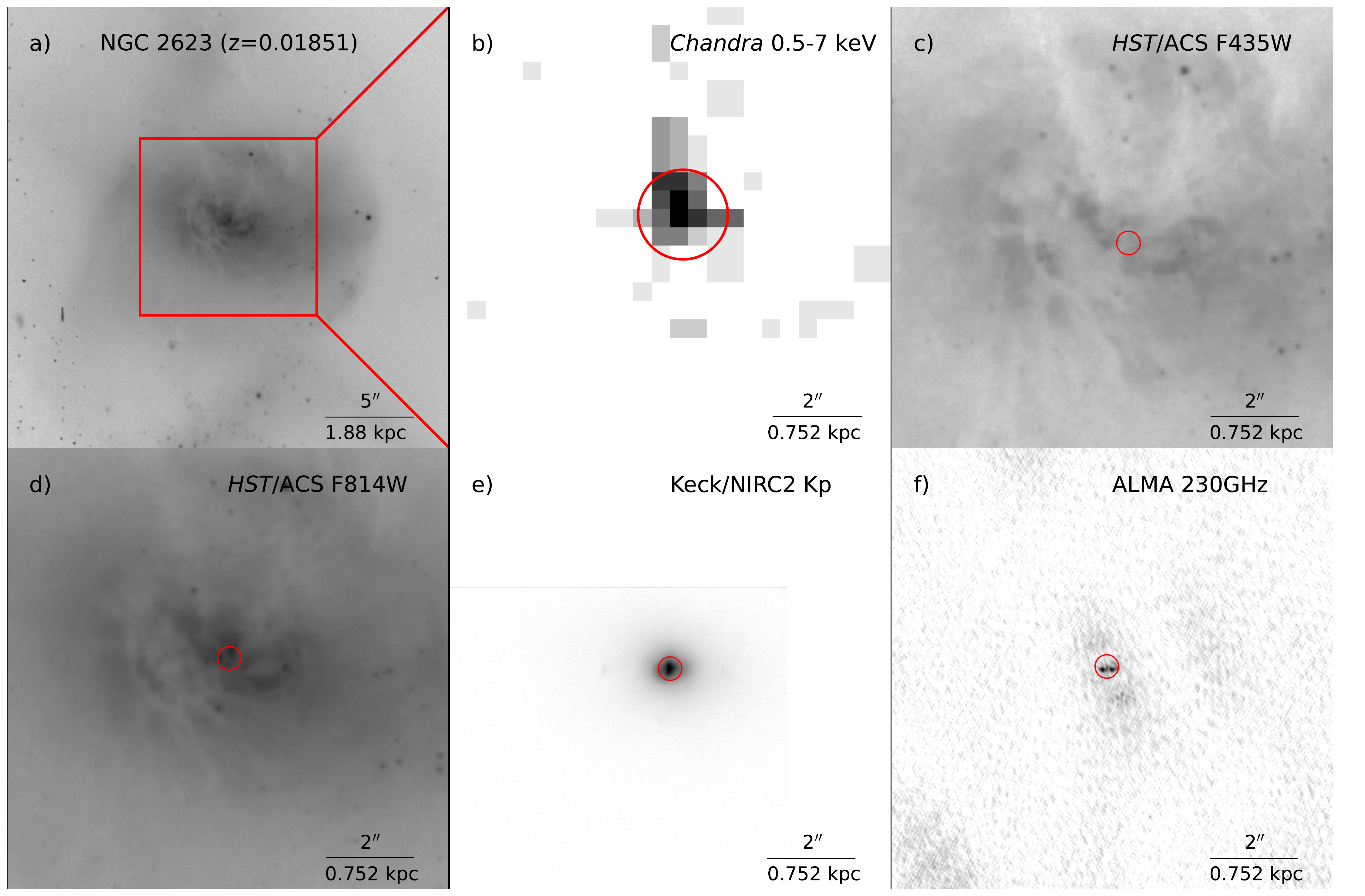}
        \includegraphics[width=0.6\textwidth]{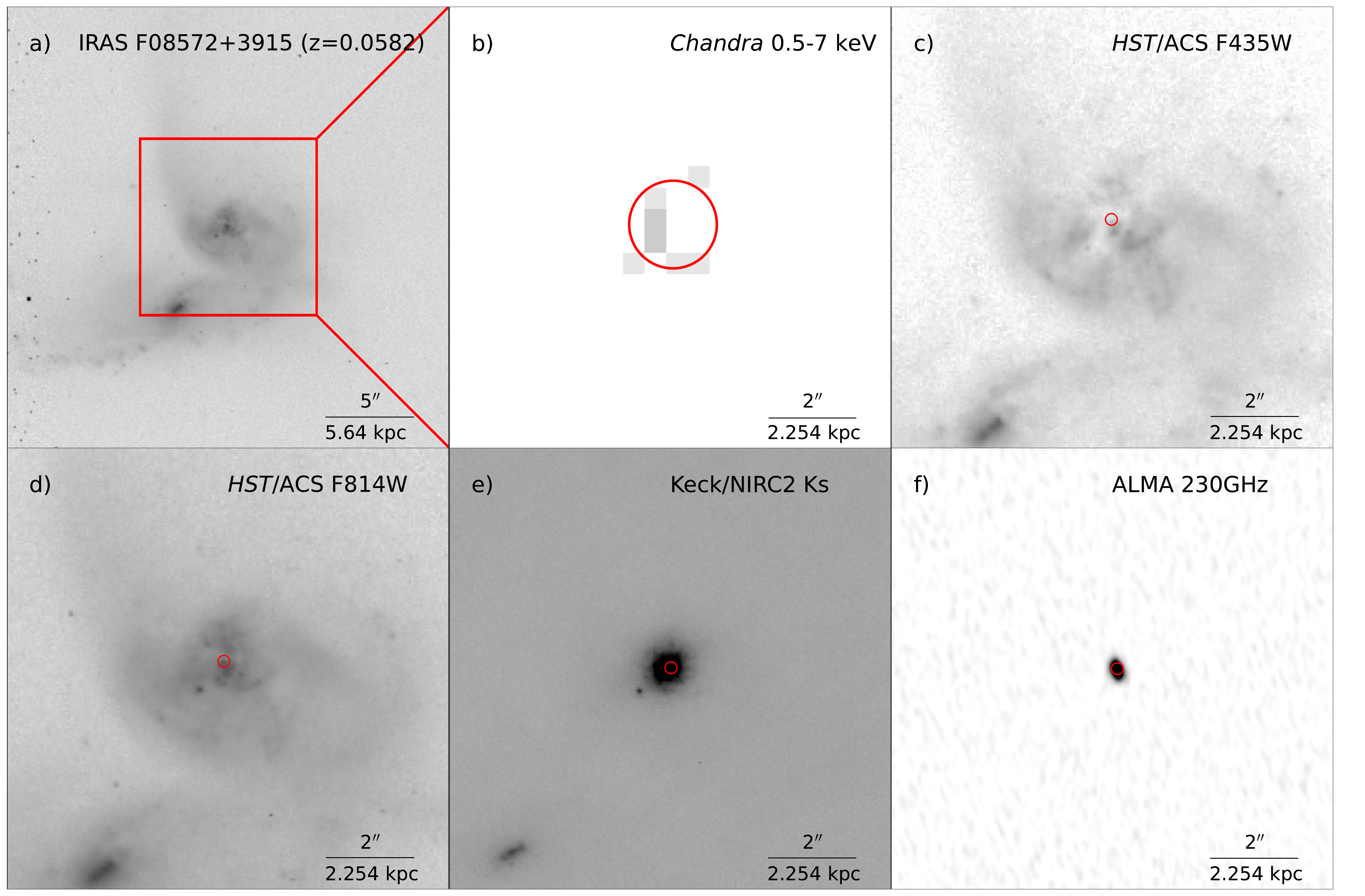}
        \includegraphics[width=0.6\textwidth]{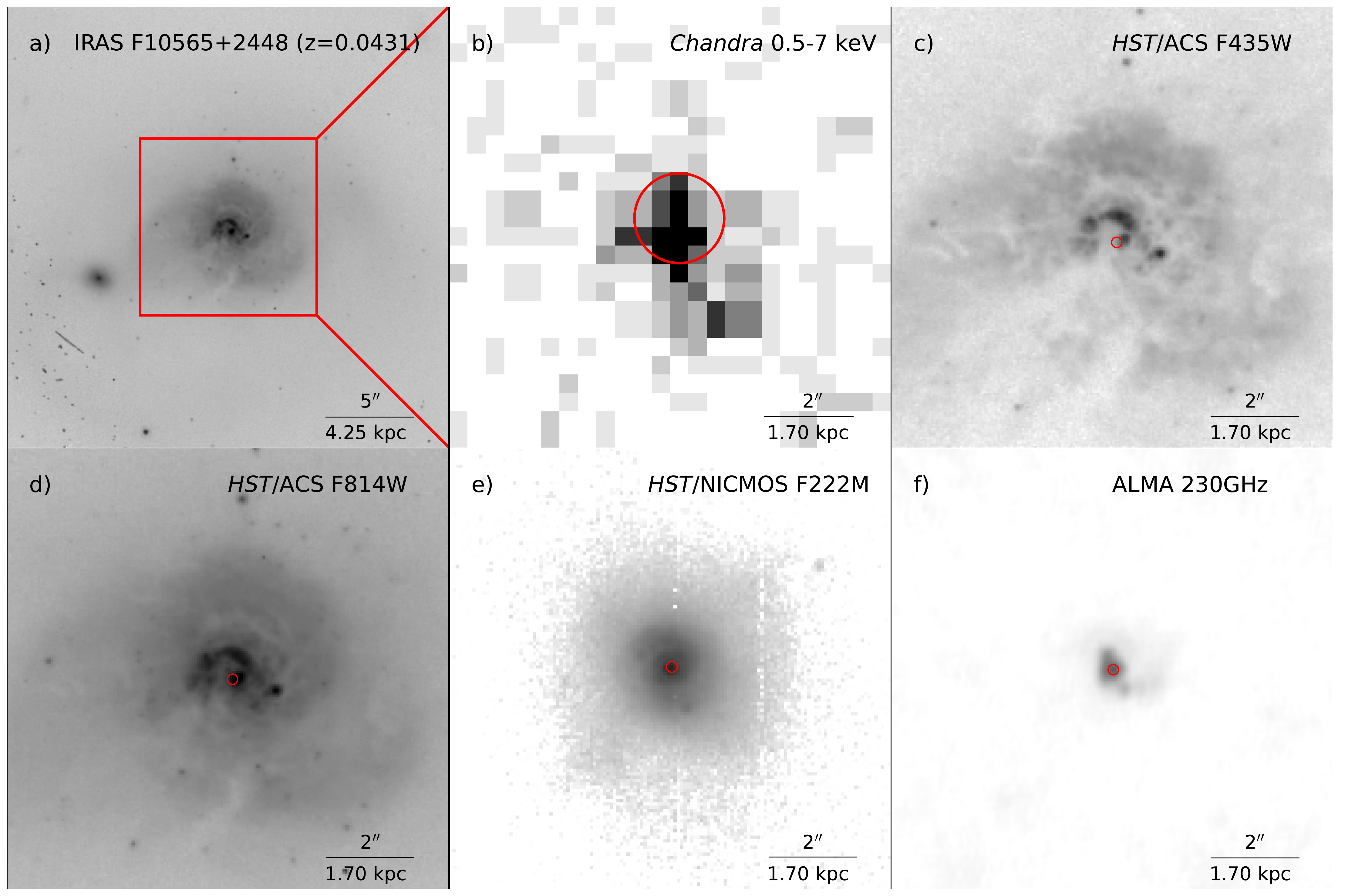}
	\caption{Continued}
	\label{fig:append_photometry4}
\end{figure*}

\addtocounter{figure}{-1}

\begin{figure*}[htbp]
	\centering
        \includegraphics[width=0.6\textwidth]{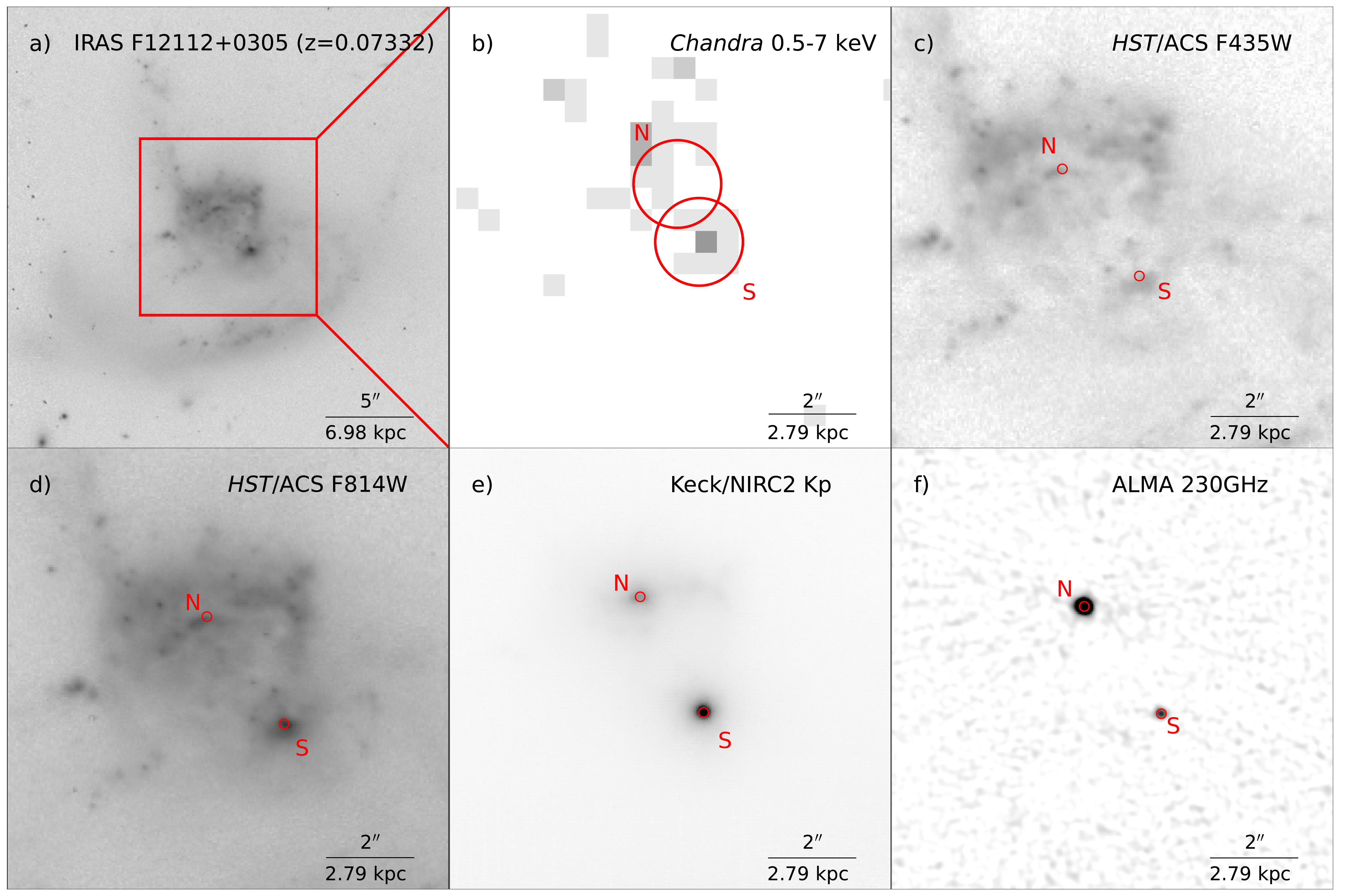}
        \includegraphics[width=0.6\textwidth]{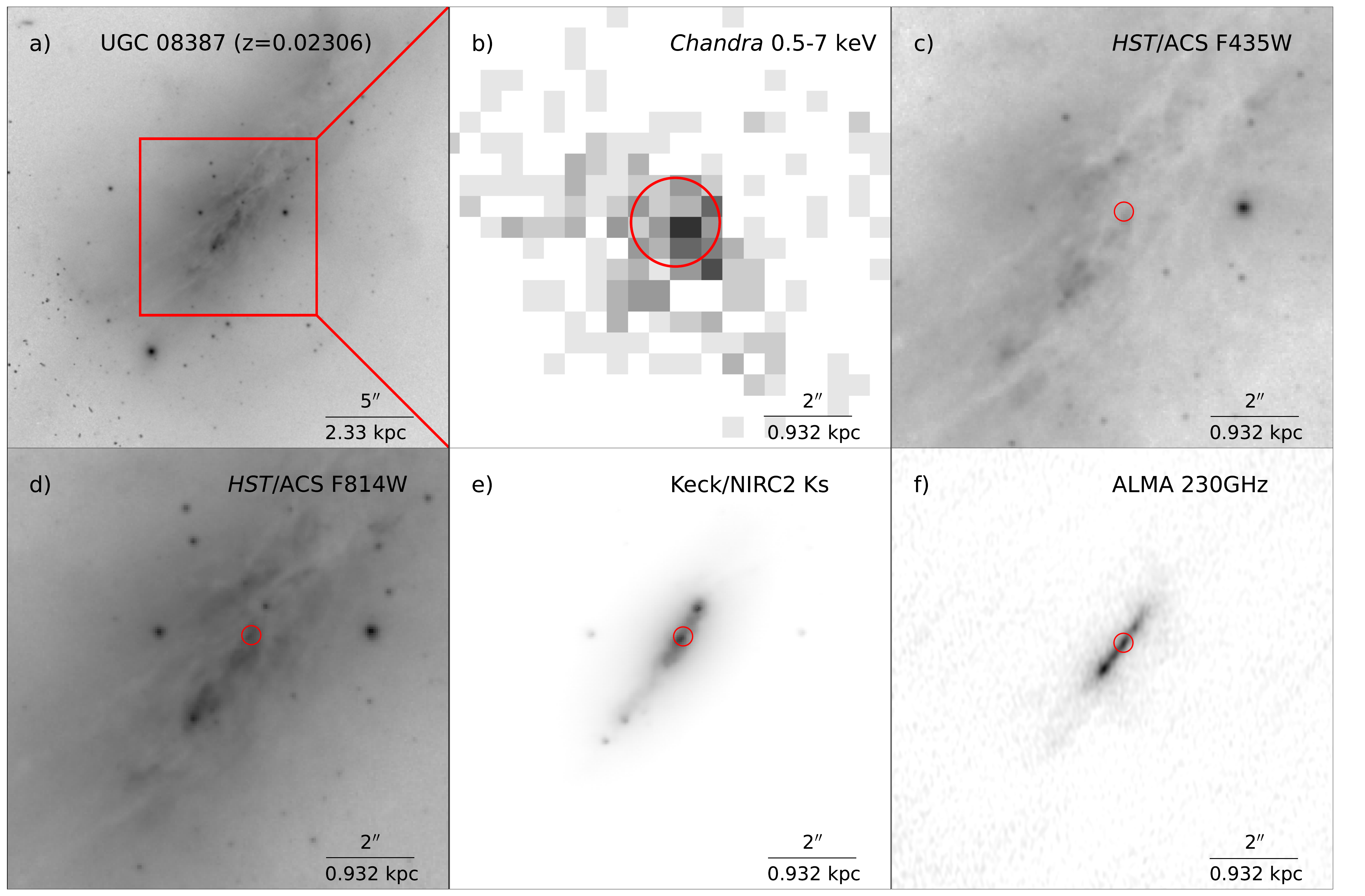}
        \includegraphics[width=0.6\textwidth]{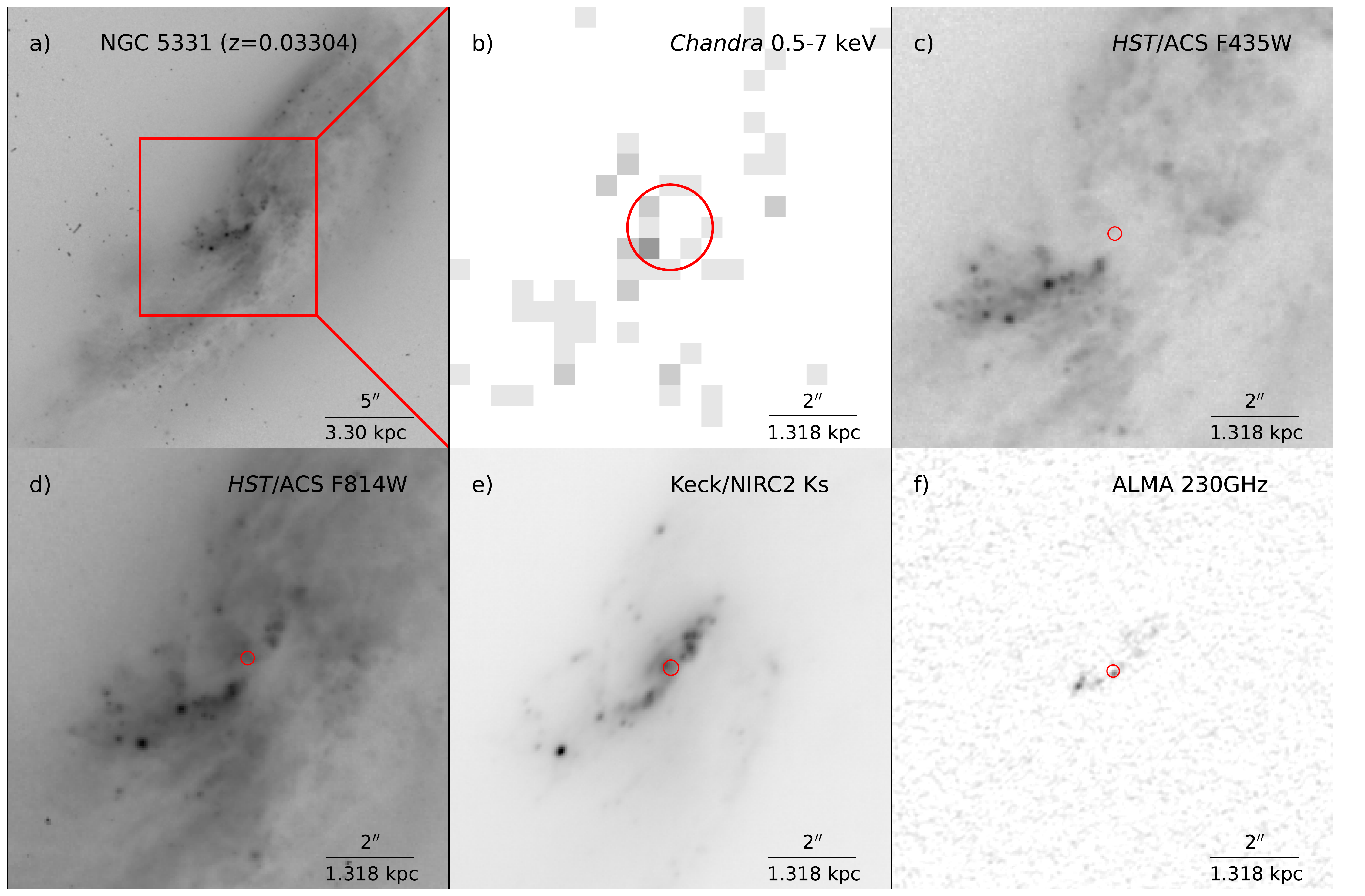}
	\caption{Continued}
	\label{fig:append_photometry5}
\end{figure*}

\addtocounter{figure}{-1}

\begin{figure*}[htbp]
	\centering
        \includegraphics[width=0.6\textwidth]{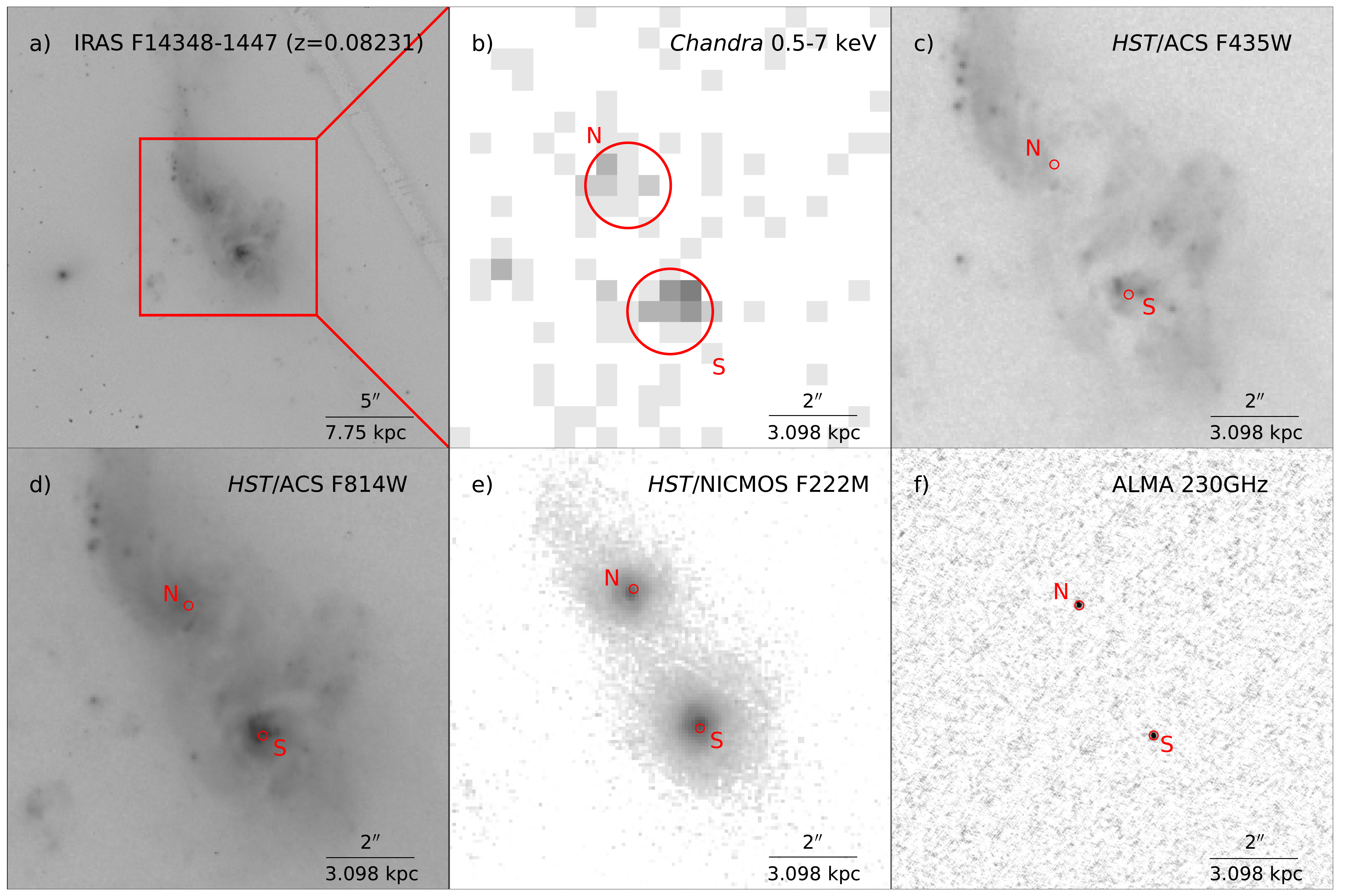}
        \includegraphics[width=0.6\textwidth]{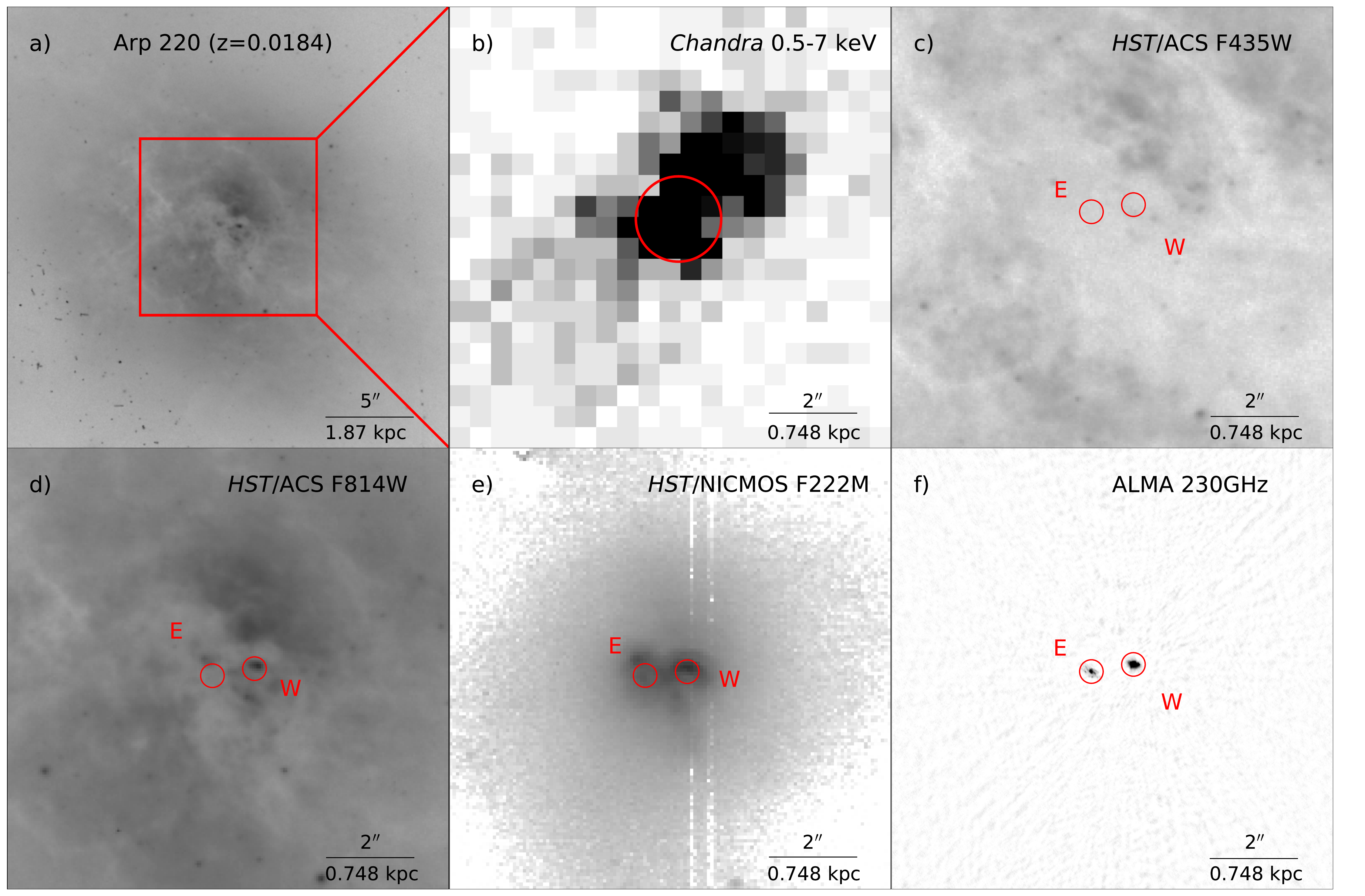}
        \includegraphics[width=0.6\textwidth]{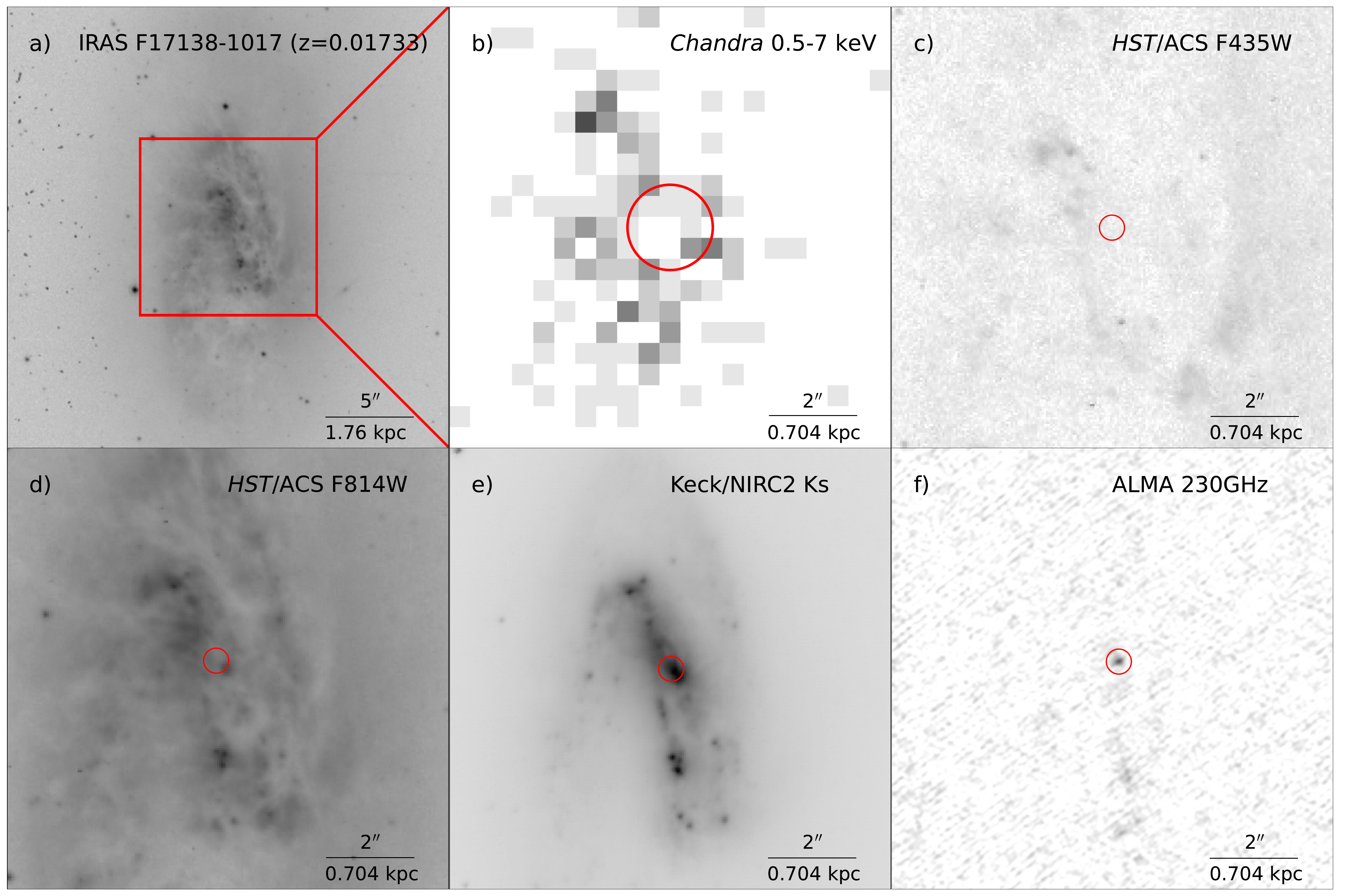}
	\caption{Continued}
	\label{fig:append_photometry6}
\end{figure*}

\addtocounter{figure}{-1}

\begin{figure*}[htbp]
	\centering
        \includegraphics[width=0.6\textwidth]{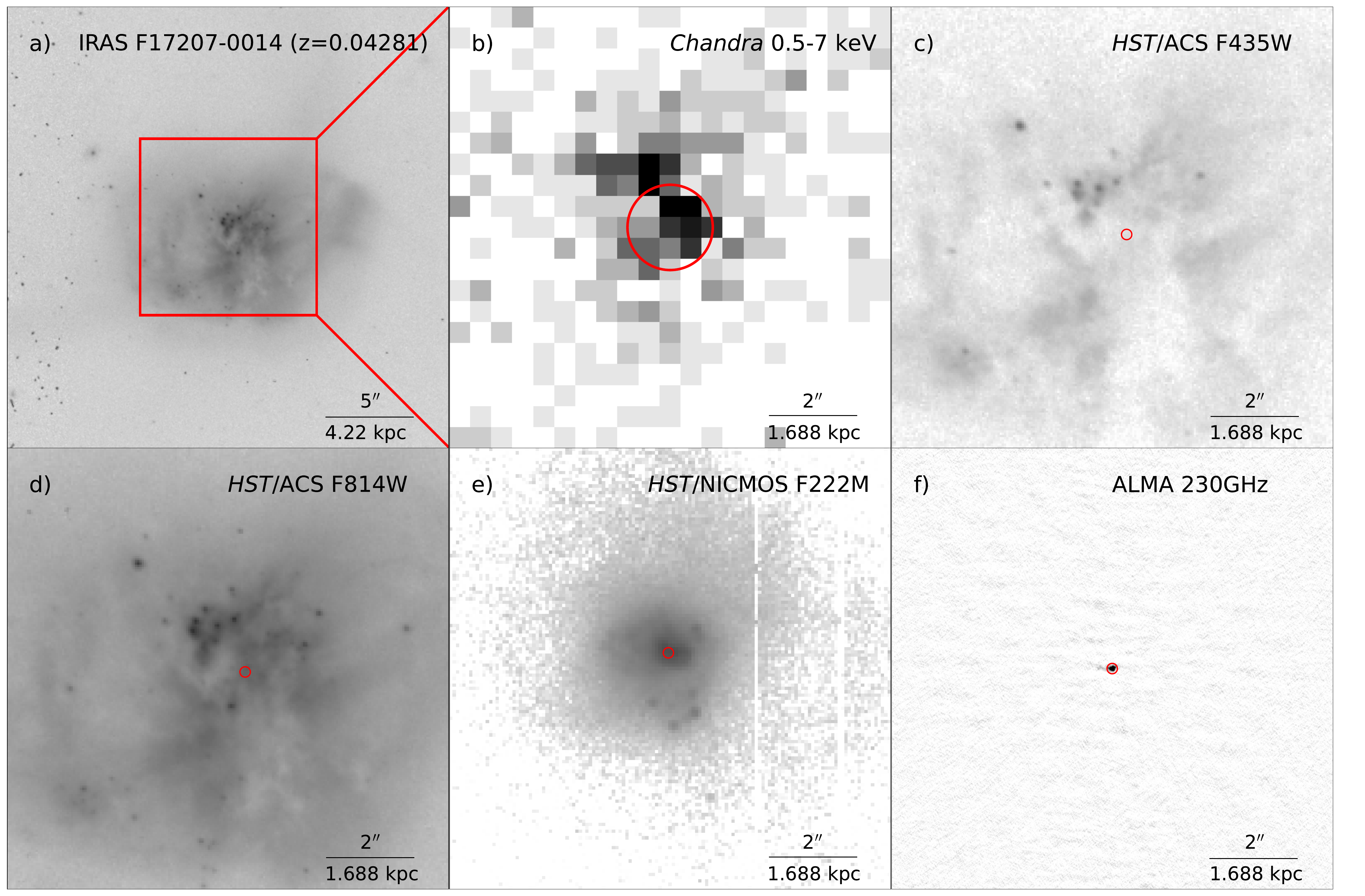}
        \includegraphics[width=0.6\textwidth]{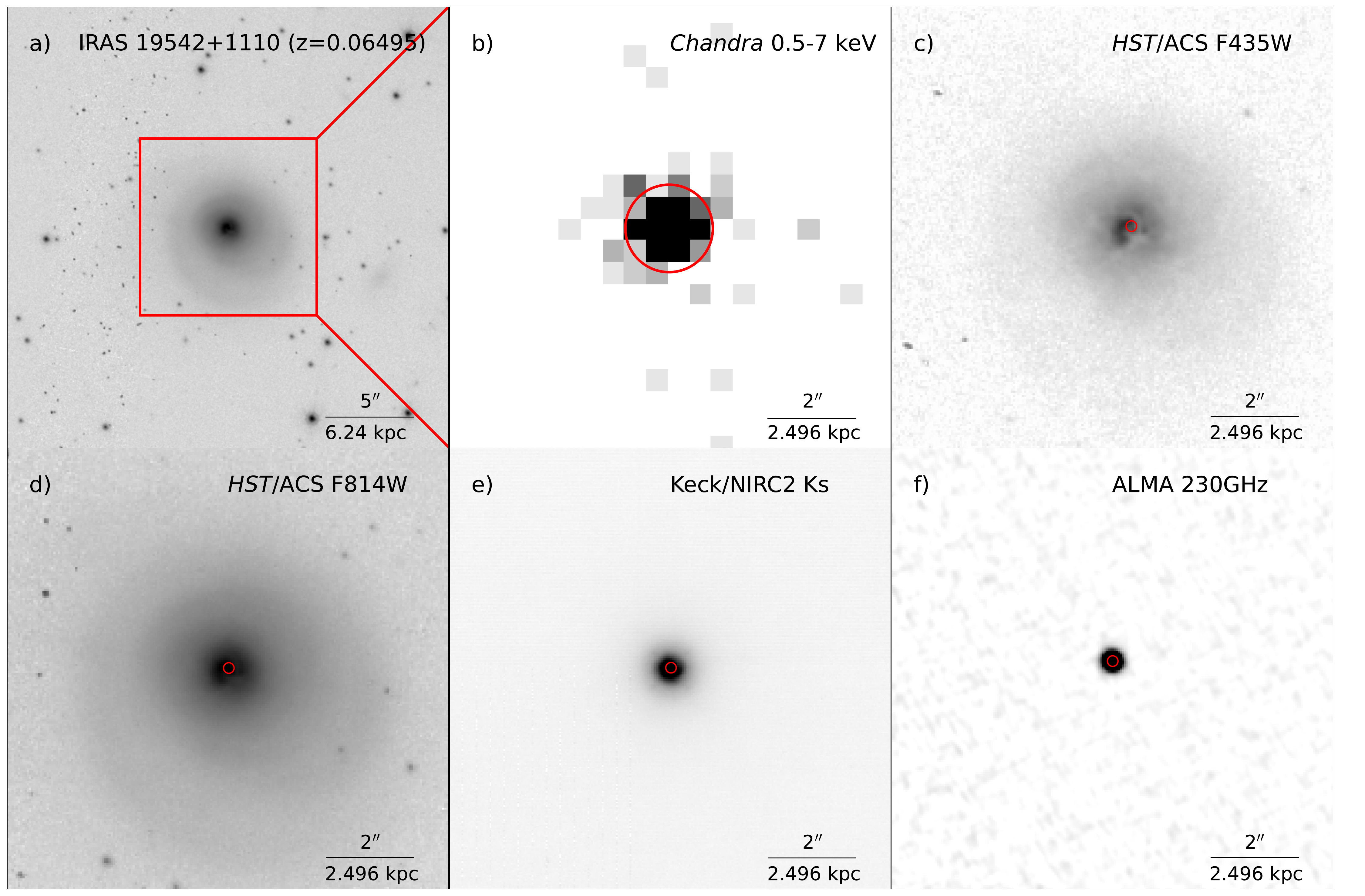}
        \includegraphics[width=0.6\textwidth]{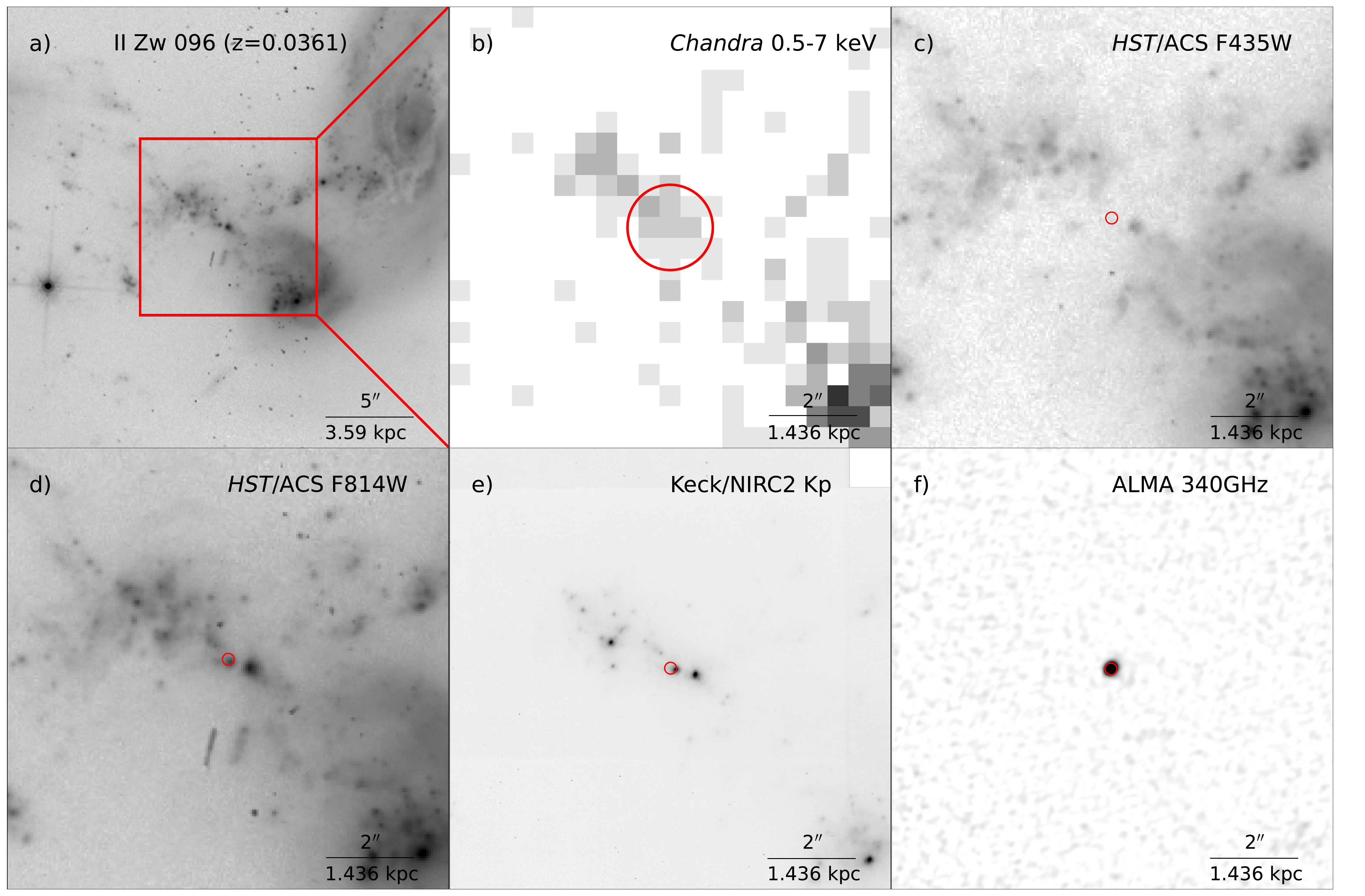}
	\caption{Continued}
	\label{fig:append_photometry7}
\end{figure*}

\addtocounter{figure}{-1}

\begin{figure*}[htbp]
	\centering
        \includegraphics[width=0.6\textwidth]{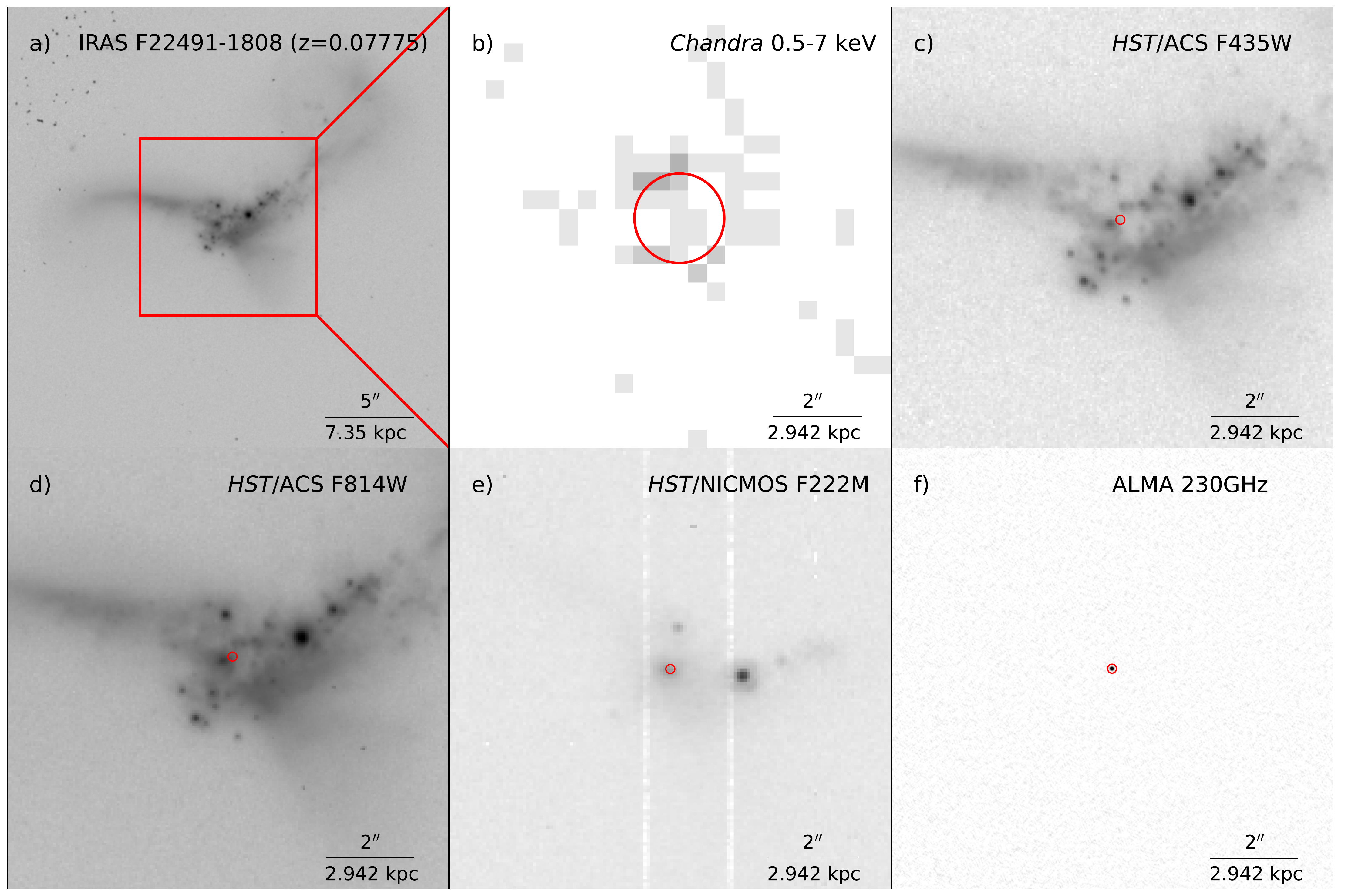}
	    \includegraphics[width=0.6\textwidth]{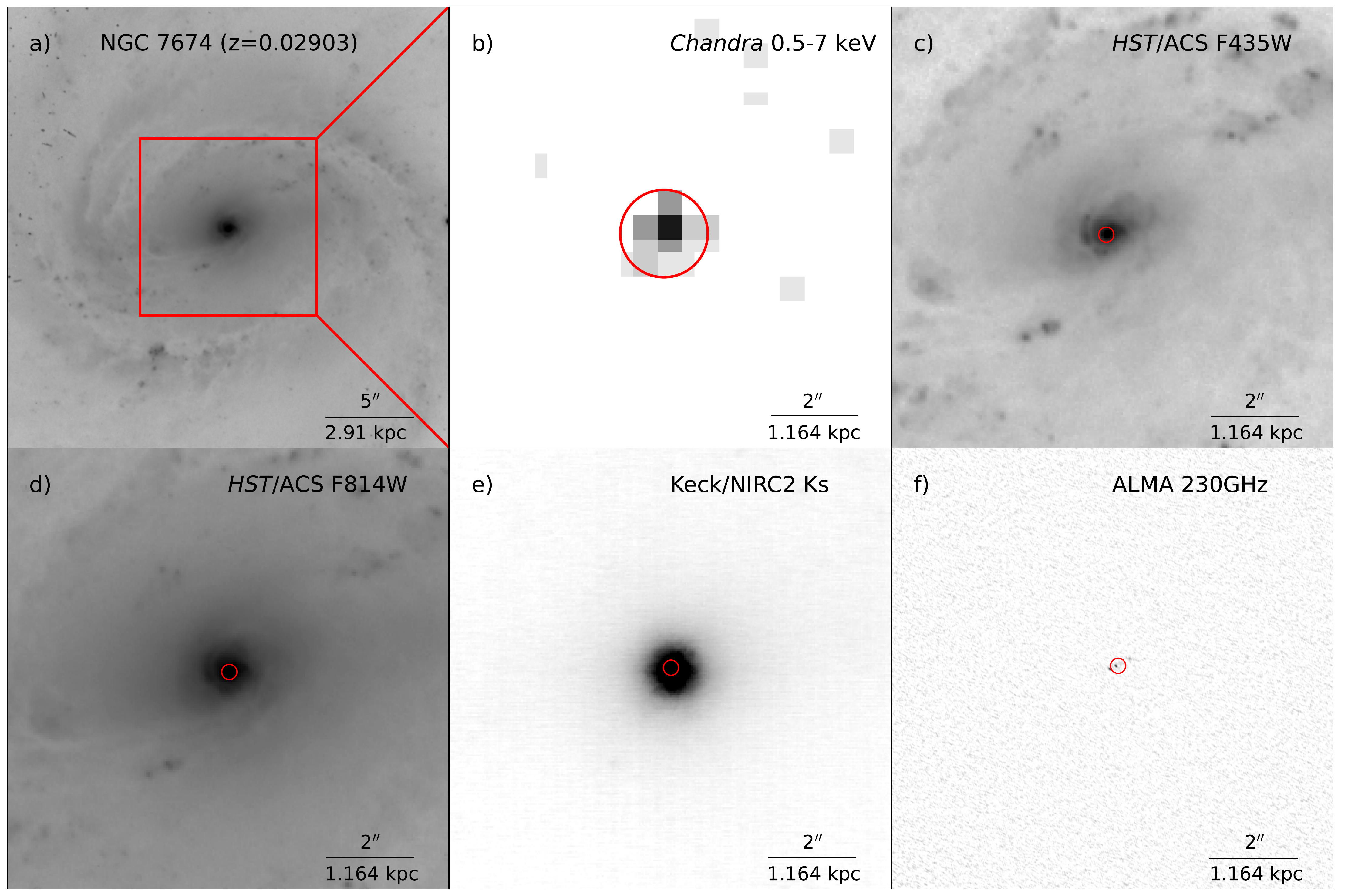}
	\caption{Continued}
	\label{fig:append_photometry8}
\end{figure*}

\FloatBarrier
\section{Best-fitting Results}
\label{sec:derived_properties}

Here we list the derived properties of each galaxy in our sample obtained through SED modeling.

\begin{deluxetable*}{lcccccc}
\tablenum{C.1}
\tablecaption{Derived Properties of the Nuclear Region(s) for Each Galaxy\label{tab:derived-properties}}
\tabletypesize{\footnotesize}
\tablewidth{0pt}
\tablehead{
\colhead{Galaxy} & \colhead{SFR($M_{\odot}\,\rm yr^{-1}$)} & \colhead{$M_{\rm star}(10^{8}\,M_{\odot})$} &
\colhead{$M_{\rm star\_old}(10^{8}\,M_{\odot})$} & \colhead{$M_{\rm star\_young}(10^{8}\,M_{\odot})$} & \colhead{$L_{\rm dust}(10^{10}\,L_{\odot})$} & \colhead{$L_{\rm AGN}(10^{10}\,L_{\odot})$}}
\startdata
CGCG 436-030 & $124.0\pm33.1$ & $10.7\pm5.19$ & $9.43\pm5.38$ & $1.27\pm0.30$ & $7.43\pm1.63$ & $0.42\pm0.30$ \\
IRAS F01364-1042 & $ 138.1\pm12.4$ & $8.22\pm2.89$ & $6.84\pm2.95$ & $1.38\pm0.12$ & $7.97\pm0.69$ & $0.43\pm0.22$ \\
III Zw 035N & $208.8\pm56.6$ & $7.92\pm3.60$ & $5.49\pm3.62$ & $2.43\pm0.15$ & $14.06\pm0.91$ & $0.00\pm0.00$ \\
NGC 0695 & $0.30\pm0.38$ & $15.5\pm8.88$ & $15.44\pm8.87$ & $0.03\pm0.02$ & $0.28\pm0.17$ & $0.43\pm0.10$ \\
NGC 1614 & $0.07\pm0.01$ & $110.7\pm5.53^{(a)}$ & $110.6\pm5.53$ & $0.01\pm0.01$ & $1.46\pm0.07$ & $0.00\pm0.00$ \\
IRAS F05189-2524 & $20.2\pm31.3$ & $0.84\pm1.92^{(a)}$ & $0.60\pm1.71$ & $0.24\pm0.30$ & $1.25\pm1.69$ & $3.46\pm4.76$ \\
IRAS F6076-2139N & $106.4\pm86.0$ & $7.15\pm4.59$ & $5.61\pm4.62$ & $1.54\pm0.81$ & $8.76\pm4.44$ & $0.12\pm0.06$ \\
IRAS 07251-0248E & $154.8\pm21.3$ & $2.17\pm0.93$ & $0.60\pm0.93$ & $1.57\pm0.13$ & $9.01\pm0.74$ & $0.81\pm1.51$ \\
IRAS 07251-0248W & $28.2\pm35.7$ & $1.52\pm0.82$ & $1.18\pm0.94$ & $0.35\pm0.34$ & $1.95\pm1.91$ & $14.99\pm9.51$ \\
NGC 2623 & $2.16\pm0.77$ & $137.7\pm66.0$ & $137.5\pm66.0$ & $0.21\pm0.08$ & $2.44\pm0.51$ & $0.07\pm0.01$ \\
IRAS F08572+3915NW & $0.56\pm0.57$ & $140.5\pm60.9$ & $140.5\pm60.9$ & $0.05\pm0.05$ & $1.31\pm0.32$ & $0.04\pm0.01$ \\
IRAS F10565+2448 & $ 1.62\pm1.85$ & $234.6\pm49.9^{(a)}$ & $234.5\pm49.9$ & $0.16\pm0.18$ & $2.19\pm1.67$ & $0.17\pm0.13$ \\
IRAS F12112+0305N & $38.2\pm28.3$ & $1.73\pm1.39$ & $0.94\pm1.38$ & $0.79\pm0.06$ & $4.39\pm0.24$ & $0.13\pm0.01$ \\
IRAS F12112+0305S & $5.52\pm2.79$ & $390.2\pm147.1^{(a)}$ & $389.6\pm147.4$ & $0.54\pm0.27$ & $6.32\pm2.25$ & $0.09\pm0.01$ \\
UGC 08387 & $29.1\pm29.5$ & $7.12\pm1.08$ & $6.32\pm1.25$ & $0.80\pm0.30$ & $4.79\pm1.59$ & $0.06\pm0.01$ \\
NGC 5331S & $0.58\pm0.87$ & $25.3\pm16.7$ & $25.2\pm16.7$ & $0.06\pm0.08$ & $0.63\pm0.55$ & $0.02\pm0.01$ \\
IRAS F14348-1447N & $33.4\pm28.6$ & $6.11\pm3.48$ & $5.09\pm3.18$ & $1.02\pm0.54$ & $5.76\pm2.90$ & $1.97\pm1.23$ \\
IRAS F14348-1447S & $60.3\pm52.7$ & $11.5\pm7.02$ & $9.21\pm6.99$ & $2.27\pm0.54$ & $12.5\pm2.90$ & $5.40\pm3.67$ \\
Arp 220E & $73.5\pm3.70$ & $2.19\pm1.61$ & $1.45\pm1.64$ & $0.74\pm0.04$ & $4.23\pm0.21$ & $0.06\pm0.05$ \\
Arp 220W & $38.8\pm4.66$ & $6.55\pm4.42$ & $4.53\pm4.48$ & $2.01\pm0.20$ & $11.0\pm1.02$ & $0.22\pm0.17$ \\
NGC 6240N & $1.18\pm0.71$ & $21.2\pm7.50$ & $21.1\pm7.53$ & $0.12\pm0.07$ & $1.22\pm0.41$ & $0.77\pm0.43$ \\
NGC 6240S & $8.25\pm1.66$ & $99.0\pm18.8$ & $98.2\pm18.9$ & $0.80\pm0.16$ & $7.05\pm0.77$ & $0.75\pm0.55$ \\
IRAS F17138-1017 & $0.17\pm0.21$ & $13.2\pm3.67$ & $13.2\pm3.67$ & $0.02\pm0.02$ & $0.23\pm0.13$ & $0.02\pm0.04$ \\
IRAS F17207-0014 & $69.8\pm35.9$ & $5.45\pm2.96$ & $4.24\pm2.98$ & $1.22\pm0.29$ & $6.87\pm1.48$ & $0.07\pm0.09$ \\
IRAS 19542+1110 & $33.7\pm6.30$ & $ 310.4\pm142.1^{(a)}$ & $307.1\pm142.5$ & $3.27\pm0.58$ & $27.0\pm3.02$ & $2.82\pm0.21$ \\
II Zw 096E & $39.6\pm14.4$ & $0.71\pm0.29$ & $0.22\pm0.29$ & $0.48\pm0.04$ & $2.73\pm0.19$ & $0.23\pm0.19$ \\
IRAS F22491-1808 & $15.7\pm1.37$ & $1.42\pm0.80$ & $0.60\pm0.81$ & $0.81\pm0.05$ & $4.41\pm0.26$ & $0.06\pm0.07$ \\
NGC 7674 & $7.99\pm2.83$ & $2.58\pm1.82$ & $2.15\pm1.85$ & $0.44\pm0.08$ & $2.27\pm0.39$ & $2.58\pm2.25$ \\
\enddata
\tablecomments{We note that caution should be taken when interpreting these values, as their reliability is primarily limited by the lack of high-resolution mid- to far-infrared data, as indicated in Appendix \ref{sec:appendix_mock_analysis}. $^{(a)}$The reliability of these values is further affected by the fact that the near-infrared emission is dominated by the PSF, i.e., the nuclear region cannot be resolved.}

\end{deluxetable*}

\section{Mock Analysis}\label{sec:appendix_mock_analysis}

\setcounter{figure}{0}

\edit1{Here, we use a 'mock analysis' to assess the reliability of the physical properties of nuclear regions obtained through SED fitting. X-CIGALE generates a series of mock photometric data points by adding Gaussian noise to each best-fit flux, with the noise standard deviation matching that of its corresponding observed flux. These mock data are then analyzed using the same approach as the observed data. By comparing the previously obtained best-fit results with the output estimates from the mock analysis, we can assess the accuracy of these derived properties.}

\begin{figure*}[htbp]
	\centering
        \includegraphics[width=0.9\textwidth]{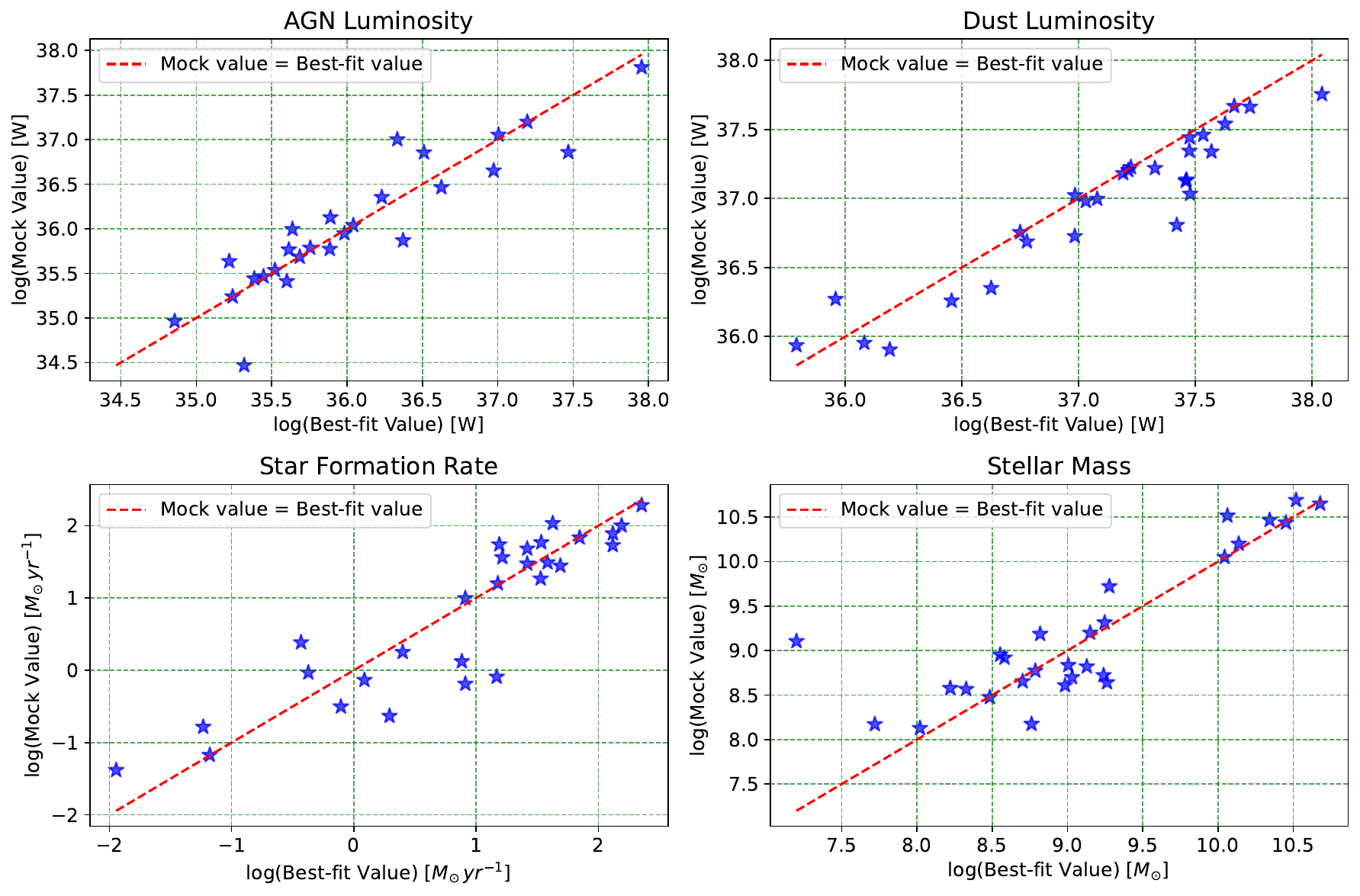}
	\caption{Comparison of the best-fit results and those derived from the mock analysis for AGN luminosity (upper left), dust luminosity (upper right), star formation rate (lower left), and stellar mass (lower right).}
	\label{fig:mock_analysis}
\end{figure*}

\edit1{Figure \ref{fig:mock_analysis} presents the results of the mock analysis for AGN luminosity, dust luminosity, SFR and stellar mass, comparing the best-fit values with the mock values. We note that although the mean differences between the best-fit values and the mock values are small ($\Delta\,L_{\mathrm{AGN}}=-0.01\pm0.31$, $\Delta\,L_{\mathrm{dust}}=-0.13\pm0.18$, $\Delta\,\mathrm{SFR}=-0.08\pm0.49$, $\Delta\,M_{*}=-0.09\pm0.46$), the mock values typically exhibit uncertainties comparable to their own magnitudes (for clarity, we have omitted the error bars of the mock values). Therefore, we conclude that the absence of mid- and far-infrared data points has indeed significantly impacted the accuracy of the SED fitting results.}

\end{document}